\documentclass[a4paper,11pt]{article}
\pdfoutput=1 % if your are submitting a pdflatex (i.e. if you have
\usepackage{jcappub} % for details on the use of the package, please
\usepackage[T1]{fontenc} % if needed
\usepackage{hyperref}
\usepackage{booktabs}
\PassOptionsToPackage{unicode}{hyperref}
\newcommand{\planck}{{\it{Planck }}\,}
\newcommand{\unwise}{{\textit{unWISE}}\,}
\newcommand{\cell}{{C_{\ell}}}

\newcommand{\mupeak}{\mu_{\mathrm{peak}, 0}}
\newcommand{\mupeakp}{\mu_{\mathrm{peak}, p}}

\newcommand{\dr}{{\mathrm{d}}}

\usepackage{tablefootnote}

\usepackage[caption=false]{subfig}

\title{\boldmath The star formation, dust, and abundance of galaxies with {\textit{unWISE}}-CIB cross-correlations}

\author[a]{Ziang Yan,}
\author[b,c,d]{Abhishek S. Maniyar,}
\author[e]{Ludovic van Waerbeke}

\affiliation[a]{Ruhr University Bochum, Faculty of Physics and Astronomy, Astronomical Institute (AIRUB), German Centre for Cosmological Lensing, 44780 Bochum, Germany}
\affiliation[b]{Department of Physics, Stanford University, 382 Via Pueblo Mall, Stanford, CA 94305, USA}
\affiliation[c]{SLAC National Accelerator Laboratory, 2575 Sand Hill Rd, Menlo Park, CA 94025, USA}
\affiliation[d]{Kavli Institute for Particle Astrophysics and Cosmology, 382 Via Pueblo Mall Stanford, CA  94305-4060, USA}
\affiliation[e]{Department of Physics and Astronomy, University of British Columbia, 6224 Agricultural Road, Vancouver, BC, V6T 1Z1, Canada}

\emailAdd{yanza21@astro.rub.de}
\abstract{The cosmic infrared background (CIB) is the accumulated infrared (IR) radiation mainly from interstellar dust heated up by early stars. In this work, we measure the cross-correlation between galaxies from the \unwise catalog and the CIB maps from the \planck satellite to simultaneously constrain the cosmic star formation rate (SFR), dust spectral energy distribution (SED), and the halo occupation distribution (HOD). The \unwise galaxy catalog is divided into three tomographic bins centered at $z\sim 0.6, 1.1, 1.5$, and the CIB maps are at 353, 545, and 857 GHz. We measure the cross-correlations between these galaxy samples and CIB maps and get a 194$\sigma$ signal within an angular scale $100<\ell<2000$, from which we constrain two CIB halo models from previous literature and one new model. The SFR, SED, and HOD model parameters are constrained consistently among the three models. Specifically, the dust temperature at $z=0$ is constrained {$T_0={21.14}^{+1.02}_{-1.34}$ K, which is slightly lower than $T_0=24.4\pm1.9$ K measured by the \planck collaboration.} The halo mass that gives the most efficient star formation is around $10^{{11.79}^{+0.73}_{-0.86}}M_{\odot}$. From the model parameters, combined with the SFR density at $z=0$ synthesized from multi-wavelength observations, we break the degeneracy between SED and SFR and recover the cosmic star formation history that is consistent with multi-wavelength surveys. We also constrain the graybody SED model in agreement with previous measurements from infrared flux stacking. From the HOD constraints, we derive an increasing trend of galaxy linear bias along redshifts that agrees with the results from cross- and auto-correlation with \unwise galaxies. This study indicates the power of using CIB-galaxy cross-correlation to study star formation, dust, and abundance of galaxies across cosmic time.}

\begin{document}
\maketitle
\flushbottom

\section{Introduction}

Star formation in galaxies encodes essential information about the evolution of galaxies in the Universe \citep[][]{1980FCPh....5..287T} and complex interactions within galaxies between gases, stars, and central black holes (for example, through feedback from supernovae or supermassive black holes). It is typically studied via multi-wavelength observations \citep{2013ruppioni, 2013magnelli,2016davies, 2016MNRAS.456.1999M} on the extragalactic emission over a wide range of wavelengths. These works derive the star formation rate (SFR) of galaxy populations by assuming the luminosity functions and the luminosity-{SFR relations} at different wavelengths from ultraviolet (UV) to far-infrared (FIR). From these studies, we learn that star formation in the Universe started between $6\lesssim z \lesssim 20$, then reached the peak at $z\sim 2$ (corresponding to a lookback time of $\sim 10.3$\, Gyr), at a rate approximately ten times higher than is observed today. Due to a subsequent deficiency of available gas fuel, star formation activity has been steadily decreasing since $z\sim 2$ (see \cite{madau_cosmic_2014} for a review).

Among multi-wavelength studies of cosmic star formation history, observations of extragalactic infrared emission have explored SFR in different redshifts \cite{2013ruppioni, 2013magnelli, 2016MNRAS.456.1999M, 2016davies}. IR emissions are mainly from interstellar dust in star-forming galaxies that are heated up by the ultraviolet (UV) emissions generated by short-lived massive stars\citep{kennicutt1998star, 2007ApJ...656..770S,2009sfrirg}. Moreover, one can also use the spectral energy distribution (SED) of infrared galaxies to study the thermodynamics of interstellar dust in these galaxies \citep{bethermin_evolution_2015, bethermin_impact_2017}. However, dusty star-forming galaxies beyond the local universe are typically highly blended, given the sensitivity and angular resolution of modern infrared observatories, because they are both faint and numerous. {This makes them more difficult to study individually as direct studies of the SFR at higher redshift. \citep[see, for example][]{Nguyen_2010}} \footnote{{On-going and} next-generation IR observatories, such as the James Webb Space Telescope, may make this direct probe more valuable, given their higher angular resolution.}. In addition, individual galaxy-based multi-wavelength studies could be biased due to selection effect and survey incompleteness. 

Therefore, the projected all-sky IR emission, known as the cosmic infrared background \citep[CIB,][]{1967ApJ...147..868P} which encodes the cumulative emission of all dusty star-forming galaxies below $z\sim 6$, serves as an alternative tool to study dusty star-forming galaxies. However, measurements of the CIB itself are complicated: imperfect removal of point sources and foreground Galactic emission can lead to bias in the measured CIB signals. Nevertheless, CIB is more robust to the various selection effects and sample variance uncertainties induced by instrumental setup and observation strategies. The CIB was first detected by the Cosmic Background Explorer (\textit{COBE}) satellite \citep{1998cobecib} and was then accurately analyzed by \textit{Spitzer}\citep{2006spitzercib}, \textit{Herschel} \citep{2010A&A...518L..30B} {via large IR galaxy samples}, and \planck\,\citep{2014planckxxx} {via extracting the CIB from its sky maps}. As the cosmic microwave background (CMB), the key to {mapping} the CIB is accurately removing the foreground Galactic emission. {However, unlike the CMB, the CIB is dominated by the thermal emission from Galactic dust.} Moreover, the CIB has no unique frequency dependence so it is more difficult to be extracted from raw sky maps than the CMB or the thermal Sunyaev-Zel'dovich effect \cite{sunyaev1972observations} and is generally restricted to high Galactic latitude. There are multiple published CIB maps that are generated by various foreground-removing methods. \cite{2014planckxxx} and \cite{lenz_large-scale_2019} remove Galactic emission by introducing a Galactic template from HI measurements, while \cite{2016planckcib} disentangles the CIB from Galactic dust by the generalized needlet internal composition \citep[GNILC,][]{Remazeilles_2011} method. The mean CIB emission measured from maps gives the mean energy emitted from star formation activities, while the anisotropies of the CIB trace the spatial distribution of star-forming galaxies and the underlying distribution of their host dark matter halos (given some galaxy bias). Therefore, analyzing the CIB anisotropies via angular power spectra and cross-correlations with other large-scale structure tracers \citep{2020caoye, maniyar_simple_2021, Yan_2022, Jego_2023, Jego_2023_2} has been proposed as an alternative method to probe cosmic star formation activity \cite{2006spitzercib, shang_improved_2012, 2014planckxxx, Serra_2014, maniyar_star_2018}. 

Due to the diversity of dust properties and star formation activities across galaxies, there is no ``standard model'' for the CIB. Nevertheless, multi-wavelength observations have summarized empirical CIB models. In general, IR luminosity is connected with SFR via the Kennicutt relation \cite{kennicutt1998star}. The CIB flux at a certain frequency is given by the spectral energy distribution (SED) which quantifies the quota of CIB flux with respect to the total CIB luminosity. The spatial distribution of CIB emissions traces the galaxy distribution. So there are three ingredients to describe the CIB. In the literature, those three ingredients are formulated according to different assumptions: 
\begin{itemize}

    \item the dust SED that describes the frequency dependence: \cite{shang_improved_2012, 2014planckxxx, 2020caoye, Serra_2014} assumed a gray-body SED and fit the dust temperature and spectral indices; \cite{maniyar_star_2018, maniyar_simple_2021,Yan_2022, Jego_2023} introduced SED obtained from multi-wavelength measurements \cite{bethermin_evolution_2015, bethermin_impact_2017};
    \item the SFR that determines the redshift dependence: \cite{2014schmidt, maniyar_star_2018, 2020caoye} assumes empirical models for SFR density (SFRD); \cite{shang_improved_2012} assumed a lognormal dependence of SFR on halo mass in the context of the halo model \cite{Seljak_2000, COORAY_2002} and a power-law dependence on $1+z$; \cite{maniyar_simple_2021} proposed a model connecting SFR with baryon accretion rate which is given by simulations and a slightly more complicated $\mathrm{SFR}-M$ connection. This model is also used in \cite{Yan_2022, Jego_2023, Jego_2023_2}.
    \item galaxy fluctuation: \cite{maniyar_star_2018, 2020caoye} assumed linear fluctuation model for CIB by introducing an effective bias parameter; \cite{shang_improved_2012, maniyar_simple_2021} and following works used the halo occupation distribution \cite[HOD]{Zheng_2005} (with slightly different implementations).
\end{itemize}

In all the models proposed/utilized by the studies listed above, the SFR and SED models are empirical and depend on external knowledge from multi-wavelength observations of galaxies. Moreover, they are statistically degenerate and cannot be constrained simultaneously with CIB anisotropies alone. Therefore, the studies mentioned above generally constrained one or two ingredients while fixing the rest. For example, \cite{shang_improved_2012} constraints dust temperature, power indices and halo mass with the most efficient SFR, but does not constrain the SFR history; \cite{Yan_2022} give a constraint on cosmic star formation history up to $z\sim 1.5$ as well as the HOD model, but fixes the dust SED with a template given by \cite{bethermin_impact_2017}. In this work, we will summarize previous CIB models and seek to constrain all three ingredients with CIB-galaxy cross-correlations and minimal introduction of external information, thus we showcase the great potential of the CIB anisotropies in comprehensively studying galaxy evolution, content, and clustering.

Compared to other large-scale structure tracers, galaxy number density fluctuations can be measured with high signal-to-noise. In addition, cross-correlating with galaxy samples in different redshifts (or ``tomographic cross-correlation'') can improve the precision of the evolution of certain tracers. {For such measurements, an essential information is galaxy redshift, which} can be determined directly through spectroscopy, although this process is costly and must be limited to specific samples of galaxies or small on-sky regions. Alternatively, wide-area photometric surveys offer more extensive and deeper samples of galaxies than can be observed through spectroscopy. The redshift distribution of this galaxy population can be accurately calibrated using various algorithms, as discussed in a review by \citep[see][for a review]{salvato2018flavours}. Successful models have been proposed to explain fluctuations in galaxy number density. On large scales, the distribution of galaxies is proportional to the underlying mass fluctuations, while on smaller scales, their nonlinear behavior can be represented using a halo occupation distribution \citep[HOD,][]{Zheng_2005} model. Galaxy density fluctuations have long been used to study various topics in large-scale structure by cross-correlation different tracers: {including 21-cm emission for reionization \cite{Lidz_2008}, CMB lensing for cosmological parameters \cite{Kuntz_2015, Krolewski_2020}, CMB map for integrated Sachs-Wolf effect \cite{Maniyar_2019, hang2020galaxy}, tSZ map for thermodynamic history \cite{pandey2019constraints,koukoufilippas2020tomographic,chiang2020cosmic,2021yanz}, and so on.} In the near future,  the Canada-France Imaging Survey \citep[CFIS,][]{2017cfis}, Rubin Observatory Legacy Survey of Space and Time \citep[LSST,][]{lsstsciencecollaboration2009lsst} and the \citep[\textit{Euclid},][]{laureijs2010euclid} mission will reach unprecedented sky coverage and depth, making galaxy number density fluctuation a ``treasure chest'' from which we will learn a lot about our Universe.

The origin of CIB is generally localized in galaxies, so it should be significantly correlated with galaxy distribution. Clustering-based CIB cross-correlation has been used to study star formation in different types of galaxies, for example: \citep{Serra_2014} analyze luminous red galaxies (LRG), \citep{2015MNRAS.449.4476W, 2014schmidt, 2018hall} analyze quasars, and \citep{2016ApJ...831...91C} analyze submillimeter galaxies (SMG), and \cite{Yan_2022, Jego_2023} cross-correlate larger galaxy populations in different tomographic bins. 

In this study, we cross-correlate the \unwise galaxy catalogs \cite{Schlafly_2019} with CIB maps constructed from the \planck sky map at 353, 545, and 857 GHz \cite{lenz_large-scale_2019}, leading to nine pairs of cross-correlation power spectra. The galaxy samples are divided into three tomographic bins with mean redshift 0.6, 1.1, 1.5 and cover $\sim 60\%$ of the sky, making it an unprecedentedly deep and wide sample for such measurement. {\cite{kusiak2023enhancing} measured \unwise-CIB cross-correlation for the first time, but they only use it to evaluate the potential CIB contamination in the CMB spectrum. In this work, we focus on constraining CIB models with such measurement.} This paper is structured as follows. In Sect.~\ref{sect:model} we describe the theoretical model we use for the cross-correlations. Sect.~\ref{sect:data} introduces the dataset that we are using and the method to measure cross-correlations, as well as our estimation of the covariance matrix, likelihood, and systematics. Sect.~\ref{sect:results} presents the results. We summarize the main conclusions in Sect.~\ref{sect:conclusions} and discuss them in Sect.~\ref{sect:discussions}. Throughout this study, we assume a flat $\Lambda$CDM cosmology with the fixed cosmological parameters from \cite{planckcosmo18} as our background cosmology: $ (h,\Omega_\mathrm{c} h^2,\Omega_\mathrm{b} h^2, \sigma_8,n_\mathrm{s}) = (0.676, 0.119, 0.022, 0.81, 0.967)$.

\section{Theoretical models for galaxy-CIB cross-correlation angular power spectra}
\label{sect:model}
For a general cosmological field $u$, its projected sky map $\Delta_u(\hat{\theta})$ can be written as a line-of-sight integral of its 3-D distribution $\delta_u$ multiplied by its ``radial kernel'' $W^u(\chi)$, a function of the comoving distance $\chi$:
\begin{equation}
    \Delta_u(\hat{\theta})=\int\mathrm{d}\chi W^{u}(\chi) \delta_u(\chi\hat{\theta}, \chi).
    \label{eq:fluctproj}
\end{equation}

The angular power spectrum of cross-correlation between field $u$ and field $v$ ($C^{uv}_{\ell}$) is the two-point correlation function of $uv$ in the harmonic space. It measures how $u$ and $v$ correlate at different angular scales. At the scales $\ell>10$, $C^{uv}_{\ell}$ is related to the 3-D power spectra $P^{uv}(k)$ via the Limber approximation \cite{limber1953analysis}:
\begin{equation}
    C_{\ell}^{uv}  =  \int _0 ^{\chi_{\mathrm{H}}} \frac{\mathrm{d}\chi}{\chi^2}W^u(\chi)W^v(\chi)P_{uv}\left(k=\frac{\ell+1/2}{\chi}, z(\chi)\right), 
\end{equation}
{where $\chi_{\mathrm{H}}$ is the comoving distance to the horizon.} The 3-D power spectrum $P^{uv}(k)$ describes the 3-D distribution of $uv$ correlation, which traces the underlying matter distribution of the LSS. According to the halo model \cite{Seljak_2000}, the matter is generally distributed in dark matter halos, so $P_{uv}(k)$ can be divided into the two-halo term, which accounts for the correlation between different halos, and the one-halo term, which accounts for correlations within the same halo. According to \cite{mead2021hmcode2020}, adding up the one- and two-halo will cause bias in the transit region, which results in an underestimation of the power spectrum. We correct this bias by introducing the smoothing factor $\alpha$.

\begin{equation}
    P_{uv}(k) = \left[P_{uv, \mathrm{1h}}(k)^{\alpha}+P_{uv, \mathrm{2h}}(k)^{\alpha}\right]^{\frac{1}{\alpha}}.
    \label{eq:smooth}
\end{equation}
The smoothing factor we use is from the fitting model proposed by \texttt{HMCODE2020}\cite{mead2021hmcode2020}.

Both one- and two-halo terms are related to the profiles of $u$ and $v$ in Fourier space:

\begin{equation}
\begin{aligned}
P_{uv, \mathrm{1h}}(k) &=\int_0^{\infty} \dr M \frac{\dr n}{\dr M}\langle \tilde{p}_u(k | M) \tilde{p}_v(k | M)\rangle \\
P_{uv,\mathrm{2h}}(k) &=\langle b_u\rangle(k)\langle b_v\rangle(k) P^{\mathrm{lin}}(k) \\
\langle b_u\rangle(k) & \equiv \int_0^{\infty} \dr M \frac{\dr n}{\dr M} b_{\mathrm{h}}(M)\langle \tilde{p}_u(k | M)\rangle,
\end{aligned}
\label{eq:halomodel_pk}
\end{equation}
where the angled brackets $\langle \cdot \rangle$ denotes the ensemble average. {At a given redshift,} $P^{\mathrm{lin}}(k)$ is the linear power spectrum, $\dr n/\dr M$ is the halo mass function (number density of dark matter halos in each mass bin), $b_{\mathrm{h}}$ is the halo bias, and $p_u(k | M)$ is the profile of the tracer $u$ with mass $M$ in Fourier space:
\begin{equation}
\tilde{p}_u(k | M) \equiv 4 \pi \int_{0}^{\infty} \dr r r^{2} \frac{\sin (k r)}{k r} p_u(r | M),
\end{equation}
where $p_u(r | M)$ is the radial profile of $u$ in real space. In this work, we employ the halo mass function and halo bias given by \cite{Tinker_2008} and \cite{Tinker_2010}, respectively.

In this subsection, we presented a general description of the angular distribution of LSS fields and their angular cross-correlation power spectra. In the next subsections, we will apply these tools to galaxy distribution and CIB intensity.

\subsection{Galaxy number density fluctuations}
\label{subsec:hod}
The projected galaxy number density fluctuation $\Delta_{\mathrm{g}}(\hat{\boldsymbol{\theta}})$ in direction $\hat{\boldsymbol{\theta}}$ is defined as the fractional difference between the galaxy surface number density $n(\hat{\boldsymbol{\theta}})$ and the mean galaxy number density $\bar{n}$:

\begin{equation}
    \Delta_{\mathrm{g}}(\hat{\boldsymbol{\theta}})\equiv \frac{n(\hat{\boldsymbol{\theta}})-\bar{n}}{\bar{n}}.
\end{equation}
Defining $\Phi_{\mathrm{g}}(z)$ the redshift distribution of the galaxy sample, $\Delta_{\mathrm{g}}$ is related to the 3-D galaxy number density fluctuation $\delta_{\mathrm{g}}(\chi(z) \hat{\boldsymbol{\theta}}, \chi)$ with:

\begin{equation}
\Delta_{\mathrm{g}}(\hat{\boldsymbol{\theta}})=
\int_0 ^{\chi_{\mathrm{H}}} \mathrm{d} \chi \frac{H(z)}{c}\Phi_{\mathrm{g}}(z(\chi)) \delta_{\mathrm{g}}(\chi(z) \hat{\boldsymbol{\theta}}, \chi)
,
\label{eq:gal_2d}
\end{equation}
The radial kernel for galaxy number density fluctuation is defined as:
\begin{equation}
W^{\mathrm{g}}(\chi) = \frac{H(\chi)}{c} {\Phi_{\mathrm{g}}\left(z(\chi)\right)}
\label{eq:wg}
.\end{equation}
{For a large galaxy sample, we use the ensemble-averaged density fluctuation $\left\langle\delta_{\mathrm{g}}\right\rangle$ to describe the 3-D galaxy distribution}: 
\begin{equation}
\begin{aligned}
        \left\langle \delta_{\mathrm{g}}(\boldsymbol{r} | M)\right\rangle&=\frac{1}{\bar{n}_{\mathrm{g}}(z)} p_{\mathrm{g}}(r| M)\\ &=\frac{1}{\bar{n}_{\mathrm{g}}(z)}\left[ \left\langle N_{\mathrm{c}}(M)\right\rangle \delta^{\mathrm{3D}}(\boldsymbol{r})+ \left\langle N_{\mathrm{s}}(M)\right\rangle p_{\mathrm{s}}(r | M)\right].
\end{aligned}
\label{eq:delta_g}
\end{equation}
{Note that the bold symbol $\boldsymbol{r}$ is the 3D position vector with respect to the halo center, and $r$ is the radial distance. We assume that the ensemble-averaged halo profiles are spherically symmetric, so the profiles only depend on $r$.} In addition, $\delta^{\mathrm{3D}}(\boldsymbol{r})$ is the 3-D Dirac delta function, $N_{\mathrm{c}}(M)$ and $N_{\mathrm{s}}(M)$ are the number of central galaxy and satellite galaxies as a function of the halo mass $M$ {respectively}, and $p_{\mathrm{s}}(r | M)$ is the number density profile of the satellite galaxies. Its Fourier transform will be given in Eq.~\eqref{eq:p_s}.  $\bar{n}_{\mathrm{g}}(z)$ is the mean galaxy number density at redshift $z$, which is given by

\begin{equation}
    \bar{n}_{\mathrm{g}} = \int \dr M \frac{\dr n}{\dr M}(\left\langle N_{\mathrm{c}}(M)\right\rangle+\left\langle N_{\mathrm{s}}(M)\right\rangle) .
\end{equation}
Though we cannot say anything about galaxy counts for individual halos, their ensemble averages can be estimated via the halo occupation distribution (HOD) model \citep{Zheng_2005,Peacock_2000}:

\begin{equation}
\begin{aligned}
     \langle N_{\mathrm{c}}(M) \rangle &= \frac{1}{2}\left[1+\operatorname{erf}\left(\frac{\ln \left(M / M_{\min }\right)}{\sigma_{\ln M}}\right)\right] \\
     \langle N_{\mathrm{s}}(M) \rangle &= N_{\mathrm{c}}(M) \Theta\left(M-M_{0}\right)\left(\frac{M-M_{0}}{M_{1}}\right)^{\alpha_{\mathrm{s}}},
\end{aligned}
\label{eq:ngal_hod}
\end{equation}
{where $M_{\min }$ is the mass scale at which half of all halos host a galaxy, $\sigma_{\ln M}$ denotes the transition smoothing scale, $M_1$ is a typical halo mass that consists of one satellite galaxy, $M_0$ is the threshold halo mass required to form satellite galaxies, and $\alpha_{\rm s}$ is the power law slope of the satellite galaxy occupation distribution. $\Theta$ is the Heaviside step function.}

The HOD parameters in Eq.~\eqref{eq:ngal_hod} depend on redshift \citep[][]{2012cfht}. In this work, we fix $\sigma_{\ln M}=0.4$ and $\alpha_\mathrm{s}=1$, consistent with simulations \citep{Zheng_2005} {and previous observational constraints \citep[][]{2012cfht, Ishikawa_2020}}, and adopt a simple relation for $\{M_0, M_1, M_{\mathrm{min}}\}$ with respect to redshift as in \cite{2013ApJ...770...57B}:

\begin{equation}
    \mu_{X}(z) = \mu_{X,0} + \mu_{X,p}\frac{z}{1+z},
    \label{eq:mp}
 \end{equation}
$\mu_{X}\equiv \log_{10}(M_{X}/M_{\odot})$ where $X \in \{0, \mathrm{min}, 1\}$; $\mu_{X,0}$ is the value at $z=0$,  while $\mu_{X,p}$ gives the `rate' of evolution. Therefore in total we constrain six HOD parameters: \{$\mu_{0,0}$, $\mu_{1,0}$, $\mu_{{\mathrm{min}},0}$, $\mu_{0,p}$, $\mu_{1,p}$, $\mu_{{\mathrm{min}},p}$\}. In practice, we find that the resolution of the CIB map is sufficiently low that this simple formalism fits the data well (Sect.~\ref{sect:results}).

In Fourier space, the galaxy fluctuation is given by:
\begin{equation}
    \left\langle\tilde{\delta}_{\mathrm{g}}(k| M)\right\rangle=\frac{1}{\bar{n}_{\mathrm{g}}(z)} \tilde{p}_{\mathrm{g}}(k| M) =\frac{1}{\bar{n}_{\mathrm{g}}(z)}\left[ \left\langle N_{\mathrm{c}}(M)\right\rangle+ \left\langle N_{\mathrm{s}}(M) \right\rangle\tilde{p}_{\mathrm{s}}(k | M)\right],
\label{eq:gprof_hod}
\end{equation}
where the dimensionless profile of satellite galaxies in Fourier space $\tilde{p}_{\mathrm{s}}(k | M)$ is generally taken as the Navarro-Frenk-White profile (NFW)\citep{van_den_Bosch_2013, Navarro_1996}:
\begin{equation}\label{eq:p_s}
\begin{aligned}
\tilde{p}_{\mathrm{s}}(k | M) = \tilde{p}_{\mathrm{NFW}}(k \mid M)=& {\left[\ln \left(1+c\right)-\frac{c}{\left(1+c\right)}\right]^{-1} } \\
& \times {\left[\cos (q)\left(\operatorname{Ci}\left(\left(1+c\right) q\right)-\operatorname{Ci}(q)\right)\right.} \\
&+\sin (q)\left(\operatorname{Si}\left(\left(1+c\right) q\right)-\operatorname{Si}(q)\right) \\
&\left.-\sin \left(cq\right) /\left(1+cq\right)\right]
\end{aligned}
\end{equation}
where $q\equiv kr_{200}(M)/c(M)$, $c$ is the concentration factor, and the functions \{Ci, Si\} are the standard cosine and sine integrals, respectively\footnote{{The cosine and sine integrals are defined as follows: \[\mathrm{Ci}(x)\equiv \int_x^{\infty}\frac{\cos t}{t}\mathrm{d}t,\]\[\mathrm{Si}(x)\equiv \int_0^{x}\frac{\sin t}{t}\mathrm{d}t.\]}}. The concentration-mass relation in this work is given by \cite{Duffy_2008}. Here $r_{200}$ is the radius that encloses a region where the average density exceeds 200 times the critical density of the Universe. We take the total mass within $r_{200}$ as the proxy of halo mass because in general $r_{200}$ is close to the virial radius of a halo \citep[][]{1998tx19.confE.533O}. \footnote{In the literature, this mass is typically denoted as $M_{200}$, but we omit the subscript here.}

\subsection{Halo model for CIB}
The intensity of the CIB (in $\mathrm{Jy/sr}$) is the line-of-sight integral of the comoving emissivity $j_{\nu'}$:
\begin{equation}
    I_{\nu}(\boldsymbol{\theta}) = \int \mathrm{d}\chi a j_{(1+z)\nu}(\chi, \boldsymbol{\theta}).
    \label{eq:cibintens}
\end{equation}
where {$\nu'=(1+z)\nu$ is the rest-frame frequency and $\nu$ the observed frequency}.
Comparing with Eq.~\eqref{eq:fluctproj}, one can define the radial kernel for the CIB to be:

\begin{equation}
    W^{\mathrm{CIB}}(\chi) = a(\chi) = \frac{1}{1+z(\chi)},
\end{equation}
which is independent of frequency. Thus the emissivity $j_{\nu'}$ is the ``$\delta_u$'' term for the CIB, which is related to the {specific} IR luminosity ({luminosity per unit rest-frame frequency}) in dark matter halos as:

\begin{equation}
\begin{aligned}
j_{(1+z)\nu}(z) &= \int \dr L \frac{\dr n}{\dr L}\frac{L_{(1+z)\nu}(z)}{4\pi} \\
&= \int \dr M \frac{\dr n}{\dr M}\frac{L_{(1+z)\nu}(M, z)}{4\pi},
\end{aligned}
\label{eq:j_l}  
\end{equation}
where $L_{(1+z)\nu}(z)$ is the specific IR luminosity at rest-frame frequency ${\nu'=(1+z)\nu}$ and redshift $z$, and ${\dr n}/{\dr L}$ is the IR luminosity function. The second equation assumes that galaxy luminosity is also a function of the mass of the host dark matter halo. The observed specific IR flux from a halo with mass $M$ at redshift $z$ and observed frequency $\nu$ is

\begin{equation}
    F_{\nu}(M,z) = \frac{L_{(1+z)\nu}(M, z)}{4\pi\chi^2(1+z)}.
    \label{eq:snu_def}
\end{equation}
The specific IR flux $F_{\nu}(M,z)$ is typically written as 

\begin{equation}
    F_{\nu}(M,z) = L_{\mathrm{IR}}(M, z)S_{\mathrm{eff}}[(1+z)\nu,z],
    \label{eq:snu_lir}
\end{equation}
where $L_{\mathrm{IR}}$ is the total IR luminosity over a wide range of frequency and $S_{\mathrm{eff}}[(1+z)\nu,z]$ is the effective spectral energy distribution (SED) which is the fraction of IR radiation at the rest-frame frequency $(1+z)\nu$. It is defined as the mean flux density per total solar luminosity, so it has a unit of $\mathrm{Jy}/L_{\odot}$. The IR luminosity $L_{\mathrm{IR}}$ can be modeled as proportional to SFR\cite{kennicutt1998star}: $\mathrm{SFR}(M,z)=K\times L_{\mathrm{IR}}(M, z)$ where $K=10^{-10}M_{\odot}\mathrm{yr}^{-1}L_{\odot}^{-1}$ in the wavelength range of 8-1000 $\mu\mathrm{m}$ {assuming a Chabrier initial mass function \citep[IMF,][]{2003PASP..115..763C}}. The SFR has the unit $M_{\odot}\mathrm{yr}^{-1}$. A more comprehensive derivation of this formula can be found in Appendix B of \cite{2014planckxxx}. 

The specific IR luminosity can also be divided into contributions from central and satellite galaxies. Assuming the central and satellite galaxies have the same SED \citep[][]{shang_improved_2012, 2014planckxxx} and combining Eqs \ref{eq:snu_def} and \ref{eq:snu_lir}, we get:
\begin{equation}
    L_{(1+z)\nu, \mathrm{c/s}}(M,z) = \frac{4\pi}{K} \times\mathrm{SFR}_{\mathrm{c/s}}(M,z)\chi^2(1+z)S_{\mathrm{eff}}[(1+z)\nu,z].
    \label{eq:lir}
\end{equation}
where the subscripts ``$\mathrm{c/s}$'' denote the central and satellite components respectively. The CIB luminosity profile in Fourier space is formulated as:
\begin{equation}
    L_{\nu}(k| M) = L_{\nu, \mathrm{c}}(M) + L_{\nu, \mathrm{s}}(M)p_{\mathrm{NFW}}(k | M).
    \label{eq:ciblcs}
\end{equation}
Comparing Eq.~\eqref{eq:ciblcs} and Eq.~\eqref{eq:j_l} with Eq.~\eqref{eq:delta_g}, one recognizes that the quantity $f_{\nu,\mathrm{c/s}}(M)\equiv L_{\nu,\mathrm{c/s}}(M)/4\pi$ is directly analogous to $N_{\mathrm{c/s}}(M)$\footnote{$f_{\nu,\mathrm{c/s}}(M)$ is called ``luminous intensity'' which has the physical meaning of luminosity per steradian.} in the HOD model for galaxy clustering, and $f_{\nu}(k| M)\equiv L_{\nu}(k| M)/4\pi$ is the Fourier space profile term $p_u(k| M)$ in {Eq.~}\eqref{eq:halomodel_pk} for CIB anisotropies. Following the standard practice of \cite{van_den_Bosch_2013}, we give the cross-correlation between the Fourier profile of galaxies and the CIB that is needed for calculating the one-halo term:

\begin{equation}
\begin{aligned}
\left\langle p_{g}(k| M) f_{\nu}(k| M)\right\rangle & =  \langle N_{s}(M) \rangle\langle f_{\nu, \mathrm{s}}(M)\rangle p^{2}(k| M)  \\ 
  & +  \langle N_{\mathrm{c}}(M)\rangle \langle f_{\nu, \mathrm{s}}(M)\rangle p(k| M) \\
  &+ \langle N_{\mathrm{s}}(M)\rangle \langle f_{\nu, \mathrm{\mathrm{c}}}(M)\rangle  p(k| M).
\end{aligned}
\label{eq:pj_1h}
\end{equation}

The ingredients in $f_{\nu,\mathrm{c/s}}$ to be modeled are SFR and SED. We will discuss their models in Subsec. \ref{subsec:jsfr}. Before that, we link the CIB emissivity to SFR and SED via:
\begin{equation}
    j_{(1+z)\nu}(z) = \frac{{\rho_{\mathrm{SFR}}(z)}(1+z)S_{\mathrm{eff}}\left[(1+z)\nu, z\right]\chi^2}{K},
    \label{eq:jsfrd}
\end{equation}
where ${\rho_{\mathrm{SFR}}(z)}$ is the star formation rate density (SFRD), defined as the stellar mass formed per year per unit comoving volume (in $M_{\odot}\mathrm{yr}^{-1}\mathrm{Mpc}^{-3}$). It is the integrated SFR over dark matter halos:

\begin{equation}
\rho_{\mathrm{SFR}}(z) =\int \mathrm{d}M \frac{\dr n}{\dr M} \mathrm{SFR}_{\mathrm{tot}}(M,z) =\int \mathrm{d}M \frac{\dr n}{\dr M} \left[\mathrm{SFR}_{\mathrm{c}}(M,z)+\mathrm{SFR}_{\mathrm{s}}(M,z)\right],
\label{eq:sfrd_z}
\end{equation}

\subsection{CIB models}
\label{subsec:jsfr}
There are several models for the specific IR luminosity $L_{(1+z)\nu}$ in the literature. Some of them are formulated {with simple mathematical expressions based on trends shown by observations. The physical pictures behind these empirical models are not clear, but they can fit the data well and give sensible predictions.} In this work, we briefly review some of those CIB models. Although their mathematical formulation can vary, they can essentially be separated into two groups, labeled as S12 and M21 models.

\subsubsection{The S12 model}

\label{subsec:S12}

In \cite{shang_improved_2012}, the authors present an empirical halo model (the S12 model hereafter) for CIB luminosity. This model or its updated version has also been used in \cite{2014planckxxx, Serra_2014, 2013ApJ...772...77V}, etc. Here we first introduce its original form,\footnote{To keep the notations consistent in this paper, we use slightly different formulations of CIB models from their original papers.} then reformulate it to be consistent with Eq.~\eqref{eq:lir}. In \cite{shang_improved_2012}, the CIB luminosity is modelled as:

\begin{equation}
    L_{(1+z)\nu,\mathrm{c/s}}(M, z) = L_0 \Phi_{\mathrm{CIB}}(z)\Theta\left[(1+z)\nu, z\right]\Sigma_{\mathrm{c/s}}(M),
    \label{eq:fs12}
\end{equation}
where $L_0$ is a normalization parameter. The other terms are defined as follows:
\begin{itemize}
    \item[-] Redshift evolution $\Phi_{\mathrm{CIB}}(z)$: $\Phi_{\mathrm{CIB}}(z)$ accounts for the redshift evolution of CIB contributed by cosmic star formation history. Motivated by previous semi-analytical models of galaxy formation \cite{2008MNRAS.383..615N, 2009ApJ...703..785D, Li_2010}, the authors adopted a power law to $\Phi_{\mathrm{CIB}}(z)$:
    \begin{equation}
        \Phi_{\mathrm{CIB}}(z) = (1+z)^{\delta}.
        \label{eq:Phi}
    \end{equation}
    The power index $\delta$ is assumed to be greater than 0 up to until $z\sim 2.5$ considering the fact that SFR increases up to that redshift. In the aforementioned works, $\delta$ is constrained with a value from 2 to 4. \cite{2013ApJ...772...77V} takes slightly different formulation for $\Phi_{\mathrm{CIB}}(z)$. They assume that $\delta=0$ beyond $z=2$. In this work, we use the same assumption as \cite{shang_improved_2012}, since our galaxy sample does not extend beyond $z\sim 2$, although we checked that the model from \cite{2013ApJ...772...77V} does not lead to significant differences.
    \item[-] SED shape $\Theta\left[(1+z)\nu, z\right]$: the frequency dependency of CIB is described by the mean spectral energy distribution (SED). Following \cite{Blain_2003, 2010ApJ...718..632H}, \cite{shang_improved_2012} adopted a gray-body spectrum at low frequencies ({therefore, the dust IR emission is assumed to be optically thin}) and power law at high frequencies. The low-frequency gray-body is assumed to account for imperfect blackbody radiation from the dust, while the high-frequency power-law behavior is assumed to account for observed SED shape \cite{Blain_2003}. Following the previous works, we assume a unified SED model for both central and subhalos:
    \begin{equation}
    \Theta(\nu') \propto\left\{\begin{array}{cc}
    \nu'^\beta B_{\nu'}\left(T_\mathrm{d}\right) & \nu'< {\nu'_0} \\
    \nu'^{-\gamma} & \nu' \geqslant {\nu'_0},
    \end{array}\right.
    \label{eq:Theta}
    \end{equation}
    
    {where $\nu'=(1+z)\nu$ is the frequency in the rest frame while $\nu$ is the observed frequency;} $B_{\nu'}$ is the Planck function for blackbody radiation; $T_{\mathrm{d}}$ is the effective dust temperature; and $\beta$ is the emissivity spectral index. The rest-frame pivot frequency ${\nu'_0}$ is defined so that the grey-body and power-law are connected smoothly:
    \begin{equation}
    \left.\frac{d \ln \left[\nu'^\beta B_{\nu'}\left(T_d\right)\right]}{d \ln {\nu'}}\right|_{\nu'={\nu'_0}}=-\gamma.
    \end{equation}
    The previously-constrained parameter values are $T\sim 24 K$, $\beta \sim 1.7$, and $\gamma \sim 1.7$. In \cite{2014planckxxx}, the authors assumed an evolving dust temperature:
    \begin{equation}
        T_{\mathrm{d}} = T_{0}(1+z)^{\alpha},
        \label{eq:T_d}
    \end{equation}
    where they constrained $T_{0}=24.4 K$ and $\alpha=0.36$. In this work, we take this assumption to slightly modify the S12 model. {In this case,} the rest-frame pivot frequency ${\nu'_0}$ depends on redshift. {In the \unwise redshift range, the rest-frame pivot} frequency is expected to be above ${\nu'_0}\gtrsim 3000$ GHz. {The observed frequencies $\nu_0\equiv\nu'_0/(1+z)$ is above $\sim 1700$ GHz, which} is much higher than the frequency range of the Planck CIB maps. Therefore we fix the high-frequency power index parameter $\gamma=1.7$ as the value given by \cite{2014planckxxx} throughout this work. {It should be noted that our SED model is the proposed SED for dusty star-forming galaxies (DSFG) which dominates the CIB \cite{2010ApJ...718..632H}, and is independent of galaxy properties, such as galaxy type and halo mass.}
    \item[-] the $L-M$ relation $\Sigma(M)$: a comparison between galaxy luminosity functions and halo mass functions indicate that star formation is efficient only within a halo mass range $10^{11}<M/M_{\odot}<10^{14}$ and that it is likely suppressed at both high (e.g., accreting black holes, \cite{2003MNRAS.345..349B, 2005MNRAS.363....2K}) and low halo masses (e.g., feedback from supernovae, photoionization heating, \cite{1986ApJ...303...39D, 1996ApJ...465..608T}). For this reason, $\Sigma(M)$ is parametrized as a log-normal function in \cite{shang_improved_2012}:
    \begin{equation}
    \Sigma(M)=M \frac{1}{\sqrt{2 \pi \sigma_{M,0}^2}} \exp \left[-\frac{\left(\ln M-\ln M_{\text {peak }}\right)^2}{2 \sigma_{M,0}^2}\right].
    \label{eq:SigmaM}
    \end{equation}
    where $M_{\text {peak }}$ and $\sigma_{M,0}^2$ are free parameters.
    To account for the redshift evolution of $M_{\mathrm{peak}}$, we follow the same method as the HOD parameters and define $\mu_{\mathrm{peak},0}$ and $\mu_{\mathrm{peak},p}$ as
    \begin{equation}
        \log_{10}\left[ \frac{M_{\mathrm{peak}}(z)}{M_{\odot}}\right] = \mu_{\mathrm{peak},0} + \mu_{\mathrm{peak},p} \frac{z}{1+z}.
        \label{eq:mpeak}
    \end{equation}
    
    Again, we assume that the galaxies from the central halo and subhalos share the same parametrization on $\Sigma(M)$. The contribution from the central galaxy is the generic $\Sigma(M)$ multiplied by the mean central {IR} galaxy number $\left\langle N^{\mathrm{IR}}_{\mathrm{c}}\right\rangle$:
    
    \begin{equation}
        \Sigma_{\mathrm{c}}(M) = \left\langle N^{\mathrm{IR}}_{\mathrm{c}}(M)\right\rangle \Sigma(M)
        \label{eq:Sigma_c}
    \end{equation}
    {The parametrization of $N^{\mathrm{IR}}_{\mathrm{c}}$ is the same as Eq.~\ref{eq:ngal_hod}, but with parameter values from that of the \unwise galaxies.} As one dark matter halo may contain multiple subhalos, the total contribution of subhalos is
    \begin{equation}
        \Sigma_{\mathrm{s}}(M)=\int_{M_{\min }}^M \mathrm{d} \ln m \frac{\mathrm{d} N_{\text {sub}}}{\mathrm{d} \ln m}(m \mid M)\Sigma(M),
    \end{equation}
    where $M$ is the mass of the host halo, $\frac{\mathrm{d} N_{\text {sub}}}{\mathrm{d} m}$ is the subhalo mass function at subhalo mass $m$, which is estimated with the formula from \cite{2010tinkershmf}. The minimum halo mass $M_{\mathrm{min}}$ that can host subhalos has been proved to be significantly lower than $M_{\mathrm{peak}}$ in terms of $\sigma_{L/M}$, so we fix it to be $10^{6}M_{\odot}$. This lower limit does not significantly affect our results.
    Previous studies give the constraints on the width of $\Sigma(M)$ function to be around $\sigma_{L / M}\sim 1.5$\footnote{Note that in \cite{shang_improved_2012}, the logarithmic in the definition of $\Sigma(M)$ is $\log_{10} M$ but we use $\ln M$ here to make the formulations more consistent. Therefore, the width parameter $\sigma_{L / M}$ in our case is $\ln 10$ times that of \cite{shang_improved_2012}. The extra $\ln 10$ factor in the normalization of $\Sigma(M)$ can be absorbed into the overall normalization parameter $L_0$. } and the most star-forming efficient halo mass to be $M_{\mathrm{peak}}\sim 10^{12}M_{\odot}$.
\end{itemize}

Comparing Eq.~\eqref{eq:lir} with Eq.~\eqref{eq:fs12} and relating each term, we can find the SFR and effective SED in the S12 model:

\begin{equation}
    \begin{aligned}
        \mathrm{SFR}_{\mathrm{c/s}}(M,z)&\propto\Sigma_{\mathrm{c/s}}(M, z)\Phi(z), \\
        S_{\mathrm{eff}}\left[(1+z)\nu, z\right]&\propto\frac{\Theta\left[(1+z)\nu, z\right]}{\chi^2(1+z)}.
    \end{aligned}
    \label{eq:s12_sfr}
\end{equation}
All the factors of proportionality in the relations above, along with $4\pi/K$ in Eq.~\ref{eq:lir} are absorbed in $L_0$.

The S12 model explicitly models the dust SED with physics-motivated parameters and introduces reasonable assumptions on the redshift and halo mass dependence of the SFR. However, neither SED nor SFR can be constrained solely with cross-correlation measurements because they share the same proportionality factors, and one can only constrain the overall normalization parameter $L_0$. In this work, we will constrain the S12 model parameters \{$L_0$, $T_0$, $\alpha$, $\beta$, $\delta$, $\mupeak$, $\mupeakp$, $\sigma_{M,0}$\} with our galaxy-CIB cross-correlation to study the dust properties of IR galaxies. By introducing external information, we can constrain the full SFR and SED.

\subsubsection{The M21 model}
\label{subsec:M21}

In \cite{maniyar_star_2018}, the authors explicitly introduced the $j_{\nu}$-$\rho_{\mathrm{SFR}}$ relationship (Eq.~\eqref{eq:jsfrd} and took a linear model for CIB power spectra by assuming an empirical $\rho_{\mathrm{SFR}}(z)$. In the following work \cite{maniyar_simple_2021}, a halo model is introduced for the SFR as Baryon Accretion Rate (called BAR, measured in solar masses per year: ${M_{\odot} \mathrm{yr}^{-1}}$) multiplied by the star formation efficiency $\eta$:
\begin{equation}
    \mathrm{SFR}(M, z)=\eta(M,z)\times \mathrm{BAR}(M, z).
\label{eq:sfr}
\end{equation}
{For a given halo mass $M$ at redshift $z$, BAR is the mean Mass Growth Rate (MGR, also measured in ${M_{\odot} \mathrm{yr}^{-1}}$) of the halo multiplied by the baryon-matter ratio:
\begin{equation}
\begin{aligned}
      \mathrm{BAR}(M, z) &=  \mathrm{MGR}(M, z) \times \frac{\Omega_\mathrm{b}}{\Omega_m},
\end{aligned}
\label{eq:bar}
\end{equation}
where MGR is given by \cite{fakhouri2010merger}:

\begin{equation}
\begin{aligned}
\mathrm{MGR}(M, z) = & 46.1 \left(\frac{M}{10^{12} M_{\odot}}\right)^{1.1} \times  (1+1.11 z) \sqrt{\Omega_{\mathrm{m}}(1+z)^{3}+\Omega_{\Lambda}}
\end{aligned}
\label{eq:mgr}
\end{equation}
where $\Omega_{\mathrm{m}}$, $\Omega_{\mathrm{b}}$ and $\Omega_{\Lambda}$ are the density parameters of total mass {(therefore $\Omega_{\mathrm{m}} = \Omega_{\mathrm{DM}} + \Omega_{\mathrm{b}}$ where $\Omega_{\mathrm{DM}}$ is the dark matter density parameter)}, baryons, and dark energy of the Universe today.

It was shown in \cite{bethermin_redshift_2013} that the star formation efficiency $\eta$ has a single-peak shape with respect to halo mass, with redshift-dependent peak and width. We again follow \cite{maniyar_simple_2021} to parametrize $\eta$ with a log-normal function\footnote{An alternative $\eta(M)$ model is presented in Appendix \ref{append:M13}.}:

\begin{equation}
\eta(M,z)=\eta_{\mathrm{max}} \exp{\left[-\frac{\left(\ln M-\ln M_{\mathrm{peak}}(z)\right)^{2}}{2 \sigma_{M}(z)^{2}}\right]},
\label{eq:eta}
\end{equation}
where $M_{\mathrm{peak}}$ represents the mass of a halo with the highest star formation efficiency $\eta_{\mathrm{max}}$, and $\sigma_{M}(z)$ is the variance of the lognormal, which represents the range of masses over which the star formation is efficient. {In this work, we allow $M_{\mathrm{peak}}$ to vary by} defining $\mu_{\mathrm{peak},0}$ and $\mu_{\mathrm{peak},p}$ following the S12 model. Also, following \cite{maniyar_simple_2021}, this parameter depends both on redshift and halo mass:

\begin{equation}
    \sigma_M(z) =
  \begin{cases}
    \sigma_{M, 0} & \text{if $M<M_{\mathrm{peak}}$} \\
    \sigma_{M, 0}\times\left(1-\tau/z_c\times\max \{0, z_c-z\}\right) & \text{if $M\geq M_{\mathrm{peak}}$} 
  \end{cases}
  \label{eq:sigma_m}
\end{equation}
where $z_c$ is the redshift {below} which the mass window for {star formation} starts to evolve, with a `rate' described by a free parameter $\tau$\footnote{Note that the parametrization in this work is slightly different from that in \cite{maniyar_simple_2021} and \cite{Yan_2022}. We choose this parametrization to avoid negative values of $\sigma_M$ when running MCMC.}. In \cite{maniyar_simple_2021}, $z_c$ is fixed to 1.5, but we choose to keep it as a free parameter given the depth of our galaxy sample.

For the central galaxy, the SFR is calculated with Eq.~\eqref{eq:sfr}, where $M$ describes the mass of the host halo, multiplied by the mean number of central {IR} galaxies $\langle N^{\mathrm{IR}}_{\mathrm{c}}\rangle$ as given by Eq.~\eqref{eq:ngal_hod}:

\begin{equation}
    \mathrm{SFR}_{\mathrm{c}}(M) = \langle N^{\mathrm{IR}}_{\mathrm{c}}(M)\rangle\times\mathrm{SFR}(M)
\end{equation}
\label{eq:sfr_c}

For satellite galaxies, the SFR depends on the masses of subhalos in which they are located \cite{bethermin_redshift_2013}:

\begin{equation}
    \mathrm{SFR}_{\mathrm{s}}(m| M) = \mathrm{min}\left\{\mathrm{SFR}(m), m/M\times\mathrm{SFR}(M)\right\},
\end{equation}
where $m$ is the subhalo mass, and  $\mathrm{SFR}$ is the general star formation rate defined given by Eq.~\eqref{eq:sfr}. The mean SFR for subhalos in a host halo with mass $M$ is then: 

\begin{equation}
    \mathrm{SFR_{s}}(M) = \int \mathrm{d}\ln{m} \left(\frac{\dr N_{\mathrm{sub}}}{\dr \ln{m}}\right)\mathrm{SFR}_{\mathrm{s}}(m|M),
\end{equation}
where ${\dr N_{\mathrm{sub}}}/{\dr \ln{m}}$ is the subhalo mass function. We take the formulation given by \cite{2010tinkershmf}, the same as the S12 model. 

The SED used in \cite{maniyar_simple_2021} and its related works are fixed with the method given by \cite{bethermin_redshift_2013} which assumes the mean SED of each type of galaxies (main sequence, starburst), and weighs their contribution to the whole population in construction of the effective SED. The SED templates and weights are given by \cite{bethermin_evolution_2015, bethermin_impact_2017}. 

In summary, the free model parameters are \{$\eta_{\mathrm{max}}$, $\mupeak$, $\mupeakp$, $\sigma_{M,0}$, $\tau$, $z_c$\}. Once the parameters are constrained, we can constrain the SFRD via \eqref{eq:sfrd_z}.

\subsubsection{The Y23 model}
\label{subsec:Y23}

According to Subsec.~\ref{subsec:S12} and Subsec.~\ref{subsec:M21}, the connections between SFR and halo mass in the S12 and M21 models are similar because the halo mass dependence in the M21 model is proportional to $M^{1.1}$ times $\eta(M)$, which gives roughly the lognormal distribution in $\Sigma(M)$ in \eqref{eq:SigmaM}. The difference is that the M21 model uses a non-trivial evolution of the width parameter. The redshift dependence $\Phi(z)$ for SFR in the S12 model is purely empirical, while that in the M21 is linked to the BAR, which is more physically sensitive. For the SED, the M21 model depends on templates from external studies while the S12 model parameterizes and constrains the SED with dust temperature and power indices. 

It is interesting to constrain the SED while using a physical SFR model, so we propose the ``Y23'' model as a combination of the S12 and M21 models by substituting the fixed SED term in the M21 model with the greybody spectrum given by \ref{eq:Theta}, therefore we have:

\begin{equation}
    \begin{aligned}
        \mathrm{SFR}(M,z)& = \eta(M,z)\times \mathrm{BAR}(M, z). \\
        S_{\mathrm{eff}}\left[(1+z)\nu, z\right]&\propto\frac{\Theta\left[(1+z)\nu, z\right]}{\chi^2(1+z)}.
    \end{aligned}
    \label{eq:s12_lir}
\end{equation}

The SFR is parametrized the same way as the M21 model. We will also constrain this model (the Y23 model hereafter) with our measurement. In this case, however, the measurement cannot constrain the maximum star-forming efficiency $\eta_{\mathrm{max}}$ given the undefined normalization parameter in the SED. Therefore, we introduce an overall normalization parameter $L_0$ just like in \ref{eq:fs12}. The free parameters of the Y23 model are \{$L_0$, $T_0$, $\alpha$, $\beta$, $\delta$, $\mupeak$, $\mupeakp$, $\sigma_{M,0}$, $\tau$, $z_c$\}.

\subsection{Discussions on CIB models}

There are a couple of simplifying assumptions in the aforementioned CIB models. First of all, the IR radiation from a galaxy is assumed to be entirely the thermal radiation from dust generated by star formation activity. {The IR radiation generated from CO emissions are lines and are irrelevant in diffuse broad-spectrum emission. Other non-dust radiations like free-free scattering, or synchrotron radiation \citep{Galametz_2014} are negligible in the frequencies that we are studying.} We also assume that central and satellite galaxies have the same dust SED, which might {not} be entirely accurate. In addition, we neglect the difference in quenching in central and satellite galaxies \citep[][]{2015MNRAS.449.4476W} {which is one of the reasons that their SEDs might be slightly different}. However, the IR radiation is dominated by central galaxies, so the differences between central and satellite galaxies will not significantly affect our conclusion. Also, we simplify the modeling of SFR by assuming a single overall model, but different populations (quenched, starburst, or main sequence) galaxies from stars all in different manner\cite{bethermin_redshift_2013}. In any case, these limitations need further investigation by future studies.

{In \cite{Yan_2022}, the authors assume the same HOD for the CIB and the KiDS galaxy sample. However, they might be different due to selections in the galaxy sample. In this work, we assume different HODs for IR galaxies generating CIB and \unwise galaxies and introduce two free HOD parameters $\{\mu_{\mathrm{min},0}^{\mathrm{IR}}, \mu_{\mathrm{min},p}^{\mathrm{IR}}\}$ for IR galaxies (see Eqs~\ref{eq:ngal_hod}, \ref{eq:Sigma_c}).}The HOD model {for IR galaxies} used in this work is slightly different from \cite{shang_improved_2012} and \cite{maniyar_simple_2021}. In \cite{shang_improved_2012}, the authors used a heaviside function for $N^{\mathrm{IR}}_{\mathrm{c}}(M)$ and effectively assume $M_0 = 0$ in $N_{\mathrm{s}}(M)$ (see Eq.~\eqref{eq:ngal_hod}). In this work, we follow the parametrization proposed by \cite{Zheng_2005} and take a smoother model for $N^{\mathrm{IR}}_{\mathrm{c}}(M)$. {By assumption, the IR luminosity from satellites depends only on halo mass, so we cannot constrain $N^{\mathrm{IR}}_{\mathrm{s}}(M)$ with our models. In addition, we assume that the HOD model is the same for all three \unwise samples based on the assumption that the \unwise galaxy samples are the same population in different redshifts. However, galaxy abundances might be different across these samples. We will see in Section \ref{sect:data} that the samples are selected by cuttings in colors and have large overlaps in redshift, so color-dependent properties might result in different HOD. In \cite{Kusiak_2022}, the HOD parameters are constrained independently for the three samples, but our measurements do not have the constraining power for sample-dependent HOD parameters due to the limited angular resolutions of the CIB maps and a large number of free parameters. Therefore, we keep using the simple unified HOD model to explore the capability of CIB $\times$ galaxy to constrain galaxy abundance as well as test this assumption by comparing with galaxy power spectra measurements. It should be noted that the redshift dependence of HOD parameters is only phenomenological and reflects a combination of evolution and redshift-dependence selection effects.}

In the CIB signal, the contribution from SFR and dust emission are strongly degenerate, therefore they cannot be constrained simultaneously without introducing external information. The S12 model and its modified version (e.g., $\Sigma(M)$ in \cite{2014planckxxx}) have been used to study the dust properties in IR galaxies \cite{2014planckxxx, Serra_2014, 2013ApJ...772...77V}, including dust temperature and power index. The SFR is studied by constraining the most active halo mass $M_{\mathrm{peak}}$ and the power of redshift dependence $\delta$. The cosmic star formation history was constrained in \cite{2014planckxxx} by fixing the SFRD at $z=0$. Studies using the M21 model\cite{Yan_2022, Jego_2023, Jego_2023_2} and its modified version (e.g., the linear model in \cite{maniyar_star_2018}, a different $\eta(M)$ model in \cite{Jego_2023}) constrain the SFR history while fixing the effective SED. {Using the Y23 model, we will for the first time constrain both these properties.}

Note, though, that the measured power spectrum will also contain shot noise resulting from the auto-correlated Poisson sampling noise. Therefore, the model for the total CIB-galaxy cross-correlation {$C_{\ell}^{\nu\mathrm{g}, \mathrm{tot}}$} is:

\begin{equation}
    C_{\ell}^{\nu\mathrm{g}, \mathrm{tot}} = C_{\ell}^{\nu\mathrm{g}, \mathrm{hm}} + S^{\nu\mathrm{g}},
    \label{eq:cltot}
\end{equation}
where $C_{\ell}^{\nu\mathrm{g}, \mathrm{hm}}$ is the cross-correlation predicted by the halo model, and $S^{\nu\mathrm{g}}$ is the scale-independent shot noise. Shot noise is generally not negligible in galaxy cross-correlations, especially on small scales. There are analytical models to predict shot noise \citep{2014planckxxx, 2018shot} but they depend on various assumptions, such as the CIB flux cut, galaxy colors, and galaxy physical evolution \citep{2012bethermin}. Therefore, in this work, we are not trying to model the shot noise for different pairs of cross-correlations, instead, we leave their amplitudes as free parameters in our model. In practice, we set $\log_{10}{S^{\nu\mathrm{g}}}$ to be free, where $S^{\nu\mathrm{g}}$ is in the unit of $10^{-8}\mathrm{MJy/sr}$.

\section{Data and measurements}
\label{sect:data}

\subsection{The {\textit{unWISE}} catalog}

\begin{center}
\begin{table}
\centering
\begin{tabular}{ccccc}
\toprule    
samples & $\bar{z}$ & $\delta_z$ & $\bar{n}\, [\mathrm{degree^{-1}}]$ & $s$     \\\hline
blue          & 0.6       & 0.3        & 3482                            & 0.455 \\
green         & 1.1       & 0.4        & 1846                            & 0.648 \\
red           & 1.5       & 0.4        & 105                             & 0.842 \\
\bottomrule
\end{tabular}

\caption{Information of the \unwise galaxy samples including $\bar{z}$, the mean redshift, $\delta_z$, the width of redshift distribution, $\bar{n}$ the mean galaxy number density and $s$, the response of the number density to lensing magnification. The different tomographic bins are labeled as `blue', `green', and `red' samples. }
\label{table:unwise}
\end{table} 
\end{center}

The galaxy sample in this study comes from the \unwise catalog \citep{Schlafly_2019, Krolewski_2020, Krolewski_2021} based on the Wide-Field Infrared Survey Explorer ({\textit{WISE}}) satellite mission, which mapped the whole sky at 3.4, 4.6, 12, and 22 $\mu$m (the W1, W2, W3, and W4 bands) with an angular resolution of a few arcseconds. {The {\textit{ALLWISE}} data release was made after the full {\textit{WISE}} cryogenic mission and the {\textit{NEOWISE}} post-cryogenic mission \citep{Mainzer_2011, 2013wise.rept....1C} which ended in February 2011. Then the instrument was placed into hibernation. In December 2013, the instrument was reactivated to continue the {\textit{NEOWISE}} mission given that W1 and W2 bands do not require cryogen \cite{Mainzer_2014}. The \unwise data are made from the deeper co-add images from the continuing {\textit{NEOWISE}} mission, which now feature more than 4 times longer exposure than the {\textit{ALLWISE}} data.}

In summary, The \unwise catalog is constructed by applying cuts on infrared galaxy color and magnitude in W1 and W2 bands. \unwise sources that have {\textit{Gaia}} DR2 within 2.75 arcsecs around them are considered as `pointlike' and are removed. Bright galaxies, planetary nebulae, and stars are also masked out. The resulting \unwise catalog contains over 500 million galaxies over the full sky up to redshift $z\sim 2$. The \unwise galaxies are further divided into three tomographic bins as shown in Table \ref{table:unwise}. {According to \citep{Krolewski_2020}, the redshift distributions of \unwise are calibrated by two methods: cross-matching with {\textit{COSMOS}} galaxies or cross-correlating with spectroscopic {\textit{BOSS}} galaxies and {\textit{eBOSS}} quasars. The latter method can only constrain the multiplication of redshift distribution and galaxy bias, and \citep{Krolewski_2020} has proved that the two methods are consistent, so we choose the cross-matching redshift distributions in this study.} The redshift distributions are shown in Figure \ref{fig:dndz}.

\begin{figure}
    \centering
    \includegraphics[width=0.8\textwidth]{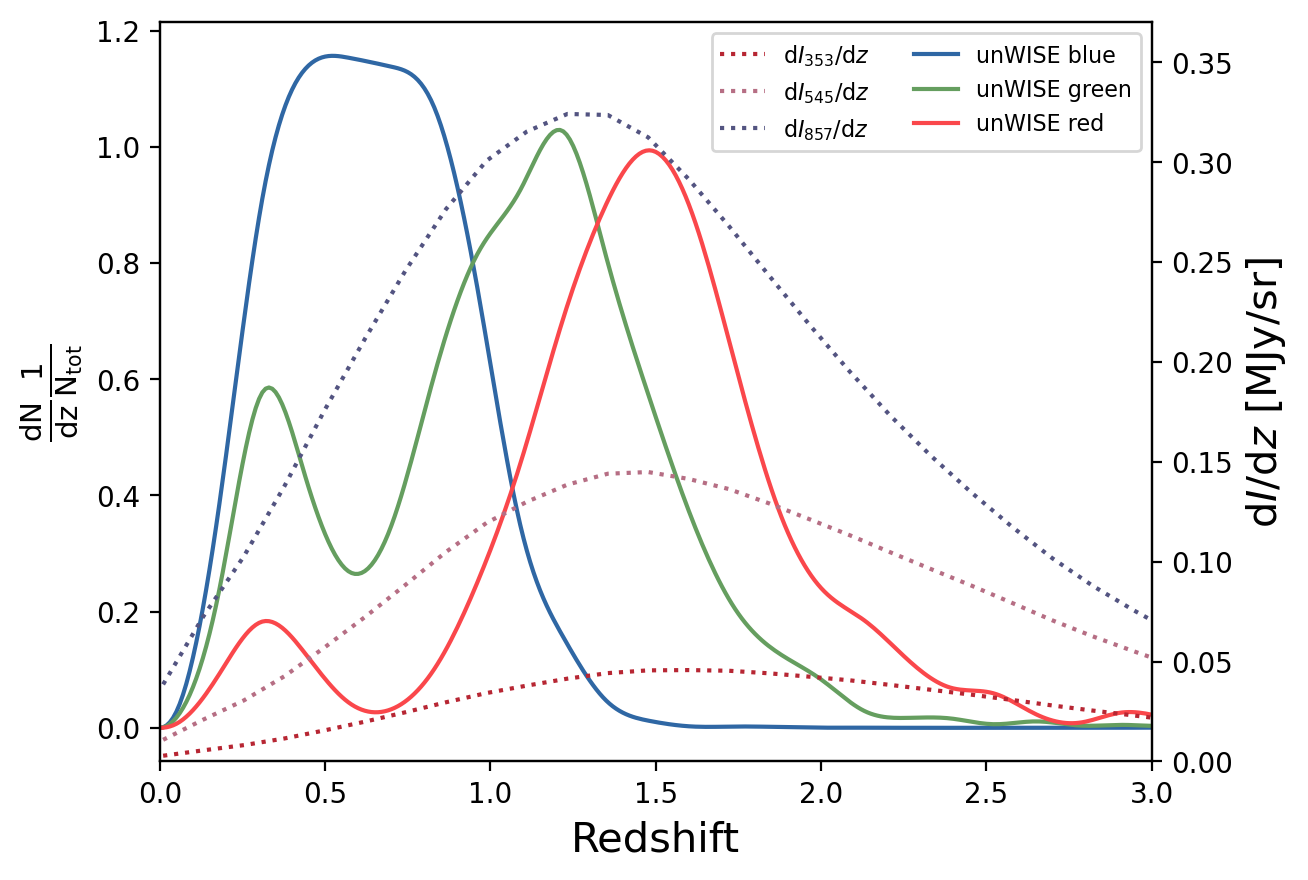}
    \caption{The normalized redshift distributions of the three \unwise samples (solid lines) and infrared intensity {(dotted lines). The y-axis on the left is for the redshift distribution while that on the right is for the CIB density distribution. The infrared intensity is calculated with the best-fit model from \citep{Yan_2022}.}}
    \label{fig:dndz}
\end{figure}

The galaxy overdensity maps are generated by the HEALPix pixelization scheme \cite{Gorski_2005} with \texttt{Nside}=2048, corresponding to an angular size of 1.7 arcmin. Therefore, the 2.75 arcsecs point-source removal will cause sub-pixel `holes', especially in regions close to the Milky Way. To account for this, we follow \cite{Kusiak_2022} and apply a fractional coverage mask that gives the unmasked fraction of each pixel on top of the binary footprint mask. Pixels with more than 20\% area lost are also masked out. This results in a 57.5\% coverage of the sky. \footnote{We note that in the overlapping regions with the CIB mask, the point-source masking causes a negligible reduction in the \unwise footprint and change in the mean galaxy number density.}

To measure the galaxy-CIB cross-correlation, we first generate galaxy overdensity maps in the HEALPix framework. The galaxy overdensity in the $p$-th HEALPix pixel is calculated as:

\begin{equation}
	\delta_p = \frac{N_p/w_p-\bar{N}}{\bar{N}},
\end{equation}
where $N_p$ is the number of galaxies in pixel $p$, $w_p$ is the coverage fraction of each pixel and $\bar{N}\equiv\sum\limits_p N_p/\sum\limits_p w_p$ is the mean galaxy number per pixel across the \unwise mask.

\subsection{The CIB map}

\begin{figure}
    \centering
    \includegraphics[width=0.8\textwidth]{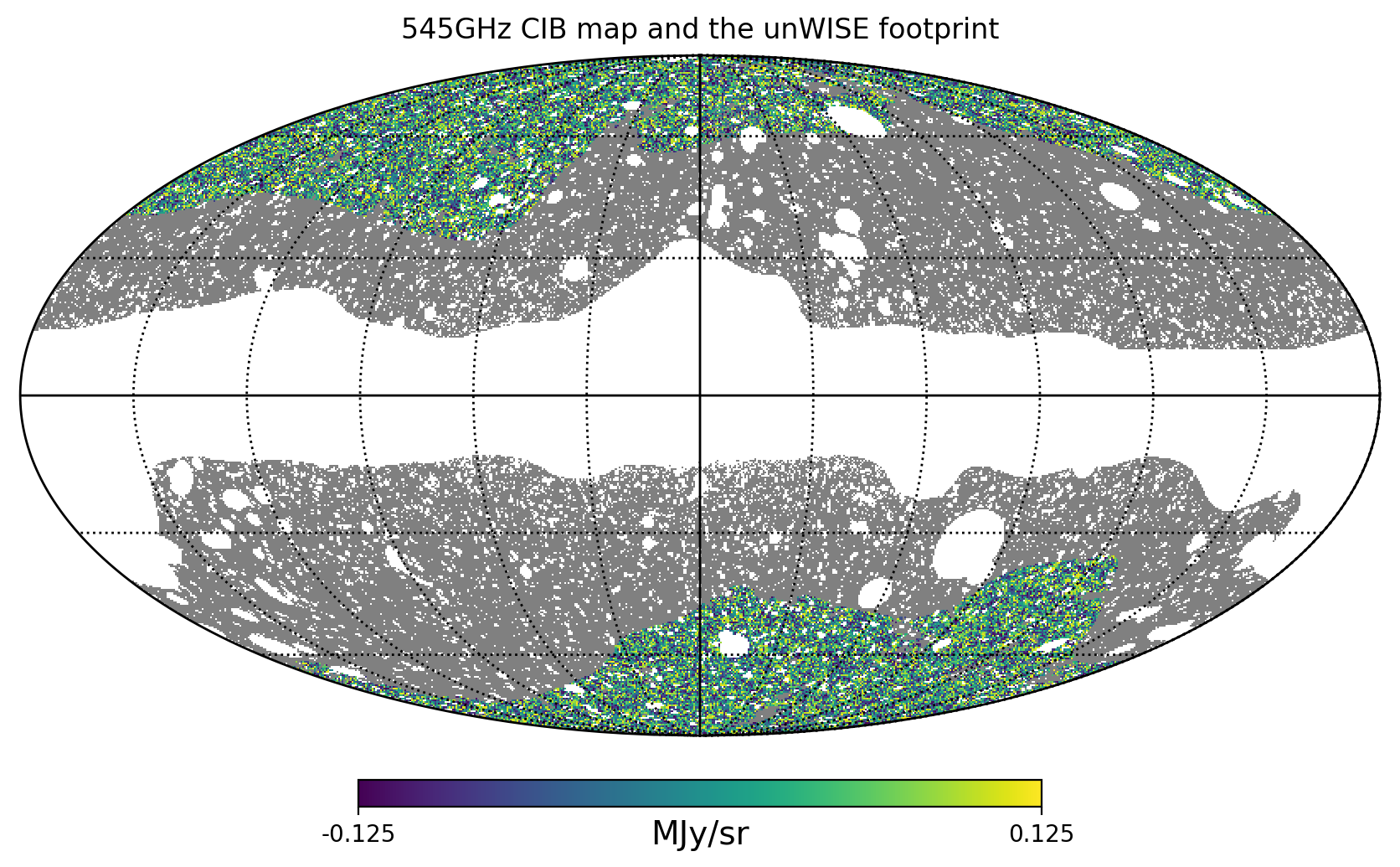}
    \caption{Mollweide projection of the \unwise footprint and the 545 GHz CIB intensity map. For the \unwise footprint, white regions are masked out Milky Way and point sources.}
    \label{fig:mask}
\end{figure}

In this work, we use the large-scale CIB maps generated by \cite{lenz_large-scale_2019} from three \textit{Planck} High-Frequency Instrument (HFI) sky maps at 353, 545, and 857 GHz (the L19 CIB map hereafter)\footnote{\href{https://github.com/DanielLenz/PlanckCIB}{https://github.com/DanielLenz/PlanckCIB}}. The infrared signal generated from Galactic dust emission is removed based on an HI column density template \citep{2011planckcib,2014planckxxx}. {The HI template is generated from the GHIGLS survey \citep{2015ApJ...809..153M}, the GASS survey \citep{McClure_Griffiths_2009}, and the EBHIS survey \citep{2010ApJS..188..488W}.} We use the CIB maps corresponding to a HI column density threshold of $2.0\times 10^{20}\mathrm{cm}^{-2}$. {The CIB mask is a combination of the \textit{Planck} 20\% Galactic mask, the \textit{Planck} extragalactic point source mask, the molecular gas mask, and the mask defined by the HI threshold.} The CIB maps have an overall sky coverage fraction of {14.6}\%. The overlapping field covers about {13.3}\% of the sky (see Fig. \ref{fig:mask} for the CIB intensity map at 545 GHz overplotted on the \textit{unWISE} footprint). The CIB signal in the maps is in the unit of $\mathrm{MJy/sr}$ with an angular resolution of 5 arcmin, as determined by the full width at half maximum (FWHM) of the beam. The original maps are in the \texttt{Healpix} format with \texttt{Nside}=2048, which corresponds to an angular resolution of 1.7 arcmins.

The \planck collaboration also provides all-sky CIB maps \citep{2016planckcib} in the three highest HFI frequencies using the GNILC method \citep{Remazeilles_2011} to disentangle the CIB signal from the Galactic dust emission. These maps have a large sky coverage (about 60\%) and have been extensively used to constrain the CIB power spectra \citep[][]{mak2017measurement,reischke2020information} and to estimate systematics for other tracers \citep{Yan_2019, chluba2017rethinking}. However, \cite{Maniyar_2019_isw, lenz_large-scale_2019} point out that when disentangling Galactic dust from CIB, there is leakage of the CIB signal into the Galactic dust map, causing biases of up to $\sim 20\%$ in the CIB map construction. Therefore, we opt to not use the \planck\, GNILC CIB map in this work at the expense of sky coverage.

\subsection{Pseudo-\texorpdfstring{$C_{\ell}$}{Cl} measurements}

In this study, we employ the pseudo-$C_{\ell}$ method \cite{Alonso_2018} to estimate the angular power spectra of cross-correlations. This approach effectively quantifies 2-point functions in harmonic space, offering a direct connection to theoretical predictions. Additionally, it adeptly accounts for the `mode coupling' arising from the footprint mask and addresses anisotropic weighting factors within the footprint. These factors may originate from various sources, such as lensing weight, partially covered pixels, and spatially varying depth.

To establish a general framework, we assume that the sky maps of two fields, denoted as $u$ and $v$, are the products of the underlying whole-sky fields and their respective weights, represented as $w^u(\hat{n})$ and $w^v(\hat{n})$, where $\hat{n}$ signifies a unit vector specifying the angular position. More precisely, for binary masks that delineate only the footprint shape, $w^u(\hat{n})$ equals 1 within the footprint and 0 elsewhere. Throughout the subsequent derivations, we use a `tilde' ($\tilde{C}_{\ell}$, for example) notation to signify quantities associated with the weighted sky.

The coupled $C_{\ell}$ is measured as:

\begin{equation}
    \tilde{C}^{uv}_{\ell} = \frac{1}{2\ell+1}\sum_{m}\tilde{a}^u_{\ell m}\tilde{a}^{v *}_{\ell m},
    \label{eq:cell_def}
\end{equation}
where $\tilde{a}^u_{\ell m}$ is the harmonic coefficient of field $u$ on weighted sky and $\tilde{a}^{u *}_{\ell m}$ is its complex conjugate. It is linked to the underlying power spectra $C_{\ell}$ via

\begin{equation}
    \tilde{C}^{uv}_{\ell} = \sum_{\ell^{\prime}} M^{uv}_{\ell\ell^{\prime}} C^{uv}_{\ell},
\end{equation}
where $M^{uv}_{\ell\ell^{\prime}}$ is the mode-mixing matrix determined by the weight maps of the two fields:

\begin{equation}
    M^{uv}_{\ell\ell^{\prime}} = \left(\frac{2\ell^{\prime}+1}{4\pi}\right)\sum_{\ell^{\prime\prime}}
    \begin{pmatrix}
  \ell^{\prime}\, & \ell^{\prime}\, & \ell^{\prime\prime}\, \\
  0\, & 0\, & 0 \,
 \end{pmatrix}^2\sum_{m}w^u_{\ell^{\prime\prime} m}w^{v *}_{\ell^{\prime\prime} m},
 \label{eq:mode-mix}
\end{equation}
where $w^u_{\ell m}$ is the harmonic coefficient of the weight map, and $\delta_{\ell\ell^{\prime}}$ is the Kronecker delta symbol. {Note that only $\ell ''$ and $m$ are summed over, the other repeated subscripts $\ell$ and $\ell '$ are free matrix indices and are not summed over.}
The true underlying power spectra can be calculated by multiplying the coupled $C_{\ell}$ with the inverse mode coupling matrix. In practice, it is more usual to bin the coupled pseudo-$C_{\ell}$ into bandpowers. A bandpower is the weighted $C_{\ell}$ within a range of $\ell$. {If we use $q$ to denote the $\ell$ bin index, $\tilde{w}_q^{\ell}$ as the weights at each $\ell$ mode in the bin, and $\vec{\ell}_q$ as the set of $\ell$'s in the $q$-th bin, the bandpower of coupled pseudo-$C_{\ell}$ in the $q$-th bin is defined as}:
\begin{equation}
\begin{aligned}
     \tilde{\boldsymbol{\textsf{C}}}^{uv}_q &= \sum_{\ell \in \Vec{\ell}_q}\tilde{w}_q^{\ell}\tilde{C}_{\ell}^{uv}   \\
     & = \sum_{\ell \in \Vec{\ell}_q}\sum_{\ell^{\prime}}\tilde{w}_q^{\ell}M^{uv}_{\ell\ell^{\prime}}{C_{\ell^{\prime}}^{uv}},     
\end{aligned}
\label{eq:bp_ps}
\end{equation}
We can also define the mode-mixing matrix of bandpower be restricting $\ell '$ in the $q'$-th $\ell$ bin:

\begin{equation}
    \boldsymbol{\textsf{M}}^{uv}_{qq^{\prime}}\equiv \sum_{\ell \in \Vec{\ell}_q}\sum_{\ell^{\prime} \in \Vec{\ell}_{q^{\prime}}}\tilde{w}_q^{\ell}M^{uv}_{\ell\ell^{\prime}}.
    \label{eq:mode-mix_binned}
\end{equation}

Then the bandpower of coupled pseudo-$C_{\ell}$ can be formally written as

\begin{equation}
    \tilde{\boldsymbol{\textsf{C}}}^{uv}_q = \sum_{q^{\prime}}\boldsymbol{\textsf{M}}^{uv}_{qq^{\prime}}\boldsymbol{\textsf{C}}_{q^{\prime}}.
    \label{eq:binned_coupled_cell}
\end{equation}
Where ${\textsf{C}}^{uv}_{q^{\prime}}$ is the bandpower of the true $C^{uv}_{\ell}$, which can be recovered with the inverse matrix of $\boldsymbol{\textsf{M}}^{uv}$:
\begin{equation}
    \boldsymbol{\textsf{C}}^{uv}_q = \sum_{q^{\prime}}\left(\boldsymbol{\textsf{M}}^{uv}\right)^{-1}_{qq^{\prime}}\tilde{\boldsymbol{\textsf{C}}}^{uv}_{q^{\prime}}.
    \label{eq:bp_true}
\end{equation}
It should be noted that ${\textsf{C}}^{uv}_q$ is not simply the true $C^{nv}_{\ell}$ binned with $\tilde{w}_q^{\ell}$. Comparing Eq.~\ref{eq:bp_true} with Eq.~\ref{eq:bp_ps}, one can connect the bandpower with the un-binned true $C^{nv}_{\ell}$:

\begin{equation}
    \boldsymbol{\textsf{C}}^{uv}_q = w_q^{\ell '} C^{uv}_{\ell '} = \sum_{q^{\prime}}\left(\boldsymbol{\textsf{M}}^{uv}\right)^{-1}_{qq^{\prime}} \sum_{\ell \in \Vec{\ell}_q}\sum_{\ell^{\prime}}\tilde{w}_q^{\ell}M^{uv}_{\ell\ell^{\prime}}{C_{\ell^{\prime}}^{uv}},
    \label{eq:binned_true_cell}
\end{equation}
where $w_q^{\ell '}\equiv \sum\limits_{q^{\prime}}\left(\boldsymbol{\textsf{M}}^{uv}\right)^{-1}_{qq^{\prime}} \sum\limits_{\ell \in \Vec{\ell}_q}\sum\limits_{\ell^{\prime}}\tilde{w}_q^{\ell}M^{uv}_{\ell\ell^{\prime}}$ is the bandpower window function for the true $C^{uv}_{\ell}$. In practice, we cannot recover $C^{uv}_{\ell}$ from $\boldsymbol{\textsf{C}}^{uv}_q$ since $w_q^{\ell}$ is generally non-invertible. Instead, we compare the measured $\boldsymbol{\textsf{C}}^{uv}_q$ with the bandpower of the theoretical $C^{uv}_{\ell}$ by binning the theoretical $C^{uv}_{\ell}$ with $w_q^{\ell}$.

In summary, the bandpower of pseudo-$C_{\ell}$ is measured with the following steps:

\begin{enumerate}
    \item directly measure the coupled power spectra $\tilde{C}^{uv}_{\ell}$ with Eq.~\eqref{eq:cell_def};
    \item calculate the mode-mixing matrix $M^{uv}_{\ell\ell'}$  with Eq.~\eqref{eq:mode-mix}
    \item bin the mode-mixing matrix with given binning weight $\tilde{w}_q^{\ell}$ and get $\boldsymbol{\textsf{M}}^{uv}_{qq^{\prime}}$ with Eq.~\eqref{eq:mode-mix_binned};
    \item decouple the bandpower with Eq.~\eqref{eq:bp_true} and get the bandpower $\boldsymbol{\textsf{C}}^{uv}_q$ for the true power spectra.
\end{enumerate}

Another consideration pertains to the smoothing effect introduced by the instrumental beam and the pixelization window function. Assuming the smoothing is isotropic, it can be characterized by a window function denoted as $b_{\ell}$ in the harmonic space. The overall window function, encompassing all the smoothing effects, results from the multiplication of the individual window functions associated with each smoothing process. Consequently, the measured pseudo-$C_{\ell}$ becomes $\tilde{C}^{uv}_{\ell}\rightarrow b^u_{\ell}b^v_{\ell}\tilde{C}^{uv}_{\ell}$, where $b^u_{\ell}$ represents the overall window function for the field $u$. In the context of the galaxy number density map, the instrumental beam has a width of approximately 10 arcseconds, significantly smaller than the pixel size employed in our analysis (1.7 arcminutes corresponding to \texttt{Nside}=2048). Therefore, we assume a flat beam window function for this case.

The preceding discussion primarily applies to spin-0 fields, including CIB maps and galaxy overdensity maps. However, for spin-2 fields, such as the CMB polarization components $Q$ and $U$ and weak lensing shear, the mode mixing becomes intricate, involving the interplay between $E$- and $B$-modes. For a more comprehensive theoretical treatment of pseudo-$C_{\ell}$ measurements, we direct the reader to \cite{Alonso_2019}. The implementation of pseudo-$C_{\ell}$ measurements is available in the publicly-accessible \texttt{NaMaster} package\footnote{\href{https://namaster.readthedocs.io/en/latest/}{https://namaster.readthedocs.io/en/latest/}}, which will be employed in this study. To maintain consistency with existing literature, throughout the remainder of this paper, we continue to use $C_{\ell}$ to denote binned angular power spectra. Here, $\ell$ should be understood as the ``effective $\ell$ mode'' corresponding to each bin.

\subsection{Covariance matrix}
To assess the uncertainty associated with the cross-correlation measurement, we adopt a methodology aligned with the standard approach for constructing the analytical covariance matrix, as documented in the existing literature \citep[see, for example,][]{Upham_2022}. In contrast to techniques based on simulations or resampling, this analytical method is devoid of sampling noise and affords us the capability to disentangle various contributing factors.

\begin{figure}
    \centering
    \includegraphics[width=\textwidth]{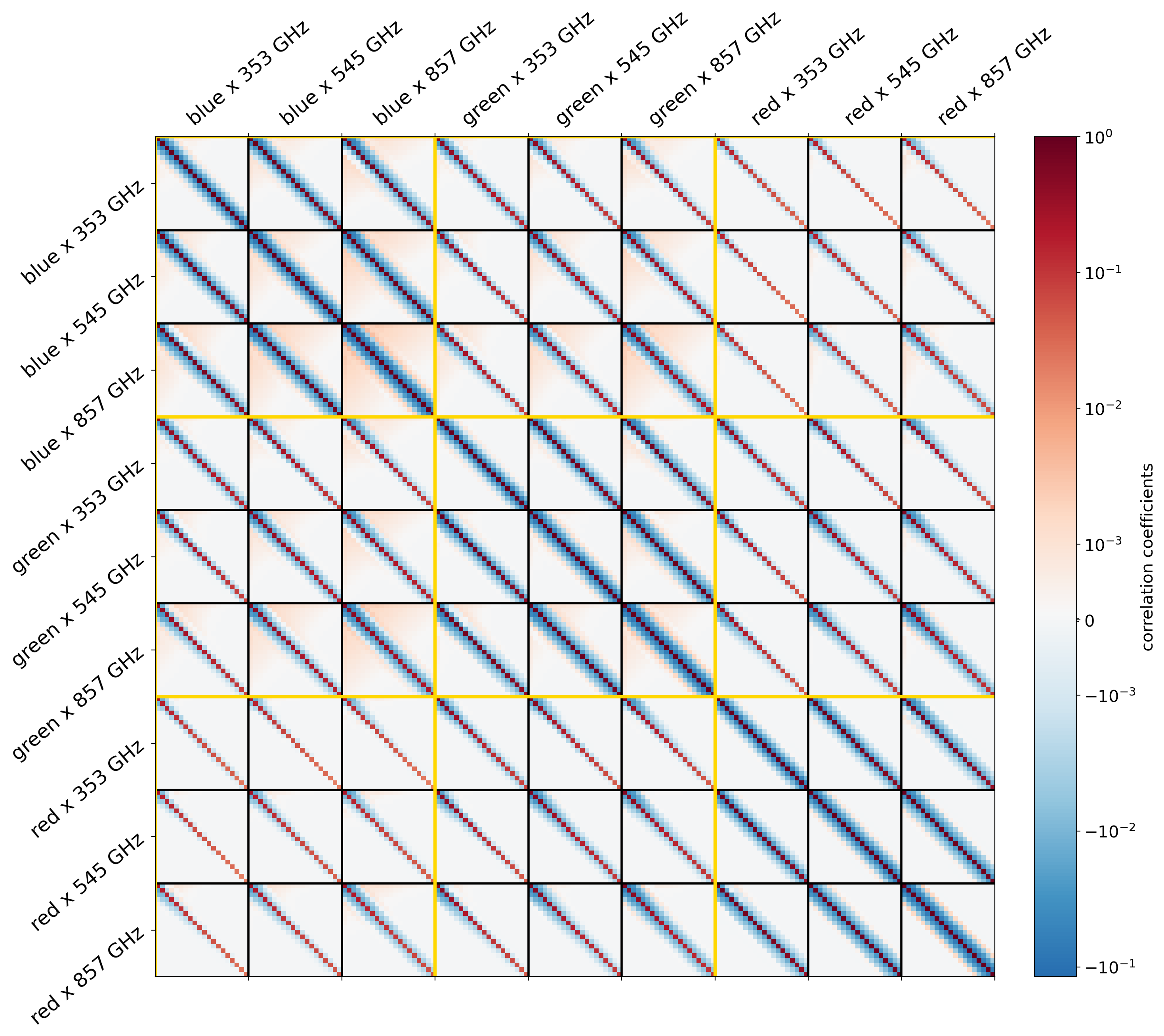}
    \caption{The correlation coefficient matrix of our cross-correlation measurements. Each block enclosed by black grids is the covariance between each pair of cross-correlations {binned in 20 linear bins from $\ell=100$ to $\ell=2000$} indicated with ticks (galaxy sample name $ \times \nu$ GHz), while that enclosed by a golden grid corresponds to the covariance between the CIB cross galaxies from each pair of galaxy samples. Note that the color bar is in the logarithmic scale.}
    \label{fig:fullcov}
\end{figure}

Following \citep[][]{troster2021joint}, we decompose the cross-correlation covariance matrix into three parts:

\begin{equation}
    \mathrm{Cov} = \mathrm{Cov^G} + \mathrm{Cov^{cNG}} + \mathrm{Cov^{SSC}}.
\end{equation}
Here $\mathrm{Cov}$ is the abbreviation of $\mathrm{Cov}_{\ell_1\ell_2}^{uv, wz}\equiv\mathrm{Cov}\left[ C^{uv}_{\ell_1}, C^{wz}_{\ell_2}\right]$. Note that both $\ell_1$ and $\ell_2$ correspond to $\ell$ bands rather than a specific $\ell$ mode. The first term $\mathrm{Cov^G}$ is the dominant `disconnected' covariance matrix corresponding to Gaussian fields, including physical Gaussian fluctuations and Gaussian noise:

\begin{equation}\operatorname{Cov}^{\mathrm{G}}\left(C_{\ell_1}^{u v}, C_{\ell_2}^{w z}\right)=\delta_{\ell_1 \ell_2} \frac{C_{\ell_1}^{u w} C_{\ell_2}^{v z}+C_{\ell_1}^{u z} C_{\ell_2}^{v w}}{(2 \ell_1+1)}
\label{eq:covg}
.\end{equation}
This is the covariance matrix for an all-sky, uniformly-weighted measurement. Sky weights or masks introduce non-zero coupling between different $\ell$ as well as enlarge the variance. To account for this, we used the method given by \cite{Efstathiou_2004} and \cite{Garc_a_Garc_a_2019}, the disconnected Gaussian covariance is

\begin{equation}
\operatorname{Cov}^{\mathrm{G}}\left(C_{\ell_1}^{u v}, C_{\ell_2}^{w z}\right)=\frac{1}{2 \ell_1+1}\left[C_{(\ell_1}^{uw} C_{\left.\ell_2\right)}^{vz} M_{\ell_1 \ell_2}\left(w_u w_w, w_v w_z\right)+C_{(\ell_1}^{uz} C_{\left.\ell_2\right)}^{vw} M_{\ell_1 \ell_2}\left(w_u w_z, w_v w_w\right)\right],
\end{equation}
where

\begin{equation}
    C_{(\ell_1}^{uw} C_{\left.\ell_2\right)}^{vz} \equiv \frac{1}{2} \left( C_{\ell_1}^{uw} C_{\ell_2}^{vz} + C_{\ell_2}^{uw} C_{\ell_1}^{vz} \right),
    \label{eq:cell_cov}
\end{equation}
and $M_{\ell_1 \ell_2}\left(w_u w_w, w_v w_z\right)$ is the mode-mixing matrix of the maps $w_u w_w$ (the multiplication of the mask corresponding to $u$ and $w$ fields) and $w_v w_z$ that is defined in Eq.~\eqref{eq:mode-mix}.This calculation is implemented in the \texttt{namaster} package \citep{2019namaster}. Following \cite{Alonso_2019}, we take $\tilde{C}^{uw}_{\ell}/\sqrt{w_uw_w}$ as the $C^{uw}_{\ell}$ in 
\eqref{eq:cell_cov}, where $\tilde{C}^{uw}_{\ell}$ is the coupled $C_{\ell}$ measured from the maps, so the contribution from noise is also included. We assume that the random noise in the map is Gaussian and independent of the signal.

The second term $\mathrm{Cov^{cNG}}$ is the connected non-Gaussian covariance from the trispectrum, which is given by:
\begin{equation}
\begin{aligned}
\operatorname{Cov}^{\mathrm{cNG}}\left(C_{\ell_1}^{u v}, C_{\ell_2}^{w z}\right)&=\int_0^{\infty} \mathrm{d} \chi \frac{W^{u}(\chi) W^{v}(\chi) W^{w}(\chi) W^{z}(\chi)}{4 \pi f_{\mathrm{sky}} \chi^{6}}\\
&\times T_{uvwz}\left(k_1=\frac{\ell_1+1 / 2}{\chi}, k_2=\frac{\ell_2+1 / 2}{\chi}, \chi \right),
\end{aligned}
\end{equation}
where $T_{uvwz}(k)$ is the trispectrum. Using the halo model, the trispectrum is decomposed into one- to four-halo terms. \cite{2018cibng} shows that the one-halo term dominates the CIB trispectrum. As galaxies have a similar spatial distribution to the CIB, we only take the one-halo term into account for our CIB-galaxy cross-correlation:
\begin{equation}
\begin{aligned}
T_{uvwz}^{\mathrm{1h}}(k_1, k_2) &\equiv \int_0^{\infty} \dr M \frac{\dr n}{\dr M} \\
&\times\langle p_u(k_1\mid M) p_v(k_1 \mid M) p_w(k_2\mid M) p_z(k_2\mid M)\rangle.
\end{aligned}
\end{equation}
We will see that this term is negligible in the covariance matrix.

The third term $\mathrm{Cov^{SSC}}$ is called the super sample covariance \citep[SSC][]{Takada_2013}, which is the sample variance that arises from modes that are larger than the survey footprint. SSC can dominate the covariance of power spectrum estimators for modes much smaller than the survey footprint and includes contributions from halo sample variance, beat coupling, and their cross-correlation. The SSC can also be calculated in the halo model framework \citep[][]{2018ssc, 2021ssc}. Note that all the covariance terms are binned into $\ell$ bins.

In this study, the non-Gaussian covariance components $\mathrm{Cov^{cNG}}$ and $\mathrm{Cov^{SSC}}$, are computed using the halo model framework as implemented in \texttt{CCL} \citep{Chisari_2019}\footnote{\href{https://github.com/LSSTDESC/CCL}{https://github.com/LSSTDESC/CCL}}. These components are subsequently combined with $\mathrm{Cov^G}$ to construct the complete covariance matrix. Specifically, we calculate the covariance between all nine pairs of cross-correlations, resulting in a square matrix with a dimension of $3\times 3\times 20=180$.

It is important to note that when computing the cNG and SSC terms of the analytical covariance matrix, we encounter a situation where we must specify the CIB model with model parameters whose values are not known \textit{a priori}. To address this, we employ an iterative approach for covariance estimation as outlined by \cite{Ando_2017}. Initially, we select reasonable values for the model parameters based on \cite{maniyar_simple_2021} to calculate these covariance matrices. Subsequently, these matrices are employed to constrain a set of best-fit model parameters. We then update the covariance matrix using these best-fit parameters and perform a new round of model fitting. In practice, the constraints derived from the second step consistently align with those from the initial round. Nevertheless, we consider the constraints obtained from the second step as our primary results. The M21 CIB model is utilized for covariance calculations. Given that the cNG and SSC terms have a subordinate impact, this choice does not exert a substantial influence on our parameter constraints.

Figure \ref{fig:fullcov} displays the correlation coefficient matrix in a logarithmic scale. Notably, the three diagonal golden blocks exhibit high off-diagonal terms, indicating that the cross-correlations between galaxies within the same tomographic bin and across three CIB channels exhibit strong correlations. This outcome stems from the fact that the CIB signals originating from different frequencies fundamentally arise from the same galaxy population and the same source (heated dust).
The correlations in the off-diagonal golden blocks, while somewhat weaker, remain non-negligible. These correlations result from the overlap of galaxy redshift distributions in distinct tomographic bins, as visually depicted in Figure \ref{fig:dndz}. The presence of negatively correlated terms can be attributed to mode-couplings associated with survey footprints. Additionally, there exist subtle positive correlations among nearly all the modes, with larger scales exhibiting more pronounced effects, primarily due to the SSC term.
It is worth noting that the SSC term contributes up to $1\%$ of the total standard deviation, which is lower compared to the SSC contribution in \cite{Yan_2022}. This discrepancy arises from the significantly larger survey area employed in this study. Furthermore, it is pertinent to mention that the trispectrum term makes a negligible contribution (less than $0.1\%$) to all covariance terms.

\subsection{Systematics}

\subsubsection{CIB color-correction and calibration error}
{
The \planck sky maps in 353, 545 and 857 GHz are given in the unit of MJy/sr. They are converted from the $K_{\mathrm{CMB}}$ after accounting for the Planck instrument response and photometric convention $\nu I_{\nu}$=constant \citep{2014planckxxx}. If the sources have different SEDs, then the modeled flux should be color-corrected to match the photometric convention in the map. According to \cite{2014hfi}, the color-correction is a multiplicative factor on the model flux, thus the angular power spectra are corrected as:

\begin{equation}
    C^{\nu\mathrm{g},\mathrm{measured}}_{\ell} = C^{\nu\mathrm{g},\mathrm{model}}_{\ell}\times cc_{\nu},
\end{equation}
where $cc_{\nu}$ is the color-correction factor at observed frequency $\nu$. In this work, we adopt the color-correction factors from \citep{2014planckxxx} calculated by implementing the CIB SED from \cite{bethermin2012unified}. These factors have values of 1.097, 1.068, and 0.995 for the 353, 545, and 857 GHz bands, respectively. We note that the color-correction depends on the SED model. However, as we will see, the SED constrained from our measurement agrees with that from \cite{bethermin2012unified}, we will keep using the aforementioned correction factors throughout.}

Additionally, in \cite{maniyar_star_2018, maniyar_simple_2021} the authors introduce a scaling factor as an additional calibration tool when working with the L19 CIB maps. However, they constrain this factor to be very close to one (at a level of $\sim \pm 1\%$). As such, in this work, we neglect the additional calibration factor.

\subsubsection{Cosmic magnification}

The measured galaxy overdensity is influenced not only by the actual distribution of galaxies but also by lensing magnification arising from the line-of-sight mass distribution \citep{1989A&A...221..221S, 1989ApJ...339L..53N}. This phenomenon, known as cosmic magnification, exerts two distinct effects on the observed galaxy overdensity:

i) Overdensities along the line-of-sight result in an increase in the local angular separation between source galaxies, causing a dilution in the spatial distribution of galaxies and a suppression of cross-correlations.

ii) Lenses situated along the line-of-sight magnify the flux emitted by source galaxies. Consequently, fainter galaxies enter the observed sample, leading to an increase in the observed overdensity.

These effects introduce biases in galaxy-related cross-correlations, particularly pronounced for high-redshift galaxies \citep{hui2007anisotropic, ziour2008magnification, hilbert2009ray}. To account for these two effects, we make adjustments to the expression for the galaxy overdensity as described by \cite{hui2007anisotropic}:

\begin{equation}
    \hat{\delta}_\mathrm{g}(z) = \delta_\mathrm{g}(z) + 2(2.5s - 1)\kappa(z)
    \label{eq:cosmicmag}
,\end{equation}
where the second term on the right-hand side of the equation is the cosmic magnification contribution. Here $\kappa(z)$ is the line-of-sight integral of the lensing convergence to the galaxy redshift $z$; $s$ is the slope of the logarithmic cumulative number counts of our galaxy sample at magnitude limit $m_{\mathrm{lim}}$
\begin{equation}
    \left.s \equiv \frac{\partial \log _{10} N(<m)}{\partial m}\right|_{m=m_{\lim }}
    \label{eq:cdf}
.\end{equation}
We correct the $C^{g\nu}_{\ell}$ model due to cosmic magnification by adding the term $2(2.5s-1)C^{\kappa\nu}_{\ell}$. The angular cross-correlation between lensing convergence and the CIB $C^{\kappa\nu}_{\ell}$ is also calculated with Limber approximation following Sect.~\ref{sect:model} with { lensing kernel given by

\begin{equation}
    W^{\kappa}(\chi) = \frac{3\Omega_{\mathrm{m}}H_0^2}{2a(\chi)c^2}\int_{\chi}^{\chi_{\mathrm{H}}} \dr \chi^{\prime} \frac{\chi(\chi^{\prime}-\chi)}{\chi^{\prime}}H(z(\chi))\Phi_{\mathrm{g}}(z(\chi)),
\end{equation}
where $\Phi_{\mathrm{g}}(z(\chi))$ is the galaxy redshift distribution and $\chi_{\mathrm{H}}$ is the comoving distance to the horizon. We use the NFW profile given in Eq~\ref{eq:p_s} for lensing convergence.
}

{To correct the cosmic magnification, it is crucial to estimate $s$ from data. Here we use the $s$ values of the three \unwise catalogues calculated from \cite{Krolewski_2020}. In summary, \cite{Krolewski_2020} estimated $s$ at high ecliptic latitude $|\lambda|>60^\circ$, where the limiting magnitude is fainter. Moreover, the WISE selections are based on magnitude-dependent color cuts. Therefore, instead of evaluating $s$ from flux histogram, \cite{Krolewski_2020} determined $s$ by shifting the magnitudes of all WISE objects by 0.02 magnitudes and re-applying the selection criteria, then calculating $s$ from the difference between the shifted and unshifted W2 histograms. See Appendix C of \cite{Krolewski_2020} for more detailed discussions on estimating $s$}. The values that we use for this study are listed in Table \ref{table:unwise}.

\subsection{Potential selection effect in the \texorpdfstring{\unwise}{unwise} samples}
\label{subsec:selection}
{Our \unwise samples are selected based on color cuts, which might result in different galaxy populations across samples. Different types of galaxies have different IR emissions, which adds to the complexity of modelling the CIB. For example, the active galactic nuclei (AGN) have different SED from non-active galaxies because the surrounding hot dust torus absorbs AGN UV/optical emission and re-radiates IR emission that usually peaks at middle IR (MIR) bands \citep{Xu_2020} \citep{1993ARA&A..31..473A}, while IR emission from normal galaxies generally peaks at FIR bands. Therefore, the hot dust emission from AGNs can boost the MIR flux in W2 and selecting the \unwise `red' W1-W2 colors may remove a fraction of low-$z$ galaxies that are intrinsically fainter than AGNs, resulting in an increasing AGN fraction from the blue to the red \unwise sample. {According to the AGN SED template fit by \cite{Laigle_2016}}, \citep{Krolewski_2020} estimates the AGN fraction in the three samples to be roughly 19\%, 30\% and 41\% for the blue, green, and red samples. This estimation is only an approximation due to the limited number of bands and AGN SED templates but it gives a hint on potential selection effects. 

The AGNs from the galaxy sample can be removed with the selection criteria given from, for example, \cite{2018ApJS..234...23A}, but we would sacrifice misidentified non-active galaxies as well as accuracy in redshift distribution. On the other hand, the \unwise galaxies are distributed up to $z\sim 2$, so the rest-frame frequencies corresponding to the \planck observed frequencies should be up to $857\times (1+2)=2571$ GHz. The corresponding lower limit of rest-frame wavelength is $\sim 117\mu\mathrm{m}$. According to \cite{jauzac_cosmic_2011, cai_hybrid_2013, roebuck_role_2016, xu_agns_2020}, the IR emission from AGNs are negligible in this wavelength range compared to dusty star-forming galaxies. Therefore, even though the AGN fraction could vary across the \unwise samples, the cross-correlation with the CIB should be dominated by the dusty star-forming galaxies. Nevertheless, given the potential selection effect in the \unwise sample, the HOD of \unwise galaxies could be different from that of the CIB, so we use separate HOD parameters for \unwise galaxy and the CIB.

Potential selection effects in the galaxy samples make the interpretation of CIB-galaxy cross-correlation difficult because the correlated signal might come from different galaxy populations across the samples. For example, the 2-halo regime traces the overall IR emission from the LSS, while the 1-halo regime could be dominated by the IR properties of the \unwise sources. The redshift-dependent \unwise HOD parameters can account for different galaxy abundance across the \unwise sample, but we are mostly interested in the SFR and SED for the general large-scale structure. Therefore, we perform two robustness tests in Subsection ~\ref{sect:syst_test} to confirm whether our constraints on these two ingredients indeed come from the large scale and are robust to possible selections that could affect the small scales. 
}

\subsection{Likelihood and MCMC}
\label{subsec:likelihood}
We fit the model parameters with angular cross-correlation power spectra within a wide $\ell$ range, so there are many degrees of freedom in each $\ell$ bin. According to the central limit theorem, the bin-averaged $C_{\ell}$'s obey a Gaussian distribution around their true values. Thus we assume that the measured power spectra follow a Gaussian likelihood:

\begin{equation}
    -2 \ln L(\boldsymbol{\tilde{C}} \mid\vec{q})=\chi^{2} \equiv(\boldsymbol{\tilde{C}}-\boldsymbol{C}(\vec{q}))^{T} \mathrm{Cov}^{-1}(\boldsymbol{\tilde{C}}-\boldsymbol{C}(\vec{q}))
,
\label{eq:like}
\end{equation}
where $\vec{q}$ stands for our model parameters presented in Section \ref{sect:model}. The data vector $\boldsymbol{\tilde{C}}$ is a concatenation of our 9 measured CIB-galaxy cross-correlations; $\boldsymbol{C}(\vec{q})$ is the cross-correlation predicted by the models described in Sect.~\ref{sect:model} with parameters $\vec{q}$. 

Here we summarize all the free parameters to be constrained:

\begin{itemize}
    \item CIB parameters in S12, M21, and Y23 models summarized in Subsects.~\ref{subsec:S12}, \ref{subsec:M21}, and \ref{subsec:Y23}.
    \item HOD parameters summarized in Subsects.~\ref{subsec:hod};
    \item Amplitudes of the shot noise power spectra:  $\left\{\log_{10}{S^{\nu\mathrm{g}}}\right\}$. Here $S^{\nu\mathrm{g}}$ is in the unit $10^{-8}$MJy/sr. The prior boundaries are $[-4, 1]$ for all the nine shot noise amplitudes.
\end{itemize}

We make constraints for the three models introduced in Sect.~\ref{sect:model} respectively with the Markov-Chain Monte-Carlo method to constrain our model parameters with the python package \texttt{emcee} \citep[][]{Foreman_Mackey_2013}. Best fit parameters are determined from the resulting chains, as is the sample with the smallest $\chi^2$ goodness-of-fit. Marginal constraints on parameters, when quoted, are marginal means and standard deviations. 

\section{Results and discussions}
\label{sect:results}

{In this section, we first present the pseudo-$C_{\ell}$ measurements along with the best-fit model in Subsect.~\ref{subsec:cell}. Then we present the MCMC results for SFR, SED, and HOD parameters and their indications in subsequent subsections \ref{subsec:sfr}, \ref{subsec:sed}, \ref{subsec:hod_constrain}, respectively. The full triangle plots of posteriors and constraints on parameters are shown in Appendix \ref{append:post}.}

\subsection{Cross-correlation measurements and fittings}
\label{subsec:cell}
In this work, we bin the angular modes into 20 linear bins from $\ell=100$ to $\ell=2000$. The modes below $\ell=100$ are omitted to avoid mask effect on the large scales. By calculating the square root of $\chi^2$ value of the null hypothesis, we get a $\sqrt{\chi^2_{\mathrm{null}}}=\sqrt{\boldsymbol{\tilde{C}}^{T} \mathrm{Cov}^{-1}\boldsymbol{\tilde{C}}}=194\sigma$ detection. %Given such a significant measurement and the likelihood and priors defined in Subsec. \ref{subsec:likelihood}, w
The MCMC is run with 200 walkers and 10,000 steps after 500 steps of burning-in. The chains converge around the 3000th step. We estimate the posterior with the last 2000 steps of all the walkers (therefore 400000 points in the parameter space). We will present the summary of the full posterior in the appendix and discuss the important parameters in the following subsections. In this section, we only discuss the goodness-of-fit of the models based on the pseudo-$\cell$ measurements.

We show the pseudo-$\cell$ measurements and the best-fit Y23 model in Fig. \ref{fig:cell}. Each panel shows one pair of cross-correlations. The same column of panels is cross-correlations with the same CIB map while the same row is with the same galaxy sample. The error bars are the square root of the diagonal terms of the covariance matrix. The solid lines are the best-fit Y23 model prediction. The best-fit parameters in each model are chosen as those given the lowest {$\chi^2$} in the MCMC chains. The 1- and 2-halo contributions are also shown with dashed and dashed-dotted lines respectively. The shot noise terms are much below the 1-halo terms and are not shown in this figure. Note that the total $\cell$ is not the sum of one and two halo terms because the smoothing in the transition region is considered (see Eq.~\eqref{eq:smooth}). The one halo term is almost a straight line because it’s contributed mainly by the central galaxy.

\begin{figure}
    \centering
    \includegraphics[width=\textwidth]{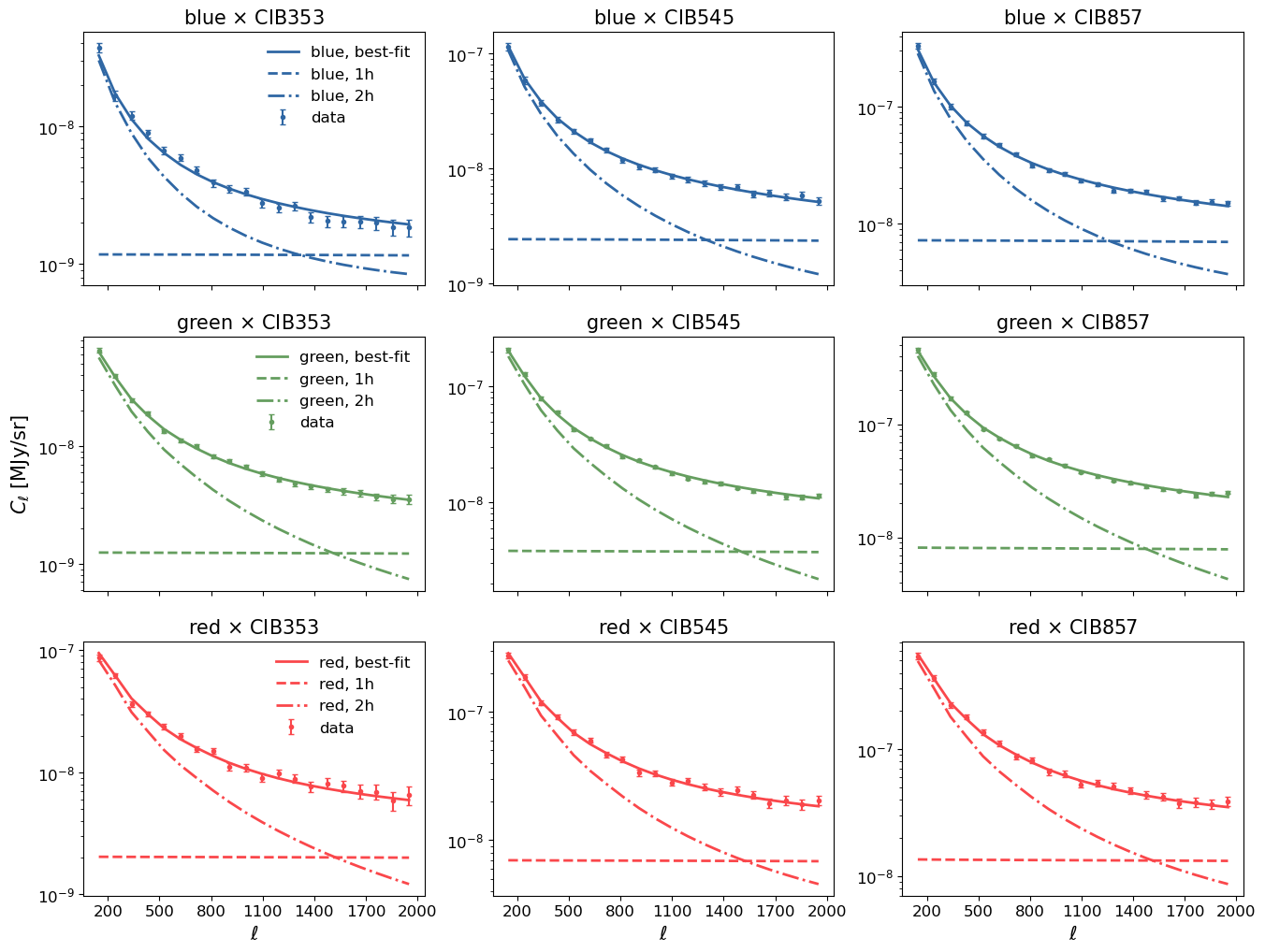}
    \caption{CIB-galaxy cross-correlations with the three \unwise samples (rows) and the three CIB maps (columns). {The three samples have mean redshifts equal to 0.6, 1.1, and 1.5, respectively}. The points are measured from data, with standard deviation error bars calculated using the square root of the diagonal terms of the covariance matrix. The solid lines show the best-fit cross-correlation signals calculated using the CIB-galaxy cross-correlation measurements. The dashed lines and dased-dotted lines are the one and two halo terms respectively. The data points and best-fit models are color-coded according to the corresponding galaxy sample. }
    \label{fig:cell}
\end{figure}

\begin{table*}

    \centering
    
\begin{tabular} { l l c c c}
\toprule
 Metric & Criteria & S12 (Sect. \ref{subsec:S12}) & M21 (Sect. \ref{subsec:M21}) & Y23 (Sect. \ref{subsec:Y23})\\
\hline

$\chi^2$/d.o.f & <1.4 & $\frac{198.919}{(180-23)}=1.267$ & $\frac{207.813}{(180-21)}=1.307$ &$\frac{198.9}{(180-24)}=1.275$\\ 
PTE & >0.001 & 0.013 & 0.006 & 0.012\\

\bottomrule
\end{tabular}

    \caption{A summary of the goodness-of-fit of each model with the best-fit parameters constrained by MCMC. The second column shows the criteria which, when satisfied, indicate the fitting is acceptable. PTE is the probability-to-exceed $\chi^2$, as defined in the text.}
    \label{tab:params_fit}
\end{table*}

We use two criteria to evaluate the goodness-of-fit. 1) The reduced $\chi^2$ is defined as $\chi^2$ divided by the degrees of freedom (the size of the data vector minus the number of free parameters, d.o.f). The fitting is considered acceptable if the reduced $\chi^2$ is lower than 1.4 \cite{abbott2018dark}; 2) The probability-to-exceed (PTE) is defined as the probability of exceeding the $\chi^2$ value assuming a $\chi^2$ distribution with the given d.o.f. The fitting is considered acceptable if PTE>0.001 (corresponding to a $\sim3\sigma$ deviation) \cite{heymans2020kids1000}. The reduced $\chi^2$ and PTE values of the best-fit parameters of each model are summarized in Table \ref{tab:params_fit}. We find that all three models pass the criteria, so we conclude that all the models fit our measurement well. We also note that all the models are quite close to failing both criteria. This is a hint that the data quality is reaching the point where more accurate new models are needed in the near future.

In the following subsections, we will discuss the constraints on star formation rate density (SFRD), spectral energy distribution (SED), and halo occupation distribution (HOD) respectively.

\begin{table}
    \centering
\begin{minipage}{\textwidth}
\centering
\begin{tabular}{llcccc} \\ 
\toprule 
 Parameters & priors& S12 & M21 & Y23 & Reference \\ 
 \hline 
$\eta_{\mathrm{max}}$ & [0.01, 1.0] & -& ${0.49}^{+0.12}_{-0.16}$& -& ${0.41}^{+0.09}_{-0.14}$\cite{Yan_2022}\\ 
$\mu_{\mathrm{peak},0}$ & [10, 14.0] & ${12.25}^{+0.76}_{-0.7}$& ${11.51}^{+0.56}_{-0.71}$& ${11.79}^{+0.73}_{-0.86}$& $12.14\pm0.36$\cite{Yan_2022}\\ 
$\mu_{\mathrm{peak},p}$ & [-5, 5] & ${0.55}^{+1.03}_{-0.91}$& ${-0.02}^{+1.08}_{-0.97}$& ${0.39}^{+1.14}_{-0.91}$& 0\\ 
$\sigma_{M,0}$ & [0.01, 4.0] & ${1.81}^{+0.69}_{-0.9}$& ${2.74}^{+0.78}_{-0.68}$& ${2.48}^{+0.77}_{-0.77}$& ${2.11}\pm 0.55$\cite{Yan_2022}\\ 
$\tau$ & [0, 1] & -& ${0.5}^{+0.18}_{-0.17}$& ${0.46}^{+0.2}_{-0.22}$& ${0.58}\pm{0.25}$\cite{Yan_2022}\footnote{Note that $\tau$ has a different definition from \cite{Yan_2022}. The reference constraints have been transformed to the new definition}\\ 
$z_{c}$ & [0.5, 3.0] & -& ${2.15}^{+0.47}_{-0.51}$& ${1.93}^{+0.51}_{-0.61}$& 1.5\\ 
$\delta$ & [2.0, 5.0] & ${2.98}^{+0.61}_{-0.71}$& -& -& ${3.2}\pm 0.2$\cite{2014planckxxx}\\ 
 \bottomrule 
\end{tabular}
\end{minipage}

    \caption{A summary of the prior ranges, the marginalized mean values, and the 68\% region of SFR parameters of our three models. {The values and errors are calculated from the posteriors marginalized over all the other parameters.} Parameters that are not included in each model are marked as a short bar. The last column shows the constraint from \cite{Yan_2022} or \cite{2014planckxxx} as a reference. Note that in \cite{Yan_2022}, $\mupeakp$ and $z_c$ are fixed.}
    \label{tab:sfr_fit}
\end{table}
\subsection{Constraints on star formation rate density (SFRD)}
\label{subsec:sfr}
\begin{figure}
    \centering
    \includegraphics[width=\textwidth]{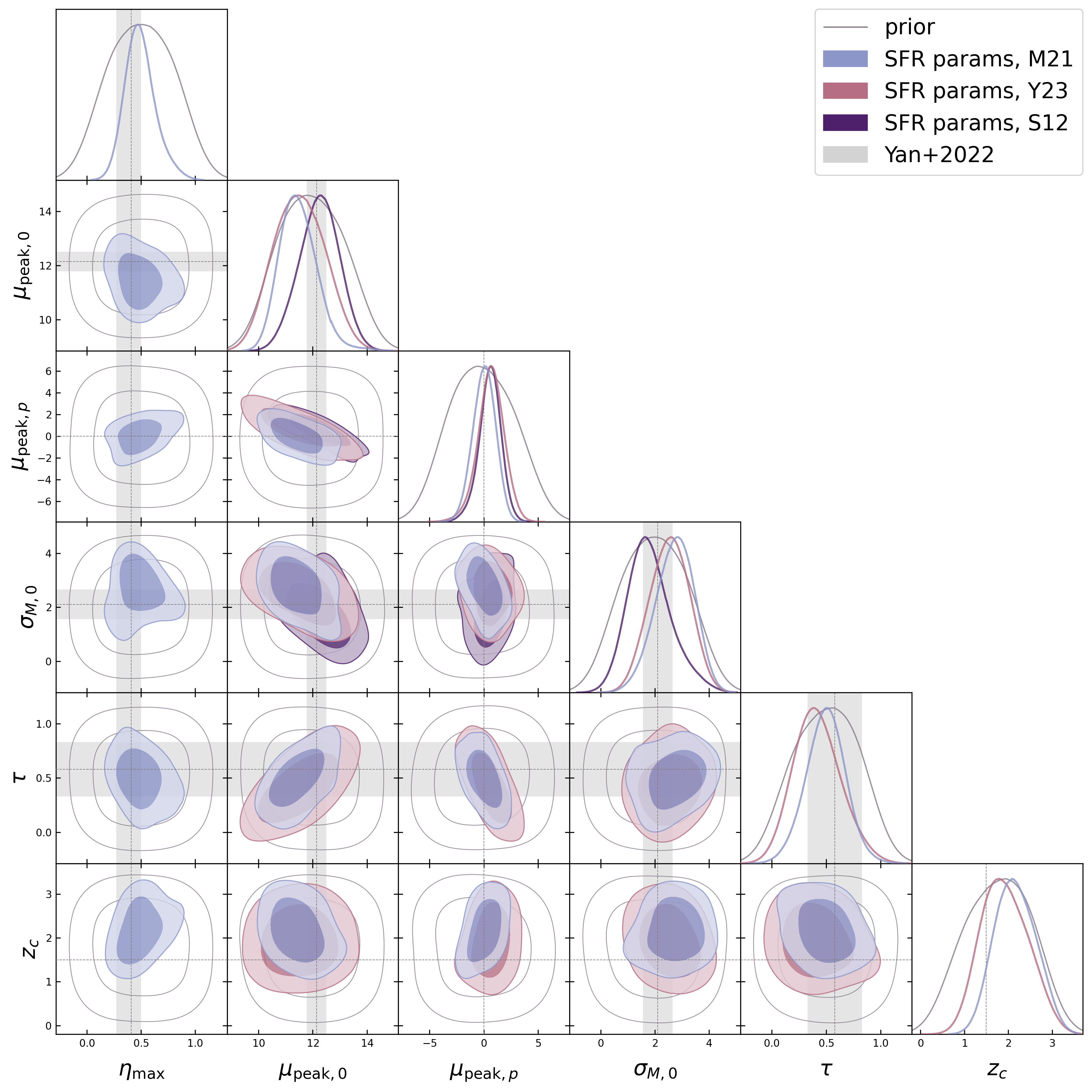}
    \caption{The MCMC constraints of model parameters related to SFRD-halo mass connection. Color-coded contours are posteriors of the three models respectively. Note that the S12 model only has three parameters $\{\mupeak,\mupeakp, \sigma_{M,0}\}$. All the contours are smoothed with a Gaussian kernel density estimate (KDE). The gray unfilled contours are the prior distribution smoothed with the same KDE. The best-fit values and 68\% credible levels from \cite{Yan_2022} are also shown with the dashed lines and gray bands respectively.}
    \label{fig:sfrd_post}
\end{figure}

Despite the distinct parameterizations employed in the three models, it's worth noting that the SFRD model comprises two primary components: one associated with halo mass dependence and the other with redshift dependence. In the case of the M21 and Y23 models, the redshift dependence components are entirely determined by Equations \eqref{eq:sfr}-\eqref{eq:eta}. Conversely, for the S12 model, the redshift dependence is {described} by the redshift power index $\delta$. Across all models, the halo mass dependencies are assumed to follow a lognormal distribution with variations in $M_{\mathrm{peak}}$, and the width parameter $\sigma_{M}$ is held constant in the S12 model, while it evolves in the M21 and Y23 models.

We present the posterior distributions of the parameters governing the connection between SFRD and halo mass in Figure \ref{fig:sfrd_post}. The marginalized 2-D posteriors for each pair of parameters are depicted using color-coded filled contours, representing the 68\% (inner part) and 95\% (outer part) regions. The diagonal panels exhibit color-coded lines representing the marginalized 1-D posteriors for each individual parameter. It is important to note that not all parameters are shared across the three models.
The contours and marginalized posteriors are smoothed using Gaussian Kernel Density Estimation (KDE) {with the standard deviation equal to 0.5 times the standard deviation of the posterior of the parameters}. To assess whether the data effectively constrains these parameters, we sample the prior distribution with 300,000 points within the parameter space and smooth its distribution using the same Gaussian KDE employed for the posterior distribution. The prior distribution is represented by unfilled gray contours.
For reference, we include the best-fit values and 68\% regions from \cite{Yan_2022}, presented as dashed lines and gray bands, respectively. If a parameter is constrained, its posterior distribution should fall within the bounds defined by the prior distribution. According to these criteria, {the halo mass parameter with the most star formation efficiency} $\mupeak$ is primarily constrained in its upper limit across all three models, while the lower bound exhibits mild constraints. The evolution parameter $\mupeakp$ is consistent with zero across all three models, in agreement with \cite{2013ApJ...770...57B}. {We note that, although this measurement has large signal-to-noise, the constraints on $\mupeak$ has comparable error as \cite{Yan_2022}. We attribute this to more free parameters and fewer tomographic bins in this study.}
The width parameter at high redshift ($\sigma_{M,0}$) and its rate of evolution ($\tau$) are both constrained at values consistent with those reported in \cite{Yan_2022}. The evolution parameter $\tau$ (see Eq.~\eqref{eq:sigma_m}) aligns with these findings. {Finally, only the lower bound of turning redshift ($z_c$) is constrained and our measurements prefer a slightly higher value than that derived in \cite{maniyar_simple_2021} and employed by \cite{Yan_2022}. However, we note that our galaxy sample does not go to higher redshift, so it is hard to interpret the constraints on $z_c$. We leave this to future studies when galaxies at higher redshifts are available.} A summary of the constraints on SFR parameters is presented in Table \ref{tab:sfr_fit}.

The constraints on SFR are not particularly strong with our dataset, primarily due to the limited number of three tomographic bins, resulting in relatively low redshift resolution compared to the KiDS data employed in \cite{Yan_2022}. Nevertheless, our data does yield reasonably sensible constraints on $\mupeak$ when using the Y23 model, compared to previous findings. Additionally, many of the empirical model parameters lack a straightforward physical interpretation. Therefore, it is more informative to examine the derived SFR density (SFRD) based on these parameter constraints.
To derive the SFRD from the Markov Chain Monte Carlo (MCMC) chains, we sample the SFRD. However, for the S12 and Y23 models, the SFRD cannot be effectively constrained using our cross-correlation measurements alone. This limitation arises because the overall amplitude factor $\eta_{\mathrm{max}}$ is degenerate with the normalization factor in the Spectral Energy Distribution (SED) and is absorbed into the overall normalization parameter $L_0$ (see Eq \eqref{eq:fs12}).
To address this constraint and effectively constrain the SFRD, we introduce $\rho_0\equiv\rho(z=0)=0.015M_{\mathrm{\odot}}\mathrm{Mpc^{-3}yr^{-1}}$, obtained from the model derived in \cite{madau_cosmic_2014}. We then convert it to the same Initial Mass Function (IMF) used in our work, following the approach outlined in \cite{2016davies}. Subsequently, we rescale the SFRD sample to ensure that the mean SFRD at $z=0$ corresponds to $\rho_0$.

Figure~\ref{fig:sfrd_t} illustrates the SFRD derived from the last 10,000 points of the Markov Chain Monte Carlo (MCMC) chains generated by our three models. The shaded regions represent the 1-sigma regions for the SFRDs, while the solid lines represent the mean SFRD values computed from the sampled points. Overlaid on the plot are SFRDs constrained by prior multi-wavelength observations \citep{2013ruppioni,2013magnelli,2016MNRAS.456.1999M,2016davies}, depicted as data points with error bars. The blue line corresponds to the best-fit model presented in \cite{madau_cosmic_2014}. {The vertical shaded regions show the redshift distribution of the whole \unwise sample where the white region means more galaxy. We note that the constraints at $z\gtrsim 2$ are beyond the redshift range of our galaxy sample, so they are likely model-driven. Within the \unwise redshifts,}
the M21 model and its analysis closely align with the findings of \cite{Yan_2022}, demonstrating consistency with previous results at the 1-sigma level. Furthermore, the S12 and Y23 models yield results that are consistent with each other, given the external constraint at $z=0$. This consistency implies that the cross-correlation method, in combination with our models and minimal reliance on external information, provides independent measurements of the SFRD.

\begin{figure}
    \centering
    \includegraphics[width=\textwidth]{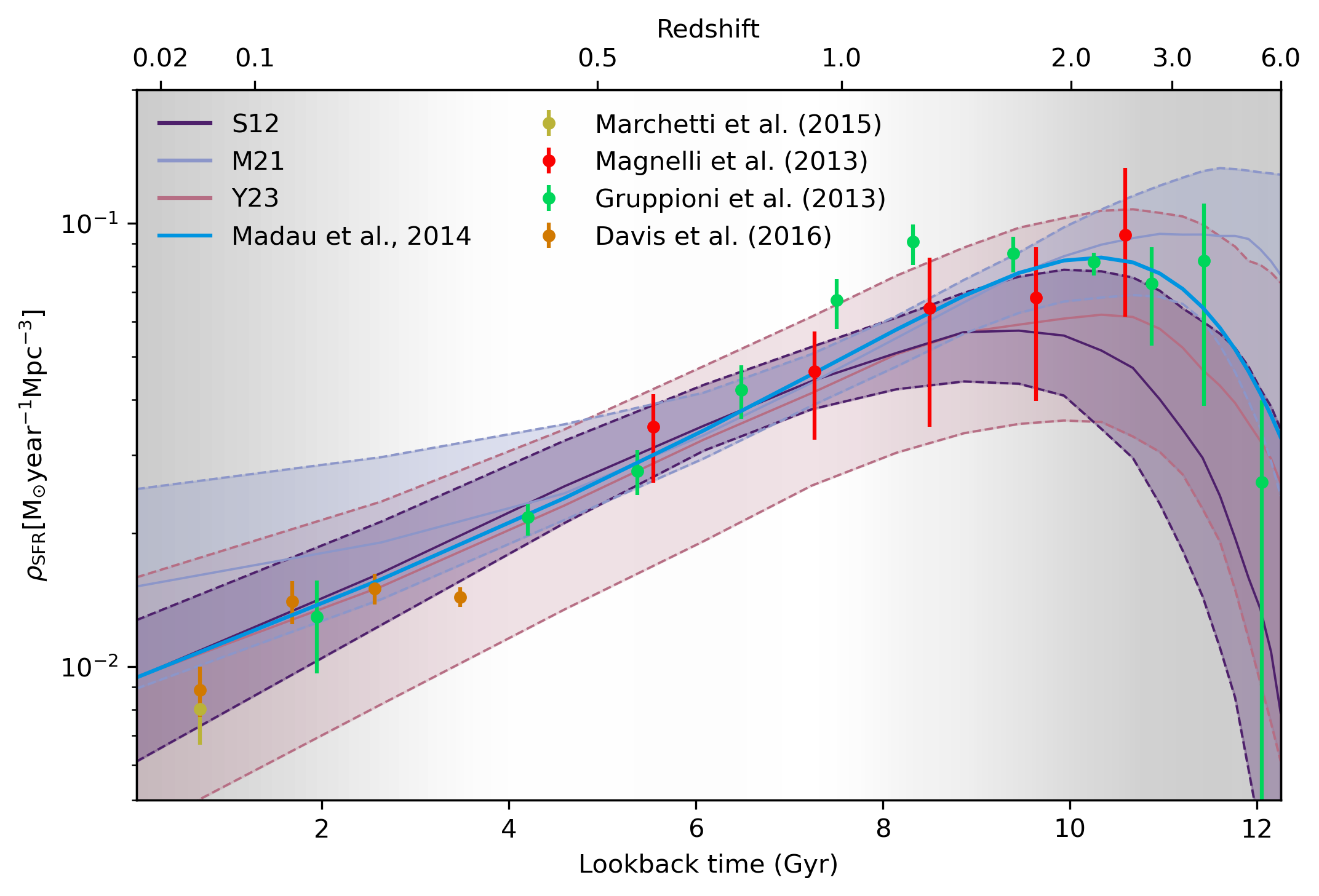}
    \caption{The evolution of SFRD with respect to lookback time (the lower x-axis) and redshift (the upper x-axis). The SFRD calculated from the three models in this work are presented as lines and shaded regions enclosing the $1\sigma$ credible region of the fits that are calculated from 10,000 realizations of SFR parameters from the posterior distributions. The blue line is the best-fit SFRD from \cite{madau_cosmic_2014}, and the points with error bars are SFRD from previous studies \citep[][]{2013ruppioni,2013magnelli, 2016MNRAS.456.1999M, 2016davies}. The gray-shaded regions indicate the redshift ranges that cannot be probed with each \unwise sample (the transparency corresponds to the redshift distribution).}
    \label{fig:sfrd_t}
\end{figure}

\subsection{Constraints on spectral energy density (SED)}
\label{subsec:sed}
\begin{figure}
    \centering
    \includegraphics[width=0.8\textwidth]{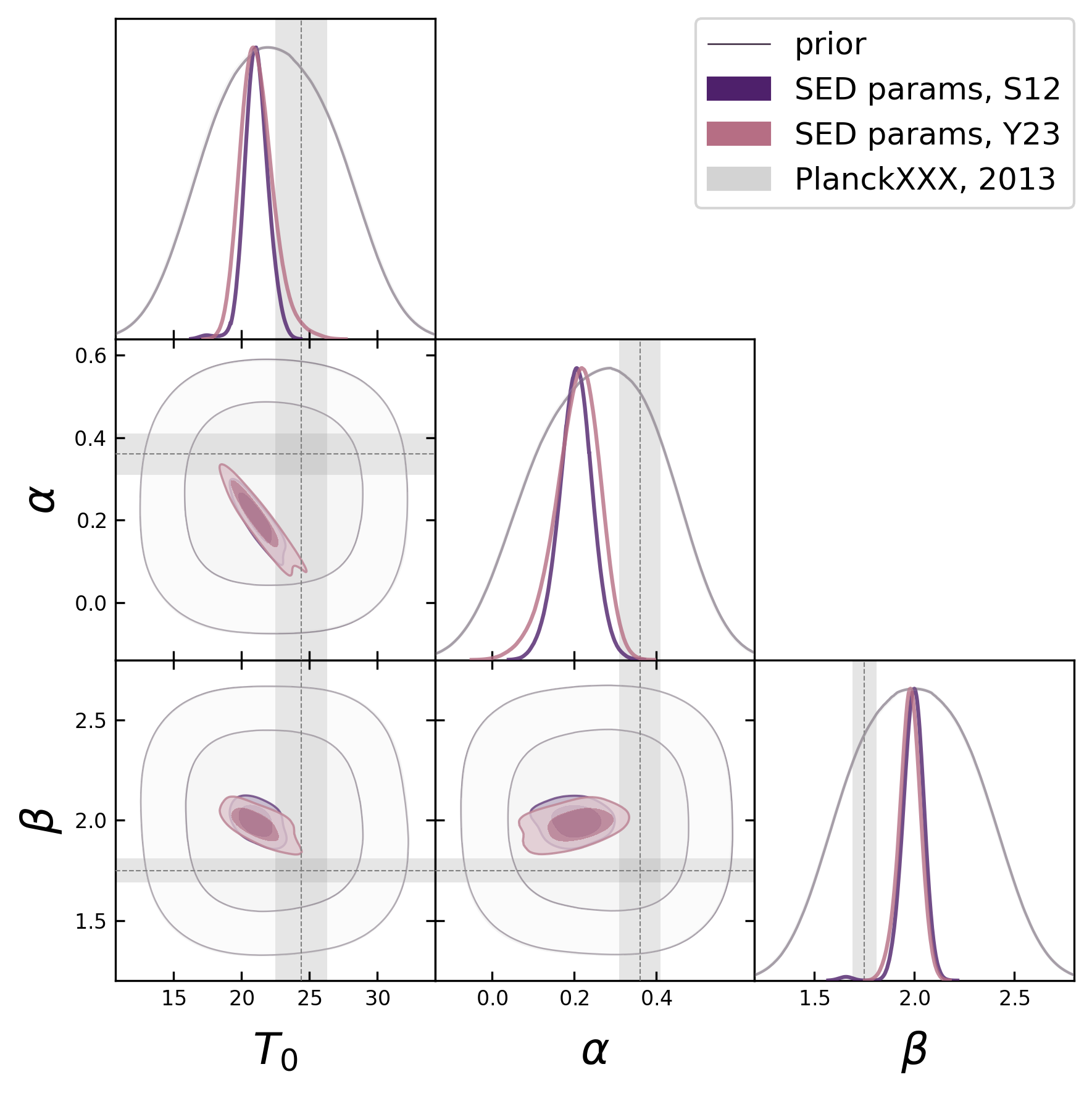}
    \caption{The MCMC constraints of model parameters related to the SED. Color-coded contours are posteriors of the S12 and Y23 models respectively. All the contours are smoothed with a Gaussian kernel density estimate (KDE). The gray unfilled contours are the prior distribution smoothed with the same KDE. The best-fit values and 68\% credible levels from \cite{2014planckxxx} are also shown with the dashed lines and gray bands respectively.}
    \label{fig:sed_post}
\end{figure}

\begin{table}
    \centering
\begin{tabular}{llccc} \\ 
\toprule 
 Parameters & priors& S12 & Y23 & Reference \cite{2014planckxxx}\\ 
 \hline 
$T_{0}$ & [15, 30] & ${21.09}^{+0.82}_{-0.86}$& ${21.14}^{+1.02}_{-1.34}$& -\\ 
$\alpha$ & [0.0, 0.5] & ${0.2}^{+0.04}_{-0.04}$& ${0.21}^{+0.06}_{-0.05}$& $0.36\pm 0.05$\\ 
$\beta$ & [1.5, 2.5] & ${1.99}^{+0.06}_{-0.05}$& ${1.98}^{+0.06}_{-0.05}$& $1.75\pm0.06$\\ 
 \bottomrule 
\end{tabular}
    \caption{A summary of the prior ranges, the marginalized mean values, and the 68\% regions of SED parameters of the S12 and Y23 models. {The values and errors are calculated from the posteriors marginalized over all the other parameters.} The last column shows the constraint from \cite{2014planckxxx} as a reference.}
    \label{tab:sed_params}
\end{table}

\begin{figure*}
    \centering
    \includegraphics[width=\textwidth]{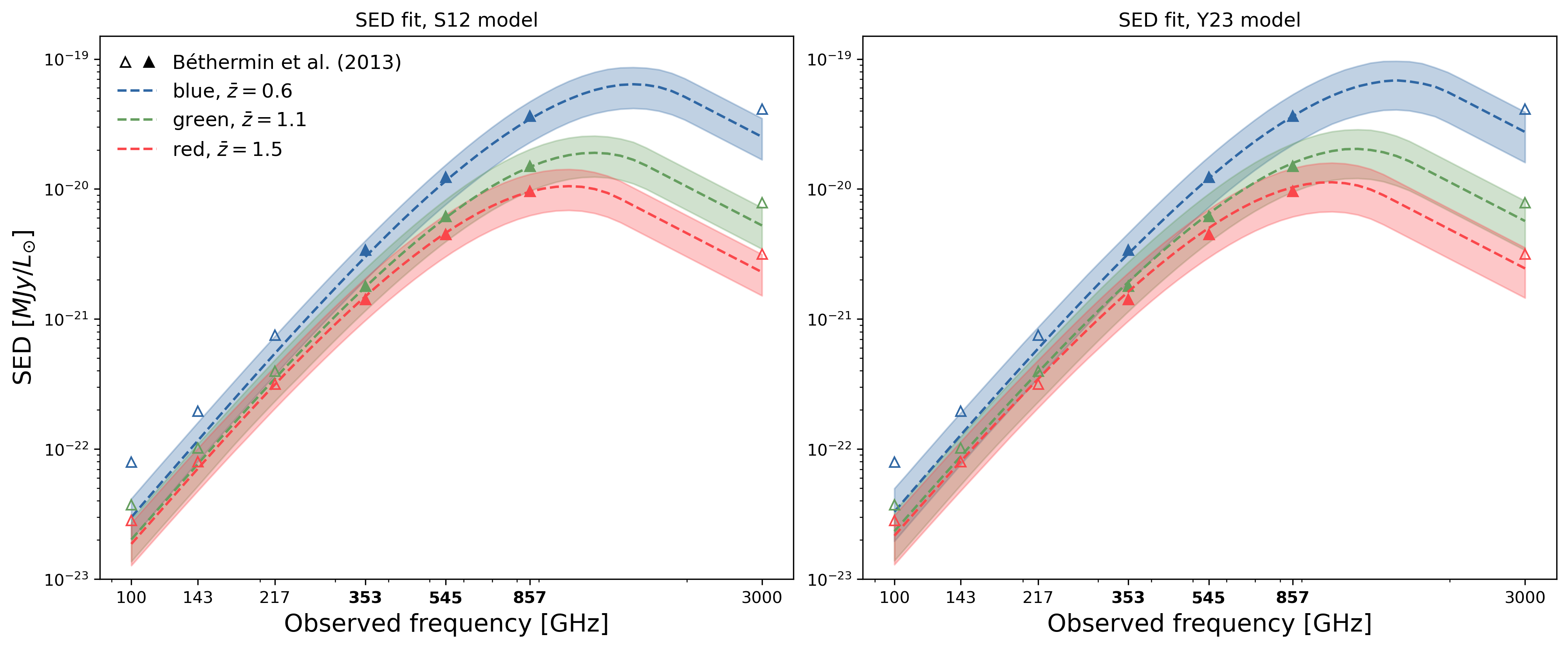}

\caption{The SED derived from the last 10,000 points of the MCMC chains of S12 and Y23 models. The bands show the 1-sigma credible region and the dashed lines show the mean of the SED samples. The SED measured by \cite{bethermin_redshift_2013} (which is used in the M21 model) {is shown with empty and filled triangular points. The frequencies that we used in this measurement are highlighted with filled triangles and bold tickers on the x-axis.}}
\label{fig:sed}
\end{figure*}

We constrain the Spectral Energy Distribution (SED) using the S12 and Y23 models, focusing on the parameters: $T_0$, the dust temperature at $z=0$; $\alpha$, the redshift-dependence parameter of dust temperature; and $\beta$, the power index of the grey-body spectrum. Figure~\ref{fig:sed_post} displays the posterior distributions derived from the last 10,000 points of the Markov Chain Monte Carlo (MCMC) chains. The plot includes gray contours representing the prior distribution and previous constraints from \cite{2014planckxxx}, indicated by gray bands around dashed lines.
A summary of the marginalized mean values and 68\% regions for these parameters is provided in Table~\ref{tab:sed_params}. Notably, the contours in the figure reveal that the SED parameters are tightly constrained and exhibit nearly identical constraints between the S12 and Y23 models.
The constraints indicate that the dust temperature at $z=0$ is slightly lower than the value reported in \cite{2014planckxxx}, and the redshift dependence is weaker. This suggests that the dust temperature for our three galaxy samples is systematically lower than the values provided by \cite{2014planckxxx} at all redshifts. Additionally, our results imply a lower dust temperature than the $26\mathrm{K}$ measurement for luminous red galaxies (LRGs) at $z\sim0.6$ presented by \cite{Serra_2014}.
Furthermore, the spectral index $\beta$ is constrained to be higher than in previous studies. \footnote{We note that $\beta$ and $T_0$ are degenerate. So to some extent, it is expected that if our $T_0$ is smaller than Planck values, $\beta$ is expected to be higher.} Given that different cross-correlations are sensitive to various infrared sources (e.g., the CIB auto power spectra in \cite{2014planckxxx} encompass the total diffused IR emission from diverse sources; the CIB-LSS cross-correlation in \cite{Serra_2014} characterizes IR emission from LRGs, whereas our study constraints IR models for general galaxy populations), the tension between our results and prior findings suggests a fascinating diversity of dust properties among different galaxy types. This topic warrants further exploration in future studies.

Due to the degeneracy between the normalization of the SED and the SFRD, we can constrain neither of them. Instead, we can only constrain the IR emission $j_{\nu}(z)$ (see Eq.~\eqref{eq:jsfrd}) which is proportional to the multiplication of SED and SFR. However, with the introduction of $\rho_0$, we can normalize the SFRD independently and factor it out of $j_{\nu}(z)$, thus we can derive the SED constraint. In Fig.~\ref{fig:sed}, we present the constrained SED constructed from the last 10,000 point of the MCMC chains. The bands show the 68\% sigma regions and the dashed line shows the mean SED of the samples. The SED measured by \cite{bethermin_redshift_2013} and used in the M21 model is shown as {filled and empty} triangular points. The three \planck HFI frequency bands that we are using are highlighted as filled triangles. {The three curves are separate from each other due to redshifts and the redshift evolution of the dust temperature.} We conclude that, with the external information of $\rho_0$, the SED is in agreement with that given by \cite{bethermin_redshift_2013} with a 1-sigma level. It should also be noted that frequency bands that are outside our measurement are slightly biased, which calls for CIB data in a wider frequency range.

\subsection{Constraints on halo occupation distributions (HOD)}
\label{subsec:hod_constrain}
\begin{figure}
    \centering
    \includegraphics[width=0.8\textwidth]{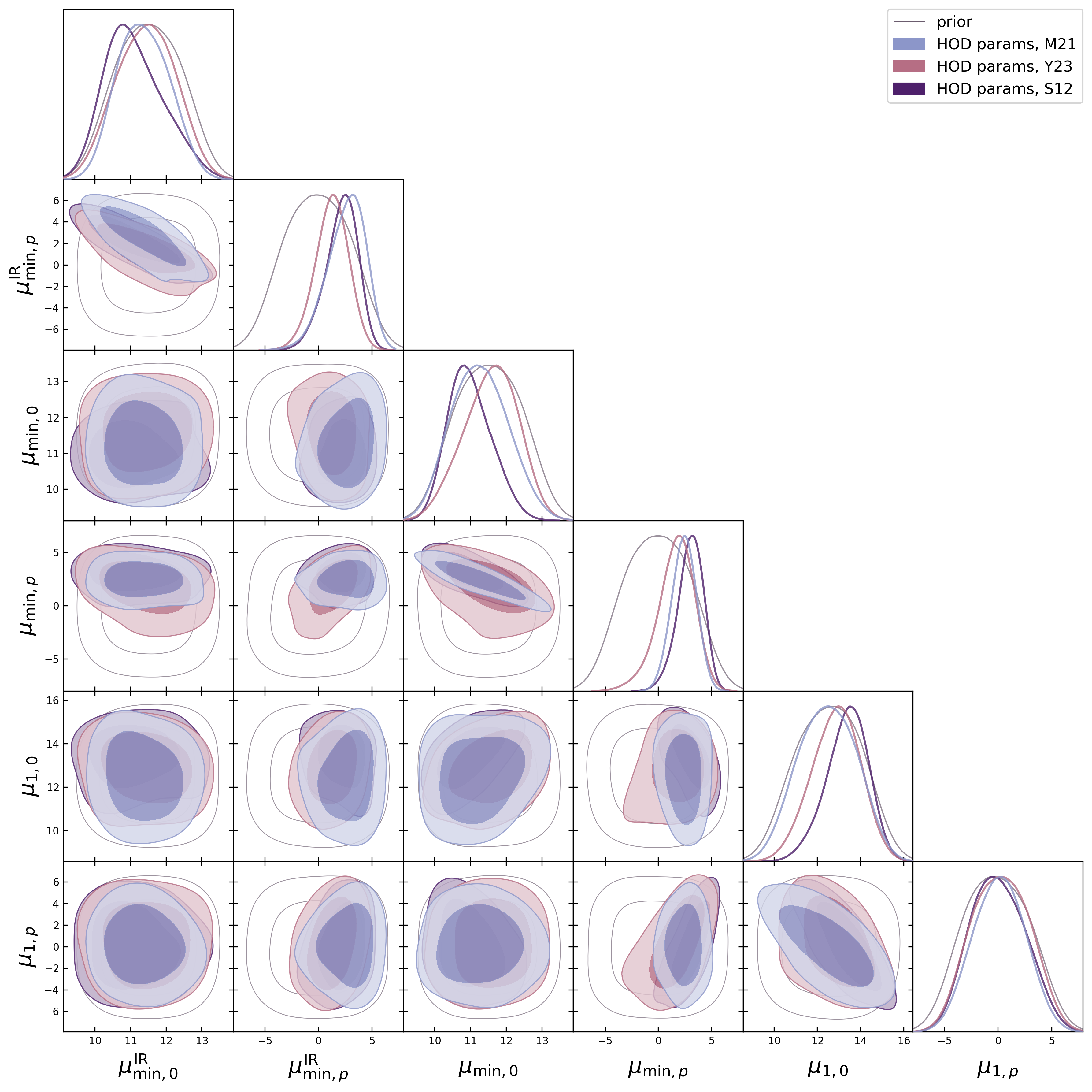}
    \caption{The MCMC constraints of model parameters related to the HOD. Color-coded contours are posteriors of the S12, M21, and Y23 models respectively. All the contours are smoothed with a Gaussian kernel density estimate (KDE). The gray unfilled contours are the prior distribution smoothed with the same KDE.}
    \label{fig:hod_post}
\end{figure}

\begin{figure}
    \centering
    \includegraphics[width=\textwidth]{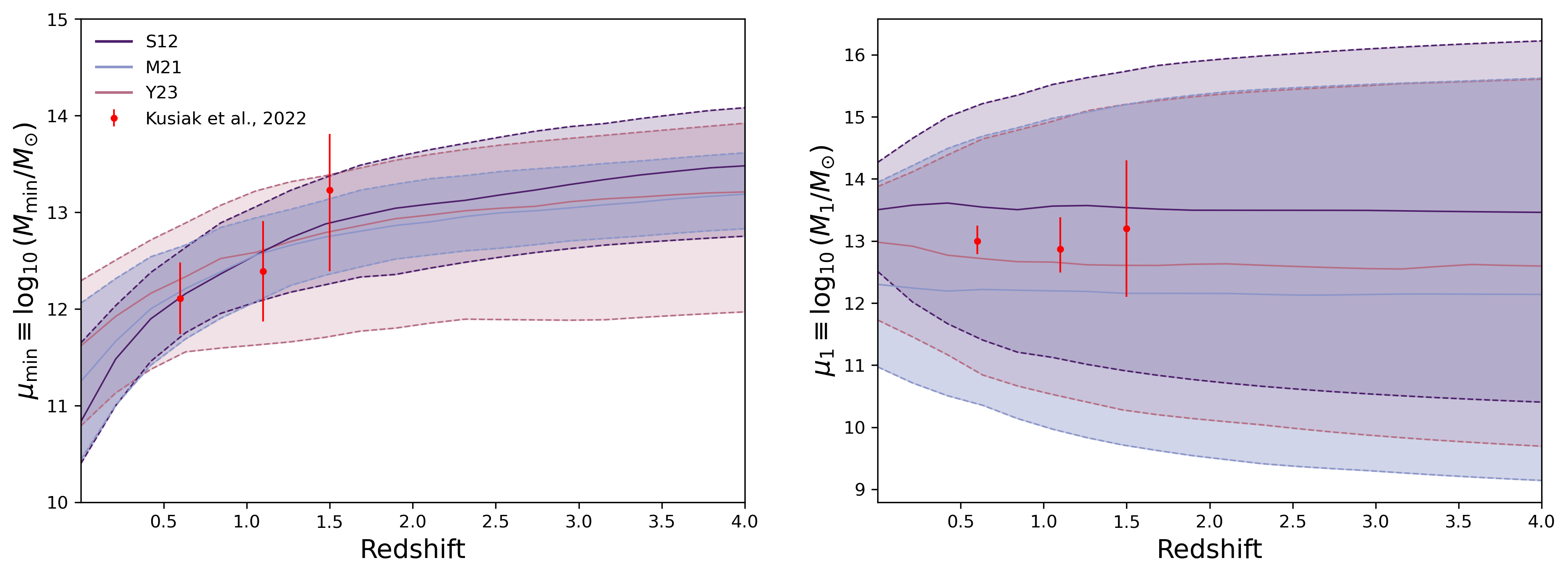}
    \caption{The evolution of HOD parameters $\mu_{\mathrm{min}}$ (the left panel) and $\mu_1$ (the right panel) with respect to redshift. The parameters calculated from the three models in this work are presented as lines and shaded regions enclosing the $1\sigma$ credible region of the fits that are calculated from 10,000 realizations of HOD parameters from the posterior distributions. The red dots with error bars are the constrained parameters from \cite{Kusiak_2022} for the blue, green, and red samples, respectively. The redshifts for these points are taken as the mean redshifts of the three galaxy samples.}
    \label{fig:lnmin}
\end{figure}

\begin{table}
    \centering
\begin{tabular}{llccc} \\ 
\toprule 
 Parameters & priors& S12 & M21 & Y23 \\ 
 \hline 
$\mu_{\mathrm{min},0}^{\mathrm{IR}}$ & [10, 13] & ${11.1}^{+0.67}_{-0.98}$& ${11.36}^{+0.71}_{-0.78}$& ${11.44}^{+0.86}_{-0.84}$\\ 
$\mu_{\mathrm{min},p}^{\mathrm{IR}}$ & [-5, 5] & ${2.21}^{+1.68}_{-1.26}$& ${2.65}^{+2.01}_{-1.5}$& ${1.28}^{+1.63}_{-1.46}$\\ 
$\mu_{\mathrm{min},0}$ & [10, 13] & ${10.98}^{+0.53}_{-0.69}$& ${11.26}^{+0.76}_{-0.86}$& ${11.58}^{+0.84}_{-0.7}$\\ 
$\mu_{\mathrm{min},p}$ & [-5, 5] & ${3.04}^{+1.32}_{-1.07}$& ${2.43}^{+1.2}_{-1.13}$& ${1.7}^{+1.85}_{-1.48}$\\ 
$\mu_{1,0}$ & [10, 15] & ${13.39}^{+1.06}_{-0.86}$& ${12.48}^{+1.4}_{-1.36}$& ${12.87}^{+1.2}_{-1.06}$\\ 
$\mu_{1,p}$ & [-5, 5] & ${0.0}^{+2.49}_{-2.79}$& ${0.11}^{+2.48}_{-2.44}$& ${0.26}^{+2.76}_{-2.79}$\\ 
 \bottomrule 
\end{tabular}
    \caption{A summary of the prior ranges, the marginalized mean values, and the 68\% regions of HOD parameters of our three models. {The parameters with superscript ``IR'' are those for IR galaxies that generate the CIB, while others are for \unwise galaxies.} {The values and errors are calculated from the posteriors marginalized over all the other parameters.}}
    \label{tab:hod_params}
\end{table}

The parameters of the Halo Occupation Distribution (HOD) model to be constrained in this study are ${\mu_{\mathrm{min},0}^{\mathrm{IR}}, \mu_{\mathrm{min},0}^{\mathrm{IR}}, \mu_{\mathrm{min},0},\mu_{\mathrm{min},p},\mu_{\mathrm{1},0},\mu_{\mathrm{1},p}}$. It's important to note that our measurement cannot effectively constrain the other HOD parameters due to limitations of the angular resolution of the CIB maps. In contrast, in \cite{Krolewski_2020}, the authors constrained HOD model parameters using the \unwise galaxy clustering power spectra, which provided constraints on the galaxy-halo connection for the entire galaxy samples. In our work, we measure the CIB-galaxy cross-correlation, {which is primarily sensitive to galaxies with strong IR emissions. Therefore, we can constrain galaxy-halo connections for IR galaxies. By comparing with the galaxy power spectrum which probes general galaxy-halo connection, we can explore the difference in clustering across different galaxy types}. As a result, we adopt an evolution model with Eq.~\eqref{eq:mp}, constraining $\mu_{\mathrm{min}}$ and $\mu_{\mathrm{1}}$ simultaneously. The other HOD parameters remain fixed since our measurements lack the statistical power required to constrain them effectively.
In Figure.~\ref{fig:hod_post}, the posterior distributions of the HOD parameters from our MCMC analysis are depicted in filled colorful contours, while the unfilled contours represent the prior distribution, smoothed using the same Gaussian KDE as the posterior distributions. We also summarize the marginalized mean values and the 68\% credible regions of HOD parameters in Table.~\ref{tab:hod_params}.
We extract the redshift evolution of the \unwise parameters $\mu_{\mathrm{min}}$ and $\mu_1$ with Eq.\eqref{eq:mp} from the posterior. The results for $\mu_{\mathrm{min}}$ and $\mu_1$ are presented in the two panels of Figure~\ref{fig:lnmin}, respectively. The lines in the figure represent the marginalized means of the parameters, and the shaded bands indicate the 1-sigma regions. Additionally, we overlay the constraints obtained from \cite{Kusiak_2022} as red points with error bars. Notably, in \cite{Kusiak_2022}, the parameters were constrained individually for each of the three galaxy samples.
From the figure, it becomes evident that our measurement does not effectively constrain $\mu_1$, which can be attributed to the limitations imposed by angular resolution. Conversely, our constraint on $\mu_{\mathrm{min}}$ reveals a slight increasing trend in $M_{\mathrm{min}}$ with redshift, consistent with the findings of \cite{Kusiak_2022}. This observation implies that the minimum mass required to host a central galaxy increases with redshift, aligning with the expectation that galaxies only form in the most massive halos in the early Universe.

In the regime of linear scales, we can express the galaxy overdensity as a linear proportion to the matter overdensity, as indicated by the equation $\delta_{\mathrm{g}}=b_{\mathrm{g}}\delta_{\mathrm{m}}$. Here, the parameter $b_{\mathrm{g}}$ represents the effective linear galaxy bias, which serves as a measure of how galaxies cluster in relation to the underlying mass distribution. Within the framework of the halo model, we define galaxy bias as the weighted linear halo bias denoted as $b_{\mathrm{h}}$. This definition incorporates considerations of the halo mass function and the distribution of galaxies in relation to halo mass.

\begin{equation}
    b_{\mathrm{g}}(z)=\frac{1}{\bar{n}_{\mathrm{g}}(z)}\int \dr M  \frac{\dr n}{\dr M}b_{\mathrm{h}}(M,z)\left[ N_{\mathrm{c}}(M, z) + N_{\mathrm{s}}(M, z)\right],
    \label{eq:bg}
\end{equation}
where the mean galaxy number density, denoted as $\bar{n}_{\mathrm{g}}$, is calculated using Eq.~\eqref{eq:ngal_hod}. To remain consistent with prior practices, we sample the redshift-dependent parameter $b_{\mathrm{g}}(z)$ from the Markov Chain Monte Carlo (MCMC) chains and present the results in Figure~\ref{fig:bg_z}. Overlaid on this figure are previous findings from galaxy cross-correlations and auto-correlations, depicted as data points with error bars (cross-correlation with spectroscopic sample) and green lines (Halo Occupation Distribution from auto-correlations as derived from the CrowCanyon2 simulation) \cite{Krolewski_2020}\cite{Krolewski_2021}. Each panel corresponds to the previous results calculated using one of the three \unwise samples. The shaded regions represent redshift ranges that contain relatively few galaxies in each sample, with the degree of transparency corresponding to the redshift distribution.
It is worth noting that our constraints are in agreement with the results reported by \cite{Krolewski_2020} and \cite{Krolewski_2021} across all three samples. This alignment indicates that CIB-galaxy cross-correlation serves as a valid tool for probing galaxy clustering on both linear and small scales. Additionally, \cite{Yan_2022} found a slight discrepancy in the linear galaxy bias measured from KiDS galaxy-CIB cross-correlations compared to galaxy-CMB lensing cross-correlations \cite{Yan_2019}. They attributed this difference to the increased clustering of infrared galaxies relative to general galaxies in the KiDS sample.
However, our work shows better consistency between the galaxy bias derived from galaxy power spectra and galaxy-CIB cross-correlations. This consistency arises because the \unwise catalog is selected in the near-infrared, meaning that the galaxy power spectra and galaxy-CIB cross-correlations essentially probe nearly identical galaxy populations. It would be of interest to compare our $b_{\mathrm{g}}(z)$ results, along with those from \cite{Krolewski_2020} and \cite{Krolewski_2021}, to those from another galaxy sample. Such a comparison would allow us to gain insight into the dependence of galaxy clustering on different galaxy types {and potentially probe the regime of the galaxy assembly bias}.

\begin{figure}
    \centering
    \includegraphics[width=\textwidth]{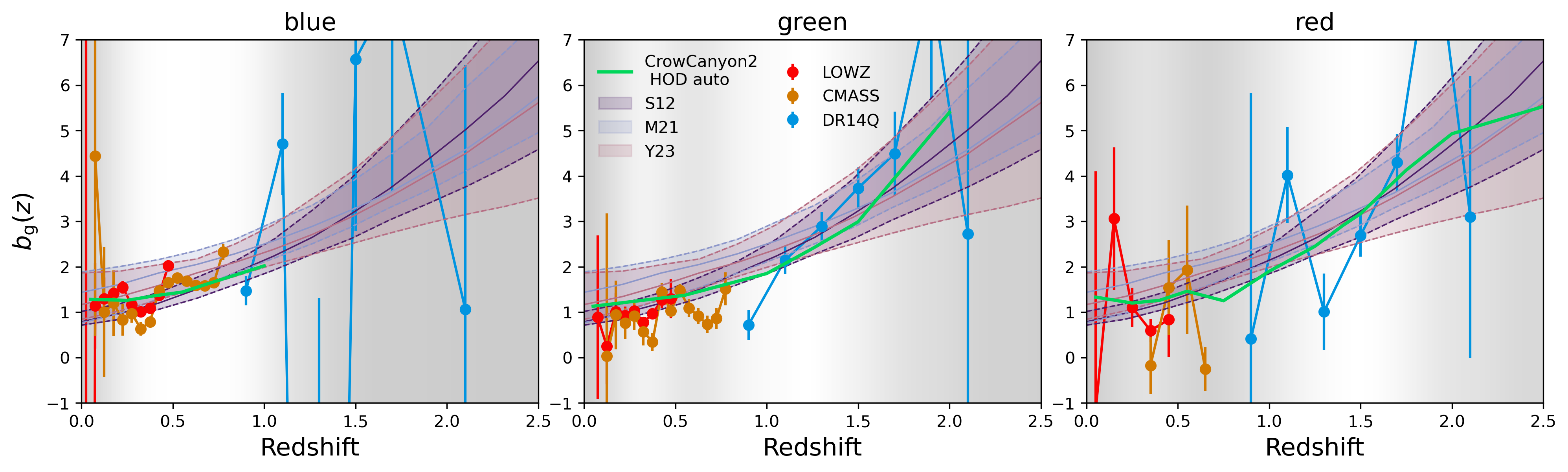}
    \caption{Effective linear galaxy bias $b_{\mathrm{g}}(z)$  versus redshift derived from the MCMC chains, overplotted with galaxy biases measured by cross-correlating each \unwise sample with LOWZ, CMASS, quasars from DR14 as data points with error bars \cite{Krolewski_2020, Krolewski_2021}, and HOD fit from auto-correlations given by the CrowCanyon2 simulation (the green solid lines). The three panels differ in the \unwise sample used to calculate $b_{\mathrm{g}}(z)$ in \cite{Krolewski_2020} (so the constraints from this work are the same in all the three panels). The gray-shaded regions indicate the redshift ranges that cannot be probed with each \unwise sample (the transparency corresponds to the redshift distribution).}
    \label{fig:bg_z}
\end{figure}

{\subsection{Systematics test of selection effect}
	\label{sect:syst_test}
	In this work, we use the halo model for the CIB-galaxy cross-correlation and fit the models with a wide range of angular scales. The contribution from the 1-halo regime is from unWISE galaxies; while the 2-halo regime is from the large-scale structure. As we mentioned in Sect.~\ref{subsec:selection}, the \unwise galaxy sample is subject to selection effects. If different galaxy populations in the \unwise catalog have significantly different IR emissions, there should be inconsistency in the parameter constraints between small and large scales, which makes the interpretation of our MCMC constraints difficult. We take different HODs for \unwise galaxy sample and the CIB to account for the selection effect in galaxy abundance, while we are interested in cosmic star formation history and the SED of the CIB. Therefore, we expect the constraints of SFR and SED to mainly come from large scales, while that of the HOD is mainly from small scales. In this subsection, we will verify this by two methods.
	
	\subsubsection{Constraint of bias-weighted SFRD}
	
	According to Eq.~\eqref{eq:halomodel_pk}, the 2-halo regime of the power spectrum is the linear power spectrum multiplied by the bias factor $\left\langle b_u\rangle\right(k)$. In the linear regime, we can further take $k\rightarrow 0$, then the bias factor for the galaxy is just the linear galaxy bias $b_{\mathrm{g}}(z)$. For CIB, combining Eq.~\eqref{eq:lir} with the third formula in Eq.~\eqref{eq:halomodel_pk}, we can derive its bias factor: }

\begin{equation}
	\begin{aligned}
		\left\langle b_{\nu} \rangle\right (k\rightarrow 0) &= \frac{\chi^2(1+z)S_{\mathrm{eff}}[(1+z)\nu, z]}{K}\int_0^{\inf}\dr M \frac{\dr n}{\dr M}b_{\mathrm{h}}(M)\mathrm{SFR}(M,z) \\
		& = \frac{\chi^2(1+z)S_{\mathrm{eff}}[(1+z)\nu, z]}{K} \left\langle b\rho_{\mathrm{SFR}}\rangle\right(z),
	\end{aligned}	
\end{equation}
{where $\left\langle b\rho_{\mathrm{SFR}}\rangle\right(z)$ is the ``bias-weighted SFRD'' that gives the biased SFRD of the large-scale structure. Note that this parameter can be constrained in the linear regime without assuming an SFR model. In \cite{Jego_2023}, the authors constrained this parameter up to $z\sim 2$ with tomographic galaxy-CIB cross-correlations.
	
We can use the SFR and HOD parameters constrained with CIB halo models to derive $\left\langle b\rho_{\mathrm{SFR}}\rangle\right(z)$. Note that it depends on the SFR parameters, the HOD parameters for the CIB, and the LSS, it does not depend on selections in the galaxy catalog to cross-correlate. Therefore, we can compare our derived $\left\langle b\rho_{\mathrm{SFR}}\rangle\right(z)$ with that measured from \cite{Jego_2023}. We show the result in Figure \ref{fig:bsfrd}. The green points with error bars are from \cite{Jego_2023} while the colored bands are the 1-$\sigma$ credible level of $\left\langle b\rho_{\mathrm{SFR}}\rangle\right(z)$ derived from the SFR parameters in our MCMC chains. We find that the results agree at 1-$\sigma$ level. The agreement between \cite{Jego_2023} and the M21 model is remarkable because they both use the same SED model.
	
	\begin{figure}
		\centering
		\includegraphics[width=\textwidth]{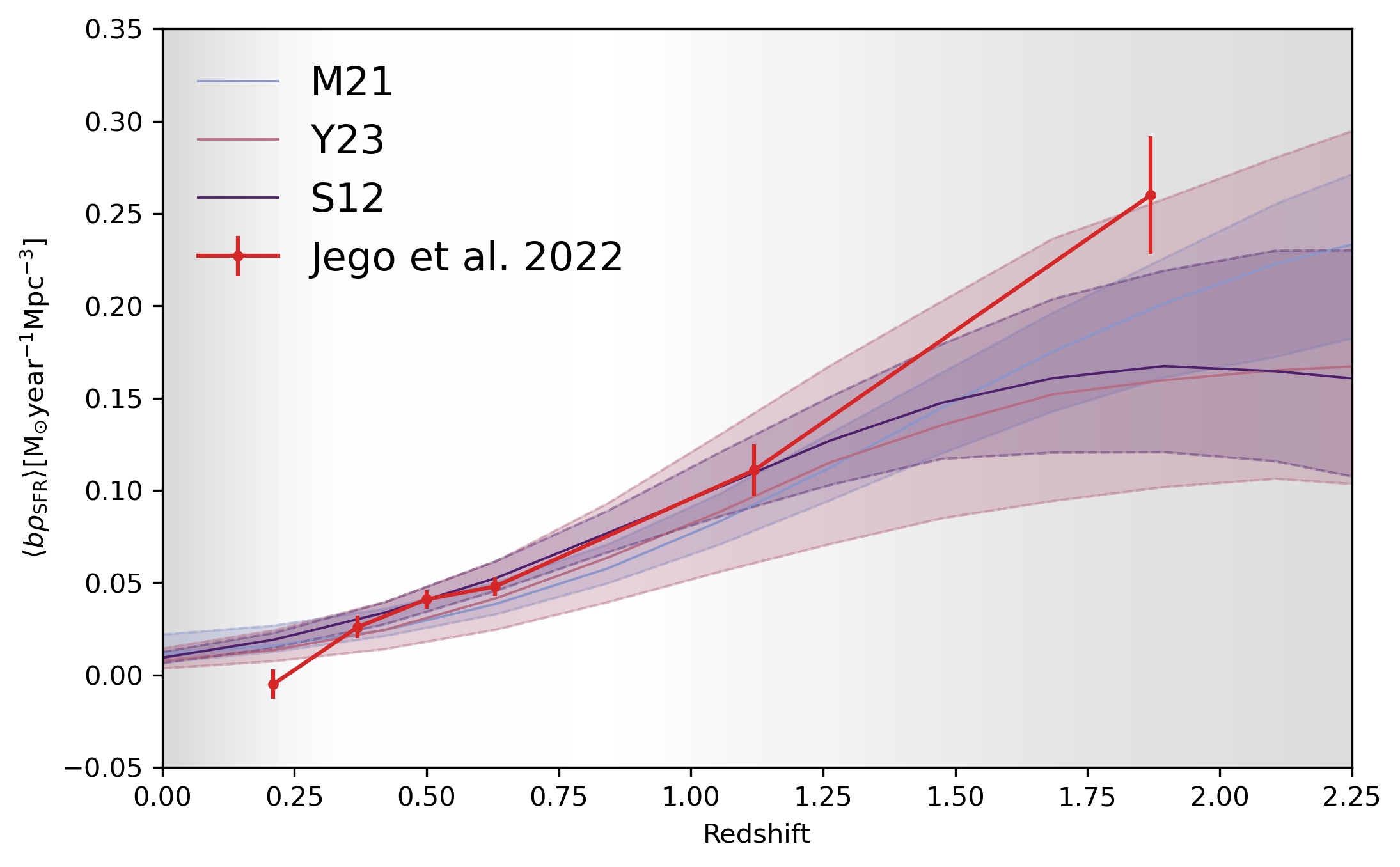}
		\caption{The evolution of bias-weighted SFRD with respect to redshift (the upper x-axis). The bias-weighted SFRD calculated from the three models in this work are presented as lines and shaded regions enclosing the $1\sigma$ credible region of the fits that are calculated from 10,000 realizations of SFR parameters from the posterior distributions. The green points with errorbars are $\left\langle b\rho_{\mathrm{SFR}}\rangle\right(z)$ constrained from \cite{Jego_2023}. The gray-shaded regions indicate the redshift ranges that cannot be probed with each \unwise sample (the transparency corresponds to the redshift distribution).}
		\label{fig:bsfrd}
\end{figure}}

{\subsubsection{Constraints on SFR and SED parameters with scale cuts}
	\label{sect:scalecut}
	
In the halo model, the large scales are dominated by the 2-halo terms contributed by the large-scale structure; while the small scales, the 1-halo terms are dominated by \unwise galaxy abundance. If the selection effect in the \unwise galaxy sample raises biases in SFR and SED parameters, it is hard to interpret the constraint with the wide angular scales that we use. In this section, we test the robustness of our constraints on the SFR and SED against the selection effect by limiting the data vector to large scales. We define the cutting scale in $k$ space such that $k_{\mathrm{cut}}=0.2\,\mathrm{Mpc}^{-1}$ where the two-halo term contributes to the power spectrum by $\gtrsim 90\%$. Due to the wide redshift distribution of the galaxy sample, $k_{\mathrm{cut}}$ does not correspond to a single value of angular scale cut. We define the angular scale cut $\ell_{\mathrm{cut}}$ such that $80\%$ of the sources have $\ell_{\mathrm{cut}}/\chi(z)<k_{\mathrm{cut}}$, where $z$ follows the \unwise redshift distribution. For simplicity, we perform the test for the Y23 model and present the SFR (Fig \ref{fig:sfr_robust}) and SED (Fig \ref{fig:sed_robust}) constraints with and without scale cut. The posterior contour of the fiducial Y23 constraint is presented with filled contours, while the constraints with scale cut are shown with unfilled contours. The light grey unfilled contours show the prior distribution. All the contours are smoothed with the same Gaussian KDE kernel as our nominal contours.
	
The contours show that the constraints from different cuts are consistent with the fiducial constraint within the $1-\sigma$ level. The consistency between the large-scale-only (the $k_{\mathrm{cut}}=0.2\,\mathrm{Mpc}^{-1}$ contours) and the fiducial means that the different galaxy samples give consistent SED and SFR constraint or the constraints come mostly from the 2-halo region. Therefore, we can conclude that the constraints on the SED and SFR are robust against galaxy selections in our sample. On the other hand, some of the HOD parameters are barely constrained with scale cuts, which is expected as their constraints are mostly driven by small scales.

	\begin{figure}
		\centering
		\includegraphics[width=0.7\textwidth]{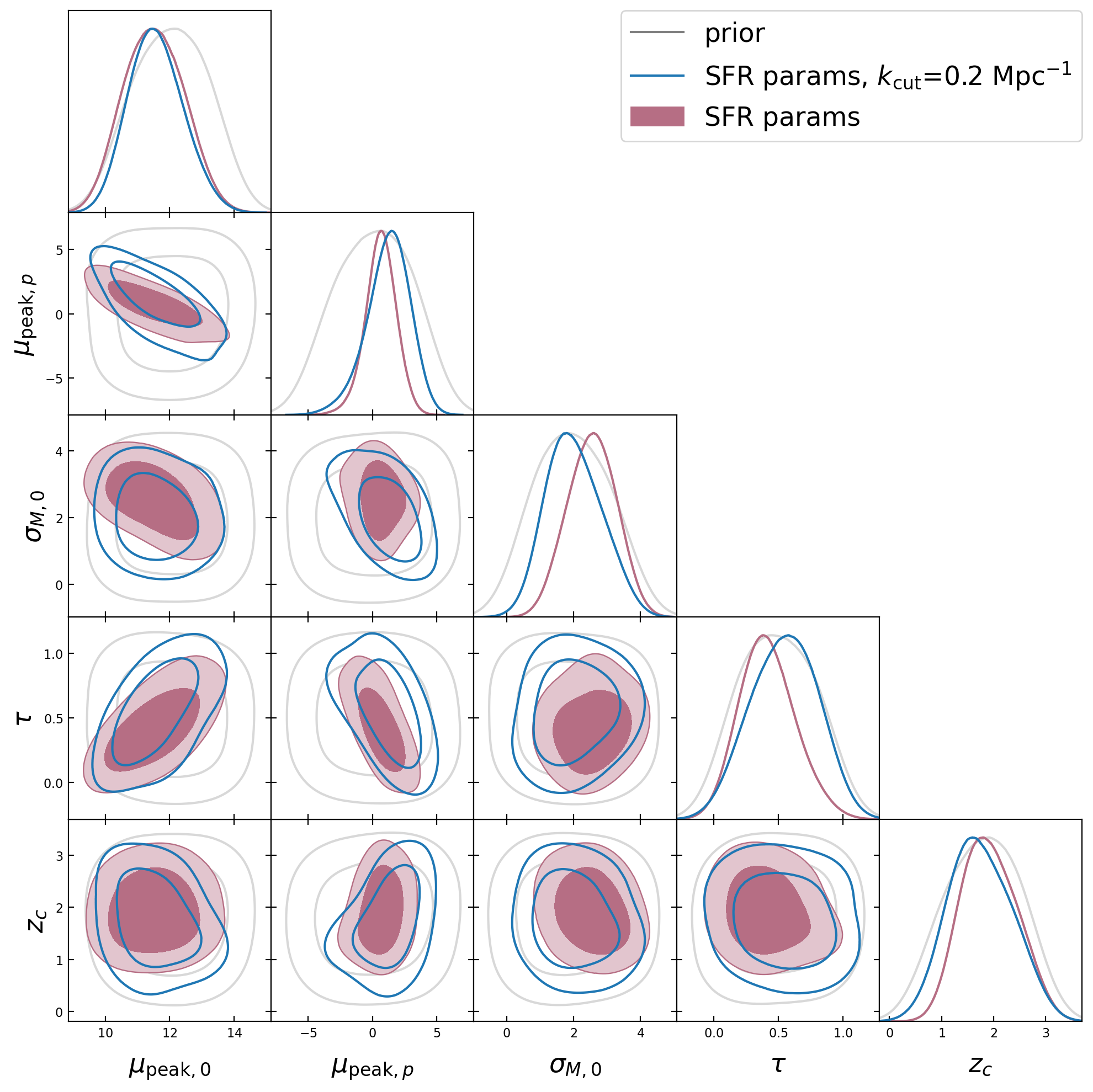}
		\caption{The posterior of SFR parameters. Contours show the marginalized 2-D posterior distributions and curves show the marginalized posterior for each parameter. The filled contour shows the fiducial constraints on the Y23 model, the light grey contour is the prior distribution and the other unfilled contours are constraints from the robustness test.}
		\label{fig:sfr_robust}
	\end{figure}
	
	\begin{figure}
		\centering
		\includegraphics[width=0.7\textwidth]{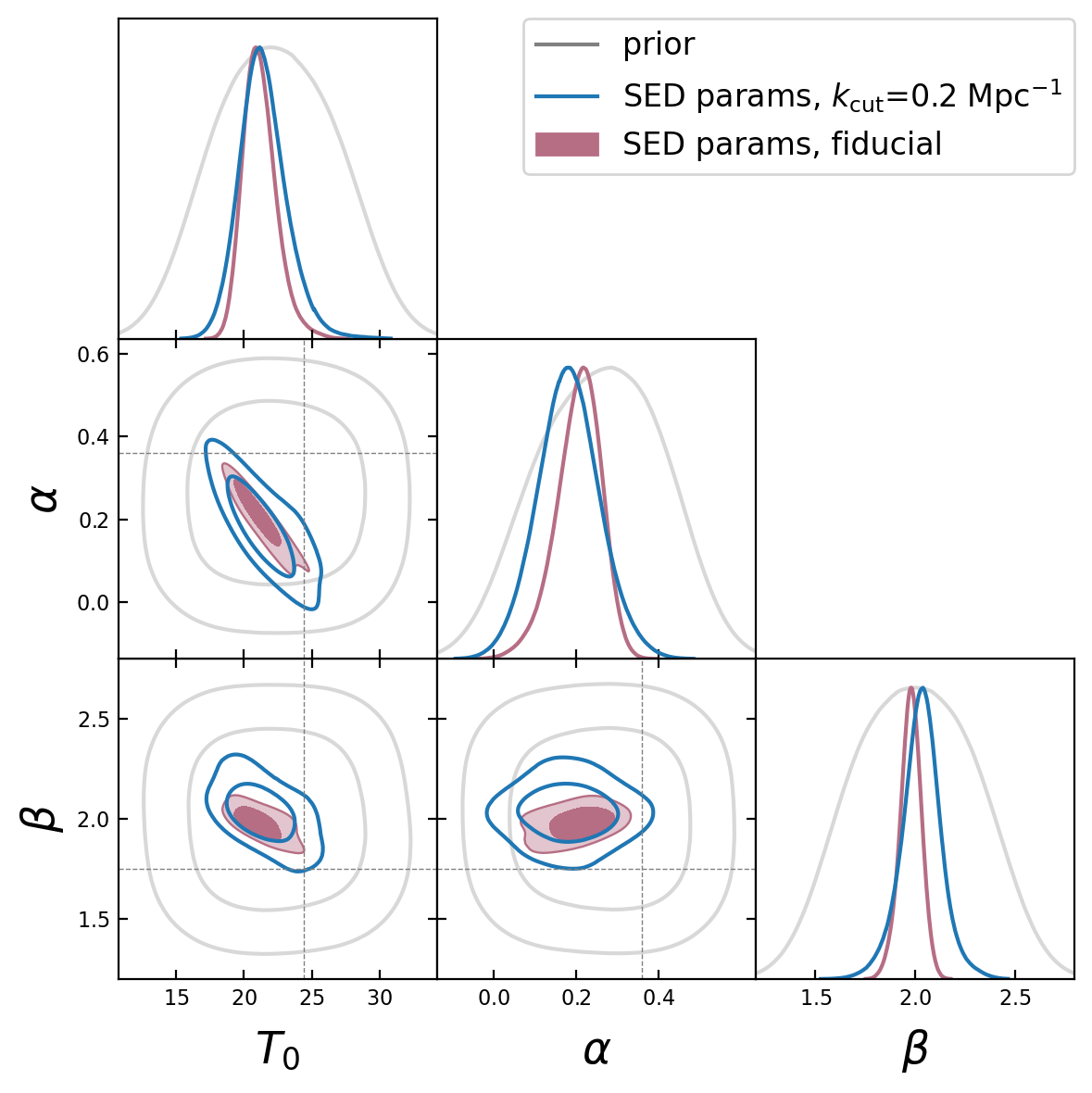}
		\caption{The posterior of SED parameters. Contours show the marginalized 2-D posterior distributions and curves show the marginalized posterior for each parameter. The filled contour shows the fiducial constraints on the Y23 model, the light gray contour is the prior distribution and the other unfilled contours are constraints from the robustness test.}
		\label{fig:sed_robust}
	\end{figure}
	
}

\section{Conclusions}
\label{sect:conclusions}
In our study, we conducted measurements of the cross-correlation between three distinct \unwise galaxy samples and the CIB maps derived from \planck, data at three different frequency bands (353, 545, and 857 GHz). Our primary objective in performing these measurements was to extract valuable insights into several key aspects, including the cosmic star formation history, the Spectral Energy Distribution (SED) of galaxies, and the relationship between galaxies and dark matter halos, as characterized by the Halo Occupation Distribution (HOD) model, utilizing three different CIB models.
We summarize our main conclusions in this section.

\begin{itemize}
    \item We measure the pseudo-$C_{\ell}$ of nine pairs of cross-correlations between three galaxy samples and three CIB maps. We calculate the covariance matrix as a combination of Gaussian covariance, connected non-Gaussian covariance, and the super sample covariance. The mode-coupling in Gaussian covariance is corrected with the \texttt{NaMaster} package, and the other two terms are calculated analytically. We yield a high significance (194$\sigma$ level) cross-correlation signal in angular scales $100<\ell<2000$. All three models give a best-fit reduced $\chi^2$ of $\lesssim1.3$, which is satisfactory given our criteria.
    
    \item The constraints on the SFRD parameters from cross-correlation agree with previous constraints from \cite{Yan_2022}. The most efficient halo mass is constrained slightly lower than the previous results. The evolution of $M_{\mathrm{peak}}$ is constrained consistent with zero, which agrees with the conclusion given by \cite{2013ApJ...770...57B}. When introducing the SFRD at $z=0$ from \cite{madau_cosmic_2014} for the S12 and Y23 model, the derived star formation history agrees with the M21 model and previous multi-wavelength studies within the redshift range covered by \unwise. This enhances the conclusion from \cite{Yan_2022} that CIB-galaxy cross-correlation is a sensible measurement of cosmic star formation history.
    
    \item The SED parameters are tightly constrained in the S12 and Y23 models. The dust temperature at $z=0$ is constrained to be around 21.4 $K$ and the temperature evolution parameter $\alpha$ is constrained to be around 0.2. Compared with the previous constraints given by CIB power spectra and CIB-LRG cross-correlations, we conclude that the dust temperature in our sample is lower and has a weaker evolution. In addition, the frequency power index $\beta$ is constrained to be around 2.0, which is higher than that given by CIB power spectra. These tensions suggest variabilities in dust properties across galaxy samples.

    \item We factor out SFRD from IR emission to recover the SED for S12 and Y23 models. In the frequency bands that we use, both models give consistent SED and are consistent with the SED given by \cite{bethermin_redshift_2013} in 1-sigma level. This indicates that our measurement captures the redshift and frequency dependence of SED consistent with previous studies.

    \item The HOD parameters are mildly constrained. Our measurement favors an increasing $M_{\mathrm{min}}$ with redshift, which indicates that in the early Universe, only massive halos could host a central galaxy. The characteristic mass for satellite galaxies $M_1$ is not constrained in our measurement due to the low angular resolution of the CIB map.

    \item We recover the effective linear galaxy bias from the HOD parameters and find that $b_{\mathrm{g}}$ increases with redshift. Comparing with the results given by cross- and auto-correlation with \unwise data, we find mildly consistent $b_{\mathrm{g}}(z)$.
    
    \item {From the bias-weighted SFRD constraint and the scale-cut tests, we conclude that our constraints on SED and SFR are mainly from the large-scale structure and are robust to the potential selection effects in the galaxy sample.} The improvements in sky coverage, redshift range, number of tomographic bins, and galaxy number density will increase the precision of the data vector, and thus the difference in galaxy population will more likely cause inconsistency in those constraints.

\end{itemize}

We will discuss the indications and prospects of these conclusions in the next section.

\section{Discussions}
\label{sect:discussions}
{Compared to other large-scale structure tracers, like the tSZ effect, weak lensing, and CMB lensing, the CIB does not have a standard model due to the complicated nature of the IR galaxies. However, the CIB contains mixed information about star formation, dust properties and galaxy abundance, consequently, different studies in the literature approach the CIB with different models. In this work, we first review the halo model-based CIB models in the literature and derive a unified formula (Eq.~\ref{eq:lir}). The most important assumption in the CIB model is the Kennicutt relation which involves the SFR. We classify the CIB models into the S12 model in Sect. \ref{subsec:S12} and the M21 model in Sect. \ref{subsec:M21} based on different assumptions about the SFR and SED. We also propose a more physically motivated Y23 model by combining the assumptions in the S12 and M21 models.}

We measure a significant cross-correlation between the CIB and galaxies from the \unwise catalog, which suggests that the physical origin of CIB is in galaxies. Moreover, from the covariance matrix, we find strong correlations between different CIB frequency bands, which means that CIB emissions in different frequencies essentially have almost the same origin. We fit our models with MCMC by assuming a Gaussian likelihood. The best-fit S12, M21, and Y23 models give $\chi^2/\lesssim 1.3$, respectively. This corresponds to PTE values of $\sim 0.01$. Given the criteria of goodness-of-fit, we conclude that our best-fit models all fit the measurement well. From the MCMC chains, we conclude consistent constraints on the most star-forming efficient halo mass around $M_{\mathrm{peak}}\sim 10^{11.8}M_{\odot}$ which is slightly lower than that from \cite{Yan_2022}. We also note that the constraints on the SFR parameters are not as tight as \cite{Yan_2022}, despite the larger sky coverage of \unwise than KiDS. According to the discussion in \cite{Yan_2022}, we attribute this to the lack of resolution in redshifts given only three tomographic bins. In this work, we assume an evolving $M_{\mathrm{peak}}$ model, and we find no significant evolution. In the \unwise redshift range, the SFRD calculated from the MCMC chains agrees with previous measurements given by multi-wavelength observations at a 2-sigma level. In addition, we measure the SED parameters for the S12 and Y23 models and get identical constraints. Our measurements favor a lower and flatter dust temperature and a higher power index than that given by CIB power spectra, suggesting varieties in dust across galaxies. The HOD parameters are only mildly constrained due to low angular resolution. The derived linear galaxy bias shows an increasing trend along redshifts, meaning that IR galaxies are more clustered at high redshifts. We find good agreement between our measurement, and \cite{Krolewski_2020} and \cite{Krolewski_2021} which measures cross- and auto-correlation with the same galaxy sample. This suggests that the spatial distribution of CIB accurately traces that of galaxies, and we can use CIB measured from diffuse extragalactic emission as an independent probe of galaxy abundance.

Although CIB cross-correlations can constrain model parameters, the overall SED and SFR cannot be constrained simultaneously without introducing external information. For example, we introduce SFRD at $z=0$ to constrain the SFRD and SED with the S12 and Y23 models; we adopt the SED given by \cite{bethermin_redshift_2013} in the M21 model; we adopt the BAR model given by \cite{fakhouri2010merger} in the M21 and Y23 models. This addresses the importance of multi-tracer and multi-probe methods in modern observational cosmology. This consistency of SED between our constraints and that given by \cite{bethermin_redshift_2013} is particularly nontrivial. After fixing the amplitude of SFRD with $\rho_0$ from \cite{madau_cosmic_2014}, our measurements agree with the reference SED in the frequency and redshift dependence as well as the overall amplitude. Given that $\rho_0$ from \cite{madau_cosmic_2014} is obtained from a synthesis of several multi-wavelength studies of SFR, and the reference SED is from stacking the flux from IR observations, our measurement on CIB-galaxy cross-correlations build a ``bridge'' between these two lines of independent probes, and the agreement of SEDs indicates that they are both valid and consistent.

Our findings reveal significant consistencies in measurements and constraints across various models, datasets, and probes, highlighting the complexity of the CIB. This result is not trivial, as it indicates that our comprehension of stars, dust, and galaxy clustering and abundance is approaching a unified understanding. An LSS tracer like CIB that depends on all of these ingredients has been proven to be a comprehensive tool for testing this picture. Previous studies, including \citep{shang_improved_2012, 2014planckxxx, Serra_2014, maniyar_star_2018, maniyar_simple_2021, Yan_2022, Jego_2023}, have studied part of these ingredients with CIB auto and cross-correlations. By probing those ingredients simultaneously with CIB cross-correlations, this work summarizes this line of cosmological and astrophysical studies and showcases the rich information that we can dig out from the cutting-edge datasets and models for the CIB and galaxies.

{Nevertheless, this study is subject to certain limitations. Our models rely on several simplified assumptions due to constraints such as limited sky coverage, survey depth, angular resolution, and the number of frequency bands available. For example, we assume that the SFR-infrared luminosity connection can be described by a single Kennicutt parameter in the IR bands \citep{kennicutt1998star}, which can be alternatively modeled by treating IR and UV components separately. We have also assumed that SFR has the same mass dependence for central and satellite galaxies. We take simplified SED and SFR models from previous studies by assuming that the CIB is dominated by star-forming galaxies and our robustness tests indicate that the constraints of SED and SFR are immune to selections of the galaxy sample. However, investigating the contributions to the cosmic infrared background (CIB) from various sources such as quasars, galaxies with different morphologies, and quenched galaxies holds significant interest. This endeavor will provide insights into the dust component as well as the formation and evolution of diverse galaxy types. Furthermore, despite the satisfactory goodness-of-fit achieved by our models, the low PTE of the best-fit model suggests a need for further improvements in model accuracy. In the near future, the fourth-generation galaxy surveys: the \textit{Euclid} survey \cite{2020euclidprep}, the Legacy Survey of Space and Time (LSST) based at Vera Rubin Observatory \cite{lsstsciencecollaboration2009lsst}, and Chinese Space Station Optical Survey (CSS-OS) \cite{2019ApJ...883..203G} will observe galaxies in a wide coverage of the sky into redshift $z\sim 3$. These datasets, combined with ground-based infrared observations with higher angular resolution such as CCAT-prime \cite{stacey2018ccatprime} and the Simons Observatory \cite{Ade_2019}, will yield more significant cross-correlation measurements, underscoring the need for more sophisticated models derived from multi-wavelength observations and hydrodynamic simulations. The enhancements in measurements and models hold promise for further refining our comprehension of star formation, dust properties, and galaxy abundance across various galaxy populations and different inter/intragalactic environments over extended cosmic time periods.}

\acknowledgments

ZY acknowledges support from the Max Planck Society and the Alexander von Humboldt Foundation in the framework of the Max Planck-Humboldt Research Award endowed by the Federal Ministry of Education and Research (Germany). LVW acknowledges the support from NSERC. We thank Dr. Alex Krolewski for sharing with us the \unwise galaxy map, redshift distribution, and galaxy bias from \cite{Krolewski_2020}. We thank Gerrit Farren, Aleksandra Kusiak Dr. David Alonso, and colleagues at the German Centre for Cosmological Lensing for useful discussions. We also thank the anonymous referee for their comments.

The data in this paper is analyzed with open-source python packages \texttt{numpy} \citep{harris2020array}, \texttt{scipy} \citep{2020SciPy-NMeth}, \texttt{astropy} \citep{astropy:2018}, \texttt{matplotlib} \citep{Hunter:2007}, \texttt{healpy} \citep{Zonca2019}, \texttt{NaMaster} \citep{2019namaster}, \texttt{CCL} \citep{Chisari_2019}, \texttt{emcee} \citep{Foreman_Mackey_2013}, and \texttt{GetDist} \citep{Lewis:2019xzd}.
We also use \texttt{WebPlotDigitizer} \citep{Rohatgi2020} to digitize some external data from plots in the literature.

\appendix

\section{The full MCMC results}
\label{append:post}
In this appendix, we present readers with the full MCMC results with the three models. For each model, the results are presented with a triangle plot showing the posterior and a table showing the constraints.

\subsection{The S12 model}

The constraints of S12 model parameters are summarized in Figure.~\ref{fig:S12_post} and Table.~\ref{tab:S12_constraints}.

\begin{figure}
    \centering
    \includegraphics[width=\textwidth]{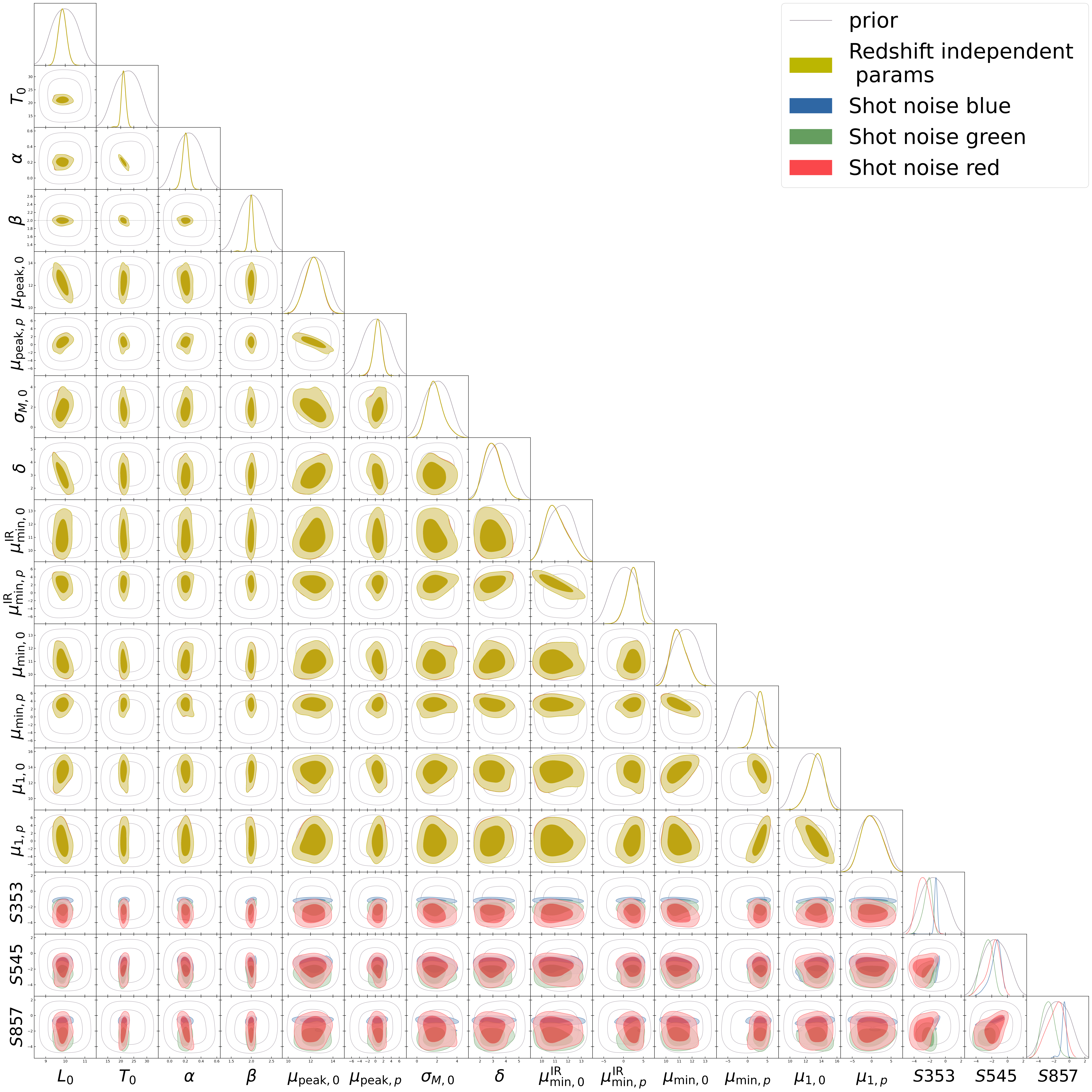}
    \caption{The posterior of all the S12 model parameters. Contours show the marginalized 2-D posterior distributions and curves show the marginalized posterior for each parameter. The shot noise parameters are color-coded according to the galaxy samples that they correspond to, and other parameters are shown with yellow contours.}
    \label{fig:S12_post}
\end{figure}

\begin{table}
    \centering
\begin{tabular}{llc} \\ 
\toprule 
 Parameters & priors& S12 \\ 
 \hline 
$\mu_{\mathrm{peak},0}$ & [10, 14] & ${12.25}^{+0.77}_{-0.71}$\\ 
$\mu_{\mathrm{peak},p}$ & [-5, 5] & ${0.57}^{+1.04}_{-0.91}$\\ 
$\sigma_{M,0}$ & [0.1, 4.0] & ${1.8}^{+0.68}_{-0.89}$\\ 
$T_{0}$ & [15, 30] & ${21.09}^{+0.82}_{-0.86}$\\ 
$L_{0}$ & [2, 4] & ${9.86}^{+0.21}_{-0.22}$\\ 
$\alpha$ & [0.0, 0.5] & ${0.2}^{+0.04}_{-0.04}$\\ 
$\beta$ & [1.5, 2.5] & ${1.99}^{+0.06}_{-0.05}$\\ 
$\delta$ & [2, 5] & ${2.98}^{+0.61}_{-0.7}$\\ 
$\mu_{\mathrm{min},0}^{\mathrm{IR}}$ & [10, 13] & ${11.1}^{+0.67}_{-0.98}$\\ 
$\mu_{\mathrm{min},p}^{\mathrm{IR}}$ & [-5, 5] & ${2.21}^{+1.68}_{-1.26}$\\ 
$\mu_{\mathrm{min},0}$ & [10, 13] & ${10.98}^{+0.53}_{-0.69}$\\ 
$\mu_{1,0}$ & [10, 15] & ${13.39}^{+1.06}_{-0.86}$\\ 
$\mu_{\mathrm{min},p}$ & [-5, 5] & ${3.04}^{+1.32}_{-1.07}$\\ 
$\mu_{1,p}$ & [-5, 5] & ${0.0}^{+2.49}_{-2.79}$\\ 
$\log_{10}(S_{\mathrm{blue}\times353}/10^{-8}\mathrm{MJy/sr})$ & [-4, 1] & ${-1.23}^{+0.21}_{-0.15}$\\ 
$\log_{10}(S_{\mathrm{green}\times353}/10^{-8}\mathrm{MJy/sr})$ & [-4, 1] & ${-2.27}^{+0.64}_{-0.42}$\\ 
$\log_{10}(S_{\mathrm{red}\times353}/10^{-8}\mathrm{MJy/sr})$ & [-4, 1] & ${-2.83}^{+0.76}_{-0.87}$\\ 
$\log_{10}(S_{\mathrm{blue}\times545}/10^{-8}\mathrm{MJy/sr})$ & [-4, 1] & ${-1.56}^{+0.86}_{-0.45}$\\ 
$\log_{10}(S_{\mathrm{green}\times545}/10^{-8}\mathrm{MJy/sr})$ & [-4, 1] & ${-2.63}^{+0.81}_{-0.72}$\\ 
$\log_{10}(S_{\mathrm{red}\times545}/10^{-8}\mathrm{MJy/sr})$ & [-4, 1] & ${-1.79}^{+1.04}_{-0.64}$\\ 
$\log_{10}(S_{\mathrm{blue}\times857}/10^{-8}\mathrm{MJy/sr})$ & [-4, 1] & ${-0.7}^{+0.32}_{-0.2}$\\ 
$\log_{10}(S_{\mathrm{green}\times857}/10^{-8}\mathrm{MJy/sr})$ & [-4, 1] & ${-2.72}^{+0.81}_{-0.87}$\\ 
$\log_{10}(S_{\mathrm{red}\times857}/10^{-8}\mathrm{MJy/sr})$ & [-4, 1] & ${-1.76}^{+1.28}_{-0.92}$\\ 
 \bottomrule 
\end{tabular}
    \caption{A summary of the prior ranges, the marginalized mean values, and the 68\% regions of the S12 model parameters. {The values and errors are calculated from the posteriors marginalized over all the other parameters.}}
    \label{tab:S12_constraints}
\end{table}

\subsection{The M13 model}
\label{append:M13}
The constraints of M21 model parameters are summarized in Figure.~\ref{fig:M21_post} and Table.~\ref{tab:M21_constraints}.

\begin{figure}
    \centering
    \includegraphics[width=\textwidth]{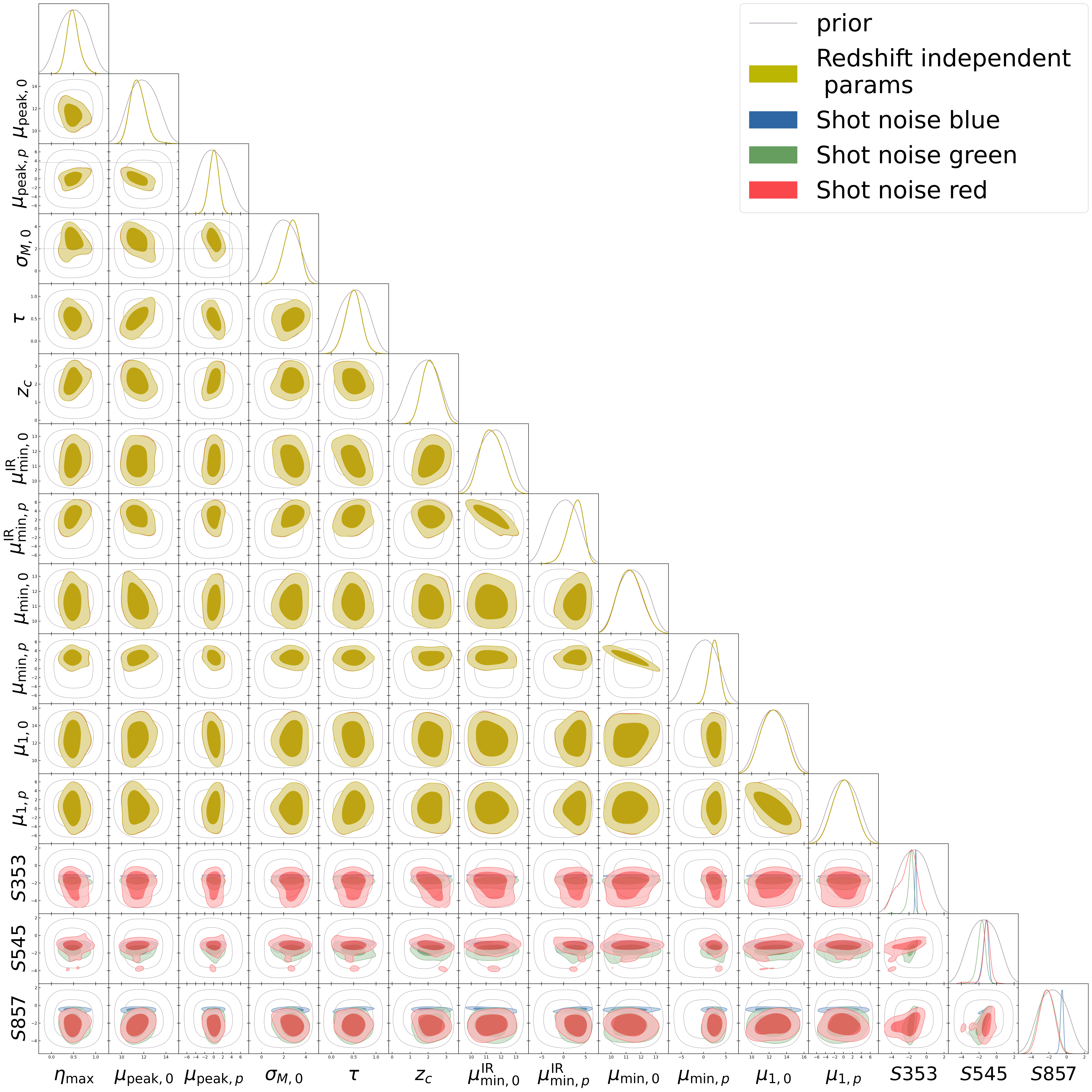}
    \caption{The posterior of all the M21 model parameters. Contours show the marginalized 2-D posterior distributions and curves show the marginalized posterior for each parameter. The shot noise parameters are color-coded according to the galaxy samples that they correspond to, and other parameters are shown with yellow contours.}
    \label{fig:M21_post}
\end{figure}

\begin{table}
    \centering
\begin{tabular}{llc} \\ 
\toprule 
 Parameters & priors& M21 \\ 
 \hline 
$\eta_{\mathrm{max}}$ & [0.1, 1.0] & ${0.49}^{+0.12}_{-0.16}$\\ 
$\mu_{\mathrm{peak},0}$ & [10, 14] & ${11.52}^{+0.56}_{-0.71}$\\ 
$\mu_{\mathrm{peak},p}$ & [-5, 5] & ${-0.02}^{+1.05}_{-0.98}$\\ 
$\sigma_{M,0}$ & [0.1, 4.0] & ${2.74}^{+0.78}_{-0.68}$\\ 
$\tau$ & [0.0, 1.0] & ${0.5}^{+0.18}_{-0.17}$\\ 
$z_{c}$ & [0.5, 3.0] & ${2.15}^{+0.46}_{-0.51}$\\ 
$\mu_{\mathrm{min},0}^{\mathrm{IR}}$ & [10, 13] & ${11.38}^{+0.72}_{-0.79}$\\ 
$\mu_{\mathrm{min},p}^{\mathrm{IR}}$ & [-5, 5] & ${2.6}^{+2.04}_{-1.56}$\\ 
$\mu_{\mathrm{min},0}$ & [10, 13] & ${11.26}^{+0.77}_{-0.84}$\\ 
$\mu_{1,0}$ & [10, 15] & ${12.5}^{+1.36}_{-1.35}$\\ 
$\mu_{\mathrm{min},p}$ & [-5, 5] & ${2.43}^{+1.19}_{-1.14}$\\ 
$\mu_{1,p}$ & [-5, 5] & ${0.12}^{+2.39}_{-2.44}$\\ 
$\log_{10}(S_{\mathrm{blue}\times353}/10^{-8}\mathrm{MJy/sr})$ & [-4, 1] & ${-1.22}^{+0.08}_{-0.08}$\\ 
$\log_{10}(S_{\mathrm{green}\times353}/10^{-8}\mathrm{MJy/sr})$ & [-4, 1] & ${-1.8}^{+0.36}_{-0.22}$\\ 
$\log_{10}(S_{\mathrm{red}\times353}/10^{-8}\mathrm{MJy/sr})$ & [-4, 1] & ${-2.16}^{+1.29}_{-0.74}$\\ 
$\log_{10}(S_{\mathrm{blue}\times545}/10^{-8}\mathrm{MJy/sr})$ & [-4, 1] & ${-1.24}^{+0.35}_{-0.25}$\\ 
$\log_{10}(S_{\mathrm{green}\times545}/10^{-8}\mathrm{MJy/sr})$ & [-4, 1] & ${-1.71}^{+0.48}_{-0.39}$\\ 
$\log_{10}(S_{\mathrm{red}\times545}/10^{-8}\mathrm{MJy/sr})$ & [-4, 1] & ${-1.19}^{+0.5}_{-0.31}$\\ 
$\log_{10}(S_{\mathrm{blue}\times857}/10^{-8}\mathrm{MJy/sr})$ & [-4, 1] & ${-0.53}^{+0.18}_{-0.15}$\\ 
$\log_{10}(S_{\mathrm{green}\times857}/10^{-8}\mathrm{MJy/sr})$ & [-4, 1] & ${-2.24}^{+0.94}_{-0.89}$\\ 
$\log_{10}(S_{\mathrm{red}\times857}/10^{-8}\mathrm{MJy/sr})$ & [-4, 1] & ${-2.2}^{+0.85}_{-0.81}$\\ 
 \bottomrule 
\end{tabular}
    \caption{A summary of the prior ranges, the marginalized mean values, and the 68\% regions of the M21 model parameters. {The values and errors are calculated from the posteriors marginalized over all the other parameters.}}
    \label{tab:M21_constraints}
\end{table}

\subsection{The Y23 model}

The constraints of Y23 model parameters are summarized in Figure.~\ref{fig:Y23_post} and Table.~\ref{tab:Y23_constraints}.

\begin{figure}
    \centering
    \includegraphics[width=\textwidth]{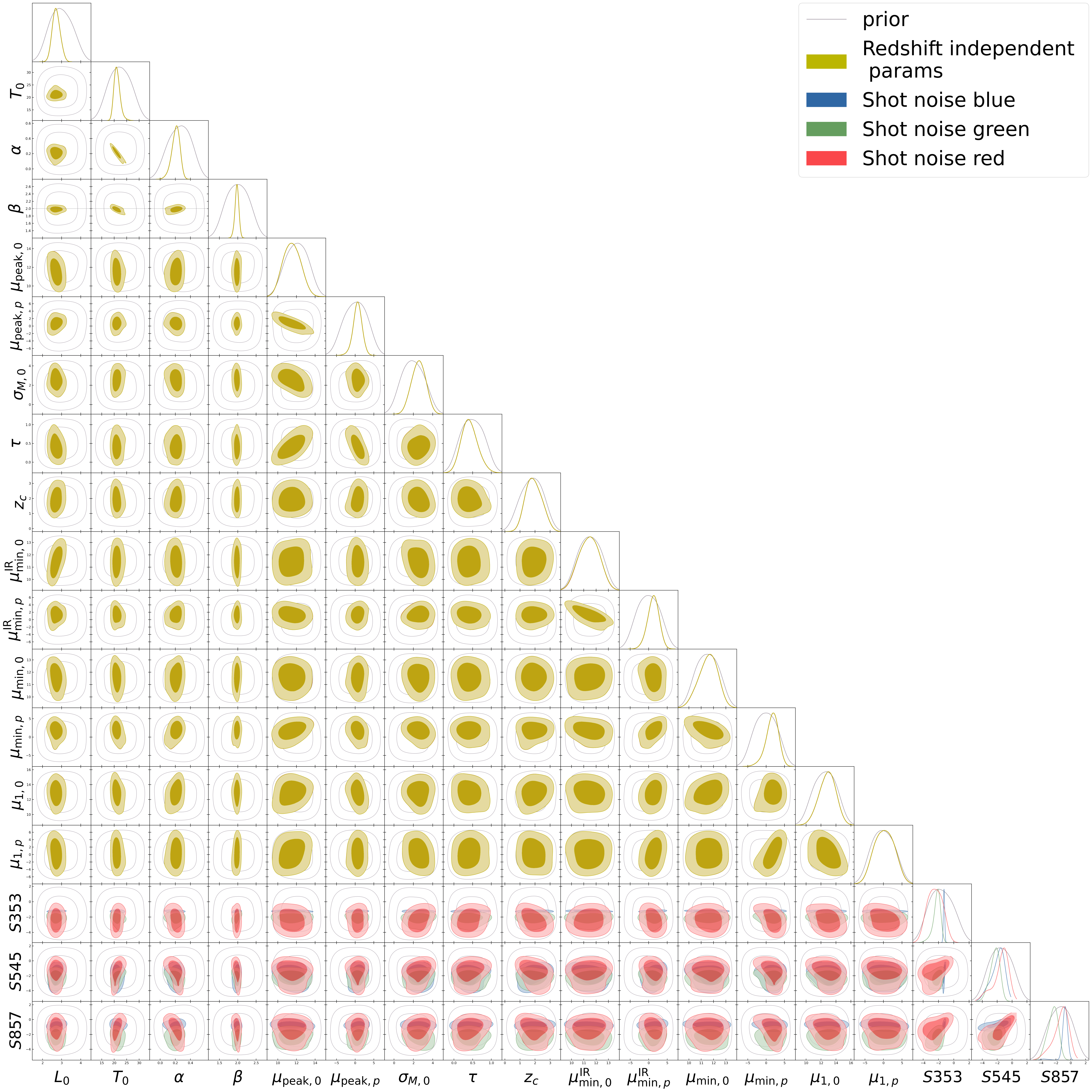}
    \caption{The posterior of all the Y23 model parameters. Contours show the marginalized 2-D posterior distributions and curves show the marginalized posterior for each parameter. The shot noise parameters are color-coded according to the galaxy samples that they correspond to, and other parameters are shown with yellow contours.}
    \label{fig:Y23_post}
\end{figure}

\begin{table}
    \centering
\begin{tabular}{llc} \\ 
\toprule 
 Parameters & priors& Y23 \\ 
 \hline 
$\mu_{\mathrm{peak},0}$ & [10, 14] & ${11.78}^{+0.73}_{-0.85}$\\ 
$\mu_{\mathrm{peak},p}$ & [-5, 5] & ${0.4}^{+1.13}_{-0.92}$\\ 
$\sigma_{M,0}$ & [0.1, 4.0] & ${2.47}^{+0.76}_{-0.76}$\\ 
$T_{0}$ & [15, 30] & ${21.13}^{+1.02}_{-1.37}$\\ 
$\tau$ & [0.0, 1.0] & ${0.45}^{+0.2}_{-0.22}$\\ 
$z_{c}$ & [0.5, 3.0] & ${1.93}^{+0.51}_{-0.6}$\\ 
$L_{0}$ & [2, 4] & ${2.7}^{+0.19}_{-0.22}$\\ 
$\alpha$ & [0.0, 0.5] & ${0.21}^{+0.06}_{-0.05}$\\ 
$\beta$ & [1.5, 2.5] & ${1.98}^{+0.06}_{-0.05}$\\ 
$\mu_{\mathrm{min},0}^{\mathrm{IR}}$ & [10, 13] & ${11.47}^{+0.86}_{-0.82}$\\ 
$\mu_{\mathrm{min},p}^{\mathrm{IR}}$ & [-5, 5] & ${1.22}^{+1.6}_{-1.39}$\\ 
$\mu_{\mathrm{min},0}$ & [10, 13] & ${11.58}^{+0.82}_{-0.69}$\\ 
$\mu_{1,0}$ & [10, 15] & ${12.96}^{+1.09}_{-0.97}$\\ 
$\mu_{\mathrm{min},p}$ & [-5, 5] & ${1.88}^{+1.75}_{-1.45}$\\ 
$\mu_{1,p}$ & [-5, 5] & ${0.43}^{+2.77}_{-2.68}$\\ 
$\log_{10}(S_{\mathrm{blue}\times353}/10^{-8}\mathrm{MJy/sr})$ & [-4, 1] & ${-1.26}^{+0.09}_{-0.09}$\\ 
$\log_{10}(S_{\mathrm{green}\times353}/10^{-8}\mathrm{MJy/sr})$ & [-4, 1] & ${-2.21}^{+0.55}_{-0.36}$\\ 
$\log_{10}(S_{\mathrm{red}\times353}/10^{-8}\mathrm{MJy/sr})$ & [-4, 1] & ${-2.49}^{+1.03}_{-1.01}$\\ 
$\log_{10}(S_{\mathrm{blue}\times545}/10^{-8}\mathrm{MJy/sr})$ & [-4, 1] & ${-1.95}^{+1.13}_{-0.67}$\\ 
$\log_{10}(S_{\mathrm{green}\times545}/10^{-8}\mathrm{MJy/sr})$ & [-4, 1] & ${-2.3}^{+0.79}_{-0.57}$\\ 
$\log_{10}(S_{\mathrm{red}\times545}/10^{-8}\mathrm{MJy/sr})$ & [-4, 1] & ${-1.5}^{+1.26}_{-0.6}$\\ 
$\log_{10}(S_{\mathrm{blue}\times857}/10^{-8}\mathrm{MJy/sr})$ & [-4, 1] & ${-0.83}^{+0.43}_{-0.26}$\\ 
$\log_{10}(S_{\mathrm{green}\times857}/10^{-8}\mathrm{MJy/sr})$ & [-4, 1] & ${-2.38}^{+0.78}_{-0.61}$\\ 
$\log_{10}(S_{\mathrm{red}\times857}/10^{-8}\mathrm{MJy/sr})$ & [-4, 1] & ${-1.34}^{+1.21}_{-0.72}$\\ 
 \bottomrule 
\end{tabular}
    \caption{A summary of the prior ranges, the marginalized mean values, and the 68\% regions of the Y23 model parameters. {The values and errors are calculated from the posteriors marginalized over all the other parameters.}}
    \label{tab:Y23_constraints}
\end{table}

\section{An alternative star formation efficiency model}

Previous research has shown that star formation activity is generally efficient within a shallow range of halo mass around $M_{\mathrm{peak}}\sim10^{12.5}M_{\odot}$\cite{2013ApJ...772...77V, bethermin_redshift_2013,2014planckxxx}, and then drops at both sides. The reason is physical. At lower masses, the gas amount is also low to maintain a high star-forming rate. In addition, the gravitational potential is shallow to balance the supernovae feedback \cite{1986ApJ...303...39D} that prevents star formation. At higher masses, energy injection from active galactic nuclei (AGN) becomes effective \cite{2008MNRAS.391..481S}, so gas cooling time becomes longer than free-falling time \cite{2005MNRAS.363....2K}, thus the star formation is suppressed. To describe this dependence, the model should include information on 1) overall star formation efficiency amplitude; 2) halo mass with the highest star formation efficiency; and 3) the decreasing slopes of star formation efficiency on both sides. In this work, we use a log-normal function for the $\eta-M$ relation as our fiducial $\eta(M)$ model. In this appendix, we use another $\eta(M)$ model proposed by \cite{2018MNRAS.477.1822M}. This model is also used in \cite{maniyar_simple_2021}, \cite{Jego_2023}, and \cite{Jego_2023_2}.

\begin{equation}
    \eta(M) = \frac{2\eta_{N}}{\left(\frac{M}{M_c}\right)^{-\beta}+\left(\frac{M}{M_c}\right)^{\gamma}},
    \label{eq:eta_M13}
\end{equation}
where $\eta_N$ is a normalization parameter, $\beta$ and $\gamma$ are the logarithmic slope at the low and high masses respectively, $M_c$ is a characteristic halo mass. We can find the most efficient halo mass:

\begin{equation}
    M_{\mathrm{peak}} = M_c\left(\frac{\beta}{\gamma} \right)^{\frac{1}{\beta+\gamma}}.
\end{equation}
We can then find the maximum efficiency $\eta_{\mathrm{max}}$ as $\eta(M_{\mathrm{peak}})$. We assume the redshift dependence of model parameters is similar to the HOD model parameters:
\begin{equation}
    X(z) = X_0 + X_p\times\frac{z}{1+z},
\end{equation}
where $X\in\{\eta_N,\beta,\gamma,M_c\}$.
In this appendix, we try to fit these SFR parameters by assuming an M21-like model, i.e., we replace the $\eta(M)$ in the M21 model with Eq.~\eqref{eq:eta_M13} and keep everything else the same, including the SED and HOD. We name it the ``M13'' model.

We fit the M13 model in the same manner as our fiducial analysis. We fix the slope parameters $\beta_0$ and $\beta_p$ for the low-mass side with the values given by \cite{2018MNRAS.477.1822M} because there are few galaxies in this halo mass range and we find that our measurements are not sensitive to them. From the posterior, we find that the parameters are degenerate, and most of them cannot be well constrained except the characteristic mass parameters, while the SFRD derived from the posterior agrees with the other models and multi-wavelength observations. In addition, the best-fit reduced $\chi^2$ is equal to 1.35, which is slightly higher than our fiducial models. Therefore, we conclude that our measurements are not suitable to constrain the SFR model with this parametrization.

%\begin{thebibliography}{99}

\bibliography{reference}{}

\providecommand{\href}[2]{#2}\begingroup\raggedright\begin{thebibliography}{100}

\bibitem{1980FCPh....5..287T}
B.M.~{Tinsley}, \emph{{Evolution of the Stars and Gas in Galaxies}},
  {\emph{Fundamentals of Cosmic Physics} {\bfseries 5} (1980) 287}.

\bibitem{2013ruppioni}
C.~Gruppioni, F.~Pozzi, G.~Rodighiero, I.~Delvecchio, S.~Berta, L.~Pozzetti
  et~al., \emph{The herschel pep/hermes luminosity function – i. probing the
  evolution of pacs selected galaxies to z $\simeq$ 4},
  \href{https://doi.org/10.1093/mnras/stt308}{\emph{MNRAS} {\bfseries 432}
  (2013) 23–52}.

\bibitem{2013magnelli}
B.~Magnelli, P.~Popesso, S.~Berta, F.~Pozzi, D.~Elbaz, D.~Lutz et~al.,
  \emph{The deepest herschel-pacs far-infrared survey: number counts and
  infrared luminosity functions from combined pep/goods-h observations},
  \href{https://doi.org/10.1051/0004-6361/201321371}{\emph{A\&A} {\bfseries
  553} (2013) A132}.

\bibitem{2016davies}
L.J.M.~Davies, S.P.~Driver, A.S.G.~Robotham, M.W.~Grootes, C.C.~Popescu,
  R.J.~Tuffs et~al., \emph{Gama/h-atlas: a meta-analysis of sfr indicators –
  comprehensive measures of the sfr–m*relation and cosmic star formation
  history atz < 0.4}, \href{https://doi.org/10.1093/mnras/stw1342}{\emph{MNRAS}
  {\bfseries 461} (2016) 458–485}.

\bibitem{2016MNRAS.456.1999M}
L.~{Marchetti}, M.~{Vaccari}, A.~{Franceschini}, V.~{Arumugam}, H.~{Aussel},
  M.~{B{\'e}thermin} et~al., \emph{{The HerMES submillimetre local and
  low-redshift luminosity functions}},
  \href{https://doi.org/10.1093/mnras/stv2717}{\emph{MNRAS} {\bfseries 456}
  (2016) 1999} [\href{https://arxiv.org/abs/1511.06167}{{\ttfamily
  1511.06167}}].

\bibitem{madau_cosmic_2014}
P.~Madau and M.~Dickinson, \emph{Cosmic {Star} {Formation} {History}},
  \href{https://doi.org/10.1146/annurev-astro-081811-125615}{\emph{Annu. Rev.
  Astron. Astrophys.} {\bfseries 52} (2014) 415}.

\bibitem{kennicutt1998star}
R.C.~Kennicutt~Jr, \emph{Star formation in galaxies along the hubble sequence},
  {\emph{ARA\&A} {\bfseries 36} (1998) 189}.

\bibitem{2007ApJ...656..770S}
J.D.T.~{Smith}, B.T.~{Draine}, D.A.~{Dale}, J.~{Moustakas}, J.~{Kennicutt},
  R.~C., G.~{Helou} et~al., \emph{{The Mid-Infrared Spectrum of Star-forming
  Galaxies: Global Properties of Polycyclic Aromatic Hydrocarbon Emission}},
  \href{https://doi.org/10.1086/510549}{\emph{ApJ} {\bfseries 656} (2007) 770}
  [\href{https://arxiv.org/abs/astro-ph/0610913}{{\ttfamily
  astro-ph/0610913}}].

\bibitem{2009sfrirg}
G.H.~Rieke, A.~Alonso-Herrero, B.J.~Weiner, P.G.~P\'erez-González,
  M.~Blaylock, J.L.~Donley et~al., \emph{Determining star formation rates for
  infrared galaxies},
  \href{https://doi.org/10.1088/0004-637x/692/1/556}{\emph{ApJ} {\bfseries 692}
  (2009) 556–573}.

\bibitem{bethermin_evolution_2015}
M.~B\'ethermin, E.~Daddi, G.~Magdis, C.~Lagos, M.~Sargent, M.~Albrecht et~al.,
  \emph{Evolution of the dust emission of massive galaxies up to \textit{z} = 4
  and constraints on their dominant mode of star formation},
  \href{https://doi.org/10.1051/0004-6361/201425031}{\emph{A\&A} {\bfseries
  573} (2015) A113}.

\bibitem{bethermin_impact_2017}
M.~Bethermin, H.-Y.~Wu, G.~Lagache, I.~Davidzon, N.~Ponthieu, M.~Cousin et~al.,
  \emph{The impact of clustering and angular resolution on far-infrared and
  millimeter continuum observations},
  \href{https://doi.org/10.1051/0004-6361/201730866}{\emph{A\&A} {\bfseries
  607} (2017) A89}.

\bibitem{Nguyen_2010}
H.T.~Nguyen, B.~Schulz, L.~Levenson, A.~Amblard, V.~Arumugam, H.~Aussel et~al.,
  \emph{{HerMES}: The {SPIRE} confusion limit},
  \href{https://doi.org/10.1051/0004-6361/201014680}{\emph{A\&A} {\bfseries
  518} (2010) L5}.

\bibitem{1967ApJ...147..868P}
R.B.~{Partridge} and P.J.E.~{Peebles}, \emph{{Are Young Galaxies Visible?}},
  \href{https://doi.org/10.1086/149079}{\emph{ApJ} {\bfseries 147} (1967) 868}.

\bibitem{1998cobecib}
E.~Dwek, R.G.~Arendt, M.G.~Hauser, D.~Fixsen, T.~Kelsall, D.~Leisawitz et~al.,
  \emph{Thecobediffuse infrared background experiment search for the cosmic
  infrared background. iv. cosmological implications},
  \href{https://doi.org/10.1086/306382}{\emph{ApJ} {\bfseries 508} (1998)
  106–122}.

\bibitem{2006spitzercib}
H.~Dole, G.~Lagache, J.-L.~Puget, K.I.~Caputi, N.~Fernández-Conde,
  E.~Le~Floc’h et~al., \emph{The cosmic infrared background resolved by
  spitzer}, \href{https://doi.org/10.1051/0004-6361:20054446}{\emph{A\&A}
  {\bfseries 451} (2006) 417–429}.

\bibitem{2010A&A...518L..30B}
S.~{Berta}, B.~{Magnelli}, D.~{Lutz}, B.~{Altieri}, H.~{Aussel}, P.~{Andreani}
  et~al., \emph{{Dissecting the cosmic infra-red background with
  Herschel/PEP}},
  \href{https://doi.org/10.1051/0004-6361/201014610}{\emph{A\&A} {\bfseries
  518} (2010) L30} [\href{https://arxiv.org/abs/1005.1073}{{\ttfamily
  1005.1073}}].

\bibitem{2014planckxxx}
{Planck Collaboration}, \emph{Planck2013 results. xxx. cosmic infrared
  background measurements and implications for star formation},
  \href{https://doi.org/10.1051/0004-6361/201322093}{\emph{A\&A} {\bfseries
  571} (2014) A30}.

\bibitem{sunyaev1972observations}
R.~Sunyaev and Y.B.~Zeldovich, \emph{The observations of relic radiation as a
  test of the nature of x-ray radiation from the clusters of galaxies},
  {\emph{Comments on Astrophysics and Space Physics} {\bfseries 4} (1972) 173}.

\bibitem{lenz_large-scale_2019}
D.~Lenz, O.~Dor\'e and G.~Lagache, \emph{Large-scale {Maps} of the {Cosmic}
  {Infrared} {Background} from {Planck}},
  \href{https://doi.org/10.3847/1538-4357/ab3c2b}{\emph{ApJ} {\bfseries 883}
  (2019) 75}.

\bibitem{2016planckcib}
{Planck Collaboration}, \emph{Planckintermediate results},
  \href{https://doi.org/10.1051/0004-6361/201629022}{\emph{A\&A} {\bfseries
  596} (2016) A109}.

\bibitem{Remazeilles_2011}
M.~Remazeilles, J.~Delabrouille and J.-F.~Cardoso, \emph{Foreground component
  separation with generalized internal linear combination},
  \href{https://doi.org/10.1111/j.1365-2966.2011.19497.x}{\emph{MNRAS}
  {\bfseries 418} (2011) 467}.

\bibitem{2020caoye}
Y.~Cao, Y.~Gong, C.~Feng, A.~Cooray, G.~Cheng and X.~Chen,
  \emph{Cross-correlation of far-infrared background anisotropies and cmb
  lensing from herschel and planck satellites},
  \href{https://doi.org/10.3847/1538-4357/abada1}{\emph{ApJ} {\bfseries 901}
  (2020) 34}.

\bibitem{maniyar_simple_2021}
A.~Maniyar, M.~B\'ethermin and G.~Lagache, \emph{Simple halo model formalism
  for the cosmic infrared background and its correlation with the thermal
  {Sunyaev} {Zel}'dovich effect},
  \href{https://doi.org/10.1051/0004-6361/202038790}{\emph{A\&A} {\bfseries
  645} (2021) A40}.

\bibitem{Yan_2022}
Z.~Yan, L.~van Waerbeke, A.H.~Wright, M.~Bilicki, S.~Gu, H.~Hildebrandt et~al.,
  \emph{Cosmic star formation history with tomographic cosmic infrared
  background-galaxy cross-correlation},
  \href{https://doi.org/10.1051/0004-6361/202243710}{\emph{Astronomy {\&}
  Astrophysics} {\bfseries 665} (2022) A52}.

\bibitem{Jego_2023}
B.~Jego, J.~Ruiz-Zapatero, C.~Garc{\'{\i} }a-Garc{\'{\i}}a, N.~Koukoufilippas
  and D.~Alonso, \emph{The star-formation history in the last 10 billion years
  from {CIB} cross-correlations},
  \href{https://doi.org/10.1093/mnras/stad213}{\emph{Monthly Notices of the
  Royal Astronomical Society} {\bfseries 520} (2023) 1895}.

\bibitem{Jego_2023_2}
B.~Jego, D.~Alonso, C.~Garc{\'{i} }a-Garc{\'{i}}a and J.~Ruiz-Zapatero,
  \emph{Constraining the physics of star formation from {CIB}-cosmic shear
  cross-correlations},
  \href{https://doi.org/10.1093/mnras/stad174}{\emph{Monthly Notices of the
  Royal Astronomical Society} {\bfseries 520} (2023) 583}.

\bibitem{shang_improved_2012}
C.~Shang, Z.~Haiman, L.~Knox and S.P.~Oh, \emph{Improved {Models} for {Cosmic}
  {Infrared} {Background} {Anisotropies}: {New} {Constraints} on the {IR}
  {Galaxy} {Population}},
  \href{https://doi.org/10.1111/j.1365-2966.2012.20510.x}{\emph{MNRAS}
  {\bfseries 421} (2012) 2832}.

\bibitem{Serra_2014}
P.~Serra, G.~Lagache, O.~Dor\'e, A.~Pullen and M.~White,
  \emph{Cross-correlation of cosmic far-infrared background anisotropies with
  large scale structures},
  \href{https://doi.org/10.1051/0004-6361/201423958}{\emph{A\&A} {\bfseries
  570} (2014) A98}.

\bibitem{maniyar_star_2018}
A.S.~Maniyar, M.~B\'ethermin and G.~Lagache, \emph{Star formation history from
  the cosmic infrared background anisotropies},
  \href{https://doi.org/10.1051/0004-6361/201732499}{\emph{A\&A} {\bfseries
  614} (2018) A39}.

\bibitem{2014schmidt}
S.J.~Schmidt, B.~M\'enard, R.~Scranton, C.B.~Morrison, M.~Rahman and
  A.M.~Hopkins, \emph{Inferring the redshift distribution of the cosmic
  infrared background},
  \href{https://doi.org/10.1093/mnras/stu2275}{\emph{MNRAS} {\bfseries 446}
  (2014) 2696–2708}.

\bibitem{Seljak_2000}
U.~Seljak, \emph{Analytic model for galaxy and dark matter clustering},
  \href{https://doi.org/10.1046/j.1365-8711.2000.03715.x}{\emph{MNRAS}
  {\bfseries 318} (2000) 203}.

\bibitem{COORAY_2002}
A.~Cooray and R.~Sheth, \emph{Halo models of large scale structure},
  \href{https://doi.org/10.1016/s0370-1573(02)00276-4}{\emph{Physics Reports}
  {\bfseries 372} (2002) 1}.

\bibitem{Zheng_2005}
Z.~Zheng, A.A.~Berlind, D.H.~Weinberg, A.J.~Benson, C.M.~Baugh, S.~Cole et~al.,
  \emph{Theoretical models of the halo occupation distribution: Separating
  central and satellite galaxies},
  \href{https://doi.org/10.1086/466510}{\emph{ApJ} {\bfseries 633} (2005) 791}.

\bibitem{salvato2018flavours}
M.~Salvato, O.~Ilbert and B.~Hoyle, \emph{The many flavours of photometric
  redshifts},  2018.

\bibitem{Lidz_2008}
A.~Lidz, O.~Zahn, S.R.~Furlanetto, M.~McQuinn, L.~Hernquist and M.~Zaldarriaga,
  \emph{Probing reionization with the 21 cm galaxy cross-power spectrum},
  \href{https://doi.org/10.1088/0004-637x/690/1/252}{\emph{ApJ} {\bfseries 690}
  (2008) 252–266}.

\bibitem{Kuntz_2015}
A.~Kuntz, \emph{Cross-correlation of cfhtlens galaxy catalogue andplanckcmb
  lensing using the halo model prescription},
  \href{https://doi.org/10.1051/0004-6361/201526940}{\emph{A\&A} {\bfseries
  584} (2015) A53}.

\bibitem{Krolewski_2020}
A.~Krolewski, S.~Ferraro, E.F.~Schlafly and M.~White, \emph{{unWISE} tomography
  of planck {CMB} lensing},
  \href{https://doi.org/10.1088/1475-7516/2020/05/047}{\emph{Journal of
  Cosmology and Astroparticle Physics} {\bfseries 2020} (2020) 047}.

\bibitem{Maniyar_2019}
A.~{Maniyar}, G.~{Lagache}, M.~{B{\'e}thermin} and S.~{Ili{\'c}},
  \emph{{Constraining cosmology with the cosmic microwave and infrared
  backgrounds correlation}},
  \href{https://doi.org/10.1051/0004-6361/201833765}{\emph{A\&A} {\bfseries
  621} (2019) A32} [\href{https://arxiv.org/abs/1809.04551}{{\ttfamily
  1809.04551}}].

\bibitem{hang2020galaxy}
Q.~{Hang}, S.~{Alam}, J.A.~{Peacock} and Y.-C.~{Cai}, \emph{{Galaxy clustering
  in the DESI Legacy Survey and its imprint on the CMB}},
  \href{https://doi.org/10.1093/mnras/staa3738}{\emph{MNRAS} {\bfseries 501}
  (2021) 1481} [\href{https://arxiv.org/abs/2010.00466}{{\ttfamily
  2010.00466}}].

\bibitem{pandey2019constraints}
S.~Pandey, E.J.~Baxter, Z.~Xu, J.~Orlowski-Scherer, N.~Zhu, A.~Lidz et~al.,
  \emph{Constraints on the redshift evolution of astrophysical feedback with
  sunyaev-zel’dovich effect cross-correlations}, {\emph{Phys. Rev. D}
  {\bfseries 100} (2019) 063519}.

\bibitem{koukoufilippas2020tomographic}
N.~Koukoufilippas, D.~Alonso, M.~Bilicki and J.A.~Peacock, \emph{Tomographic
  measurement of the intergalactic gas pressure through galaxy--tsz
  cross-correlations}, {\emph{MNRAS} {\bfseries 491} (2020) 5464}.

\bibitem{chiang2020cosmic}
Y.-K.~Chiang, R.~Makiya, B.~M{\'e}nard and E.~Komatsu, \emph{The cosmic thermal
  history probed by sunyaev--zeldovich effect tomography}, {\emph{ApJ}
  {\bfseries 902} (2020) 56}.

\bibitem{2021yanz}
Z.~Yan, L.~van Waerbeke, T.~Tröster, A.H.~Wright, D.~Alonso, M.~Asgari et~al.,
  \emph{Probing galaxy bias and intergalactic gas pressure with kids
  galaxies-tsz-cmb lensing cross-correlations},
  \href{https://doi.org/10.1051/0004-6361/202140568}{\emph{A\&A} {\bfseries
  651} (2021) A76}.

\bibitem{2017cfis}
R.A.~Ibata, A.~McConnachie, J.-C.~Cuillandre, N.~Fantin, M.~Haywood,
  N.F.~Martin et~al., \emph{The canada–france imaging survey: First results
  from the u-band component},
  \href{https://doi.org/10.3847/1538-4357/aa855c}{\emph{ApJ} {\bfseries 848}
  (2017) 128}.

\bibitem{lsstsciencecollaboration2009lsst}
{LSST Science Collaboration}, \emph{Lsst science book, version 2.0},  2009.

\bibitem{laureijs2010euclid}
R.J.~Laureijs, L.~Duvet, I.E.~Sanz, P.~Gondoin, D.H.~Lumb, T.~Oosterbroek
  et~al., \emph{The euclid mission},  in \emph{Space Telescopes and
  Instrumentation 2010: Optical, Infrared, and Millimeter Wave}, vol.~7731,
  p.~77311H, International Society for Optics and Photonics, 2010.

\bibitem{2015MNRAS.449.4476W}
L.~{Wang}, M.~{Viero}, N.P.~{Ross}, V.~{Asboth}, M.~{B{\'e}thermin}, J.~{Bock}
  et~al., \emph{{Co-evolution of black hole growth and star formation from a
  cross-correlation analysis between quasars and the cosmic infrared
  background}}, \href{https://doi.org/10.1093/mnras/stv559}{\emph{MNRAS}
  {\bfseries 449} (2015) 4476}
  [\href{https://arxiv.org/abs/1406.7181}{{\ttfamily 1406.7181}}].

\bibitem{2018hall}
K.R.~Hall, D.~Crichton, T.~Marriage, N.L.~Zakamska and R.~Mandelbaum,
  \emph{Downsizing of star formation measured from the clustered infrared
  background correlated with quasars},
  \href{https://doi.org/10.1093/mnras/sty1843}{\emph{MNRAS} {\bfseries 480}
  (2018) 149–181}.

\bibitem{2016ApJ...831...91C}
C.-C.~{Chen}, I.~{Smail}, A.M.~{Swinbank}, J.M.~{Simpson}, O.~{Almaini},
  C.J.~{Conselice} et~al., \emph{{Faint Submillimeter Galaxies Identified
  through Their Optical/Near-infrared Colors. I. Spatial Clustering and Halo
  Masses}}, \href{https://doi.org/10.3847/0004-637X/831/1/91}{\emph{ApJ}
  {\bfseries 831} (2016) 91}
  [\href{https://arxiv.org/abs/1609.00388}{{\ttfamily 1609.00388}}].

\bibitem{Schlafly_2019}
E.F.~Schlafly, A.M.~Meisner and G.M.~Green, \emph{The {unWISE} catalog: Two
  billion infrared sources from five years of wise imaging},
  \href{https://doi.org/10.3847/1538-4365/aafbea}{\emph{The Astrophysical
  Journal Supplement Series} {\bfseries 240} (2019) 30}.

\bibitem{kusiak2023enhancing}
A.~Kusiak, K.M.~Surrao and J.C.~Hill, \emph{Enhancing measurements of the cmb
  blackbody temperature power spectrum by removing cib and thermal
  sunyaev-zel'dovich contamination using external galaxy catalogs},  2023.

\bibitem{planckcosmo18}
{Planck Collaboration}, \emph{Planck 2018 results},
  \href{https://doi.org/10.1051/0004-6361/201833910}{\emph{A\&A} {\bfseries
  641} (2020) A6}.

\bibitem{limber1953analysis}
D.N.~Limber, \emph{The analysis of counts of the extragalactic nebulae in terms
  of a fluctuating density field.}, {\emph{ApJ} {\bfseries 117} (1953) 134}.

\bibitem{mead2021hmcode2020}
A.~Mead, S.~Brieden, T.~Tr{\"o}ster and C.~Heymans, \emph{Hmcode-2020: Improved
  modelling of non-linear cosmological power spectra with baryonic feedback},
  2021.

\bibitem{Tinker_2008}
J.~Tinker, A.V.~Kravtsov, A.~Klypin, K.~Abazajian, M.~Warren, G.~Yepes et~al.,
  \emph{Toward a halo mass function for precision cosmology: The limits of
  universality}, \href{https://doi.org/10.1086/591439}{\emph{ApJ} {\bfseries
  688} (2008) 709}.

\bibitem{Tinker_2010}
J.L.~Tinker, B.E.~Robertson, A.V.~Kravtsov, A.~Klypin, M.S.~Warren, G.~Yepes
  et~al., \emph{The large-scale bias of dark matter halos: Numerical
  calibration and model tests},
  \href{https://doi.org/10.1088/0004-637x/724/2/878}{\emph{ApJ} {\bfseries 724}
  (2010) 878}.

\bibitem{Peacock_2000}
J.A.~Peacock and R.E.~Smith, \emph{Halo occupation numbers and galaxy bias},
  \href{https://doi.org/10.1046/j.1365-8711.2000.03779.x}{\emph{MNRAS}
  {\bfseries 318} (2000) 1144}.

\bibitem{2012cfht}
J.~Coupon, M.~Kilbinger, H.J.~McCracken, O.~Ilbert, S.~Arnouts, Y.~Mellier
  et~al., \emph{Galaxy clustering in the cfhtls-wide: the changing relationship
  between galaxies and haloes sincez $\sim$ 1.2},
  \href{https://doi.org/10.1051/0004-6361/201117625}{\emph{A\&A} {\bfseries
  542} (2012) A5}.

\bibitem{Ishikawa_2020}
S.~Ishikawa, N.~Kashikawa, M.~Tanaka, J.~Coupon, A.~Leauthaud, J.~Toshikawa
  et~al., \emph{The subaru {HSC} galaxy clustering with photometric redshift.
  i. dark halo masses versus baryonic properties of galaxies at $0.3\leq z \leq
  1.4$}, \href{https://doi.org/10.3847/1538-4357/abbd95}{\emph{The
  Astrophysical Journal} {\bfseries 904} (2020) 128}.

\bibitem{2013ApJ...770...57B}
P.S.~{Behroozi}, R.H.~{Wechsler} and C.~{Conroy}, \emph{{The Average Star
  Formation Histories of Galaxies in Dark Matter Halos from z = 0-8}},
  \href{https://doi.org/10.1088/0004-637X/770/1/57}{\emph{ApJ} {\bfseries 770}
  (2013) 57} [\href{https://arxiv.org/abs/1207.6105}{{\ttfamily 1207.6105}}].

\bibitem{van_den_Bosch_2013}
F.C.~van~den Bosch, S.~More, M.~Cacciato, H.~Mo and X.~Yang, \emph{Cosmological
  constraints from a combination of galaxy clustering and lensing – i.
  theoretical framework},
  \href{https://doi.org/10.1093/mnras/sts006}{\emph{MNRAS} {\bfseries 430}
  (2013) 725–746}.

\bibitem{Navarro_1996}
J.F.~Navarro, C.S.~Frenk and S.D.M.~White, \emph{The structure of cold dark
  matter halos}, \href{https://doi.org/10.1086/177173}{\emph{ApJ} {\bfseries
  462} (1996) 563}.

\bibitem{Duffy_2008}
A.R.~Duffy, J.~Schaye, S.T.~Kay and C.~Dalla~Vecchia, \emph{Dark matter halo
  concentrations in thewilkinson microwave anisotropy probeyear 5 cosmology},
  \href{https://doi.org/10.1111/j.1745-3933.2008.00537.x}{\emph{MNRAS: Letters}
  {\bfseries 390} (2008) L64–L68}.

\bibitem{1998tx19.confE.533O}
R.~{Opher}, \emph{{What is the Virial Radius of a Galaxy Cluster?}},  in
  \emph{19th Texas Symposium on Relativistic Astrophysics and Cosmology},
  J.~{Paul}, T.~{Montmerle} and E.~{Aubourg}, eds., p.~533, Dec., 1998.

\bibitem{2003PASP..115..763C}
G.~{Chabrier}, \emph{{Galactic Stellar and Substellar Initial Mass Function}},
  \href{https://doi.org/10.1086/376392}{\emph{PASP} {\bfseries 115} (2003) 763}
  [\href{https://arxiv.org/abs/astro-ph/0304382}{{\ttfamily
  astro-ph/0304382}}].

\bibitem{2013ApJ...772...77V}
M.P.~{Viero}, L.~{Wang}, M.~{Zemcov}, G.~{Addison}, A.~{Amblard}, V.~{Arumugam}
  et~al., \emph{{HerMES: Cosmic Infrared Background Anisotropies and the
  Clustering of Dusty Star-forming Galaxies}},
  \href{https://doi.org/10.1088/0004-637X/772/1/77}{\emph{ApJ} {\bfseries 772}
  (2013) 77} [\href{https://arxiv.org/abs/1208.5049}{{\ttfamily 1208.5049}}].

\bibitem{2008MNRAS.383..615N}
E.~{Neistein} and A.~{Dekel}, \emph{{Constructing merger trees that mimic
  N-body simulations}},
  \href{https://doi.org/10.1111/j.1365-2966.2007.12570.x}{\emph{MNRAS}
  {\bfseries 383} (2008) 615}
  [\href{https://arxiv.org/abs/0708.1599}{{\ttfamily 0708.1599}}].

\bibitem{2009ApJ...703..785D}
A.~{Dekel}, R.~{Sari} and D.~{Ceverino}, \emph{{Formation of Massive Galaxies
  at High Redshift: Cold Streams, Clumpy Disks, and Compact Spheroids}},
  \href{https://doi.org/10.1088/0004-637X/703/1/785}{\emph{ApJ} {\bfseries 703}
  (2009) 785} [\href{https://arxiv.org/abs/0901.2458}{{\ttfamily 0901.2458}}].

\bibitem{Li_2010}
I.H.~Li, K.~Glazebrook, D.~Gilbank, M.~Balogh, R.~Bower, I.~Baldry et~al.,
  \emph{Dependence of star formation activity on stellar mass and environment
  from the redshift one {LDSS}-3 emission line survey},
  \href{https://doi.org/10.1111/j.1365-2966.2010.17816.x}{\emph{Monthly Notices
  of the Royal Astronomical Society} {\bfseries 411} (2010) 1869}.

\bibitem{Blain_2003}
A.W.~Blain, V.E.~Barnard and S.C.~Chapman, \emph{Submillimetre and far-infrared
  spectral energy distributions of galaxies: the luminosity-temperature
  relation and consequences for photometric redshifts},
  \href{https://doi.org/10.1046/j.1365-8711.2003.06086.x}{\emph{Monthly Notices
  of the Royal Astronomical Society} {\bfseries 338} (2003) 733}.

\bibitem{2010ApJ...718..632H}
N.R.~{Hall}, R.~{Keisler}, L.~{Knox}, C.L.~{Reichardt}, P.A.R.~{Ade},
  K.A.~{Aird} et~al., \emph{{Angular Power Spectra of the Millimeter-wavelength
  Background Light from Dusty Star-forming Galaxies with the South Pole
  Telescope}}, \href{https://doi.org/10.1088/0004-637X/718/2/632}{\emph{ApJ}
  {\bfseries 718} (2010) 632}
  [\href{https://arxiv.org/abs/0912.4315}{{\ttfamily 0912.4315}}].

\bibitem{2003MNRAS.345..349B}
Y.~{Birnboim} and A.~{Dekel}, \emph{{Virial shocks in galactic haloes?}},
  \href{https://doi.org/10.1046/j.1365-8711.2003.06955.x}{\emph{MNRAS}
  {\bfseries 345} (2003) 349}
  [\href{https://arxiv.org/abs/astro-ph/0302161}{{\ttfamily
  astro-ph/0302161}}].

\bibitem{2005MNRAS.363....2K}
D.~{Kere{\v{s}}}, N.~{Katz}, D.H.~{Weinberg} and R.~{Dav{\'e}}, \emph{{How do
  galaxies get their gas?}},
  \href{https://doi.org/10.1111/j.1365-2966.2005.09451.x}{\emph{MNRAS}
  {\bfseries 363} (2005) 2}
  [\href{https://arxiv.org/abs/astro-ph/0407095}{{\ttfamily
  astro-ph/0407095}}].

\bibitem{1986ApJ...303...39D}
A.~{Dekel} and J.~{Silk}, \emph{{The Origin of Dwarf Galaxies, Cold Dark
  Matter, and Biased Galaxy Formation}},
  \href{https://doi.org/10.1086/164050}{\emph{ApJ} {\bfseries 303} (1986) 39}.

\bibitem{1996ApJ...465..608T}
A.A.~{Thoul} and D.H.~{Weinberg}, \emph{{Hydrodynamic Simulations of Galaxy
  Formation. II. Photoionization and the Formation of Low-Mass Galaxies}},
  \href{https://doi.org/10.1086/177446}{\emph{ApJ} {\bfseries 465} (1996) 608}
  [\href{https://arxiv.org/abs/astro-ph/9510154}{{\ttfamily
  astro-ph/9510154}}].

\bibitem{2010tinkershmf}
J.L.~Tinker and A.R.~Wetzel, \emph{What does clustering tell us about the
  buildup of the red sequence?},
  \href{https://doi.org/10.1088/0004-637x/719/1/88}{\emph{ApJ} {\bfseries 719}
  (2010) 88–103}.

\bibitem{fakhouri2010merger}
O.~Fakhouri, C.-P.~Ma and M.~Boylan-Kolchin, \emph{The merger rates and mass
  assembly histories of dark matter haloes in the two millennium simulations},
  {\emph{MNRAS} {\bfseries 406} (2010) 2267}.

\bibitem{bethermin_redshift_2013}
M.~B\'ethermin, L.~Wang, O.~Dor\'e, G.~Lagache, M.~Sargent, E.~Daddi et~al.,
  \emph{The redshift evolution of the distribution of star formation among dark
  matter halos as seen in the infrared},
  \href{https://doi.org/10.1051/0004-6361/201321688}{\emph{A\&A} {\bfseries
  557} (2013) A66}.

\bibitem{Galametz_2014}
M.~Galametz, M.~Albrecht, R.~Kennicutt, G.~Aniano, F.~Bertoldi, D.~Calzetti
  et~al., \emph{Dissecting the origin of the submillimetre emission in nearby
  galaxies with herschel and {LABOCA}},
  \href{https://doi.org/10.1093/mnras/stu113}{\emph{Monthly Notices of the
  Royal Astronomical Society} {\bfseries 439} (2014) 2542}.

\bibitem{Kusiak_2022}
A.~Kusiak, B.~Bolliet, A.~Krolewski and J.C.~Hill, \emph{Constraining the
  galaxy-halo connection of infrared-selected unwise galaxies with galaxy
  clustering and galaxy-cmb lensing power spectra},
  \href{https://doi.org/10.1103/physrevd.106.123517}{\emph{Physical Review D}
  {\bfseries 106} (2022) }.

\bibitem{2018shot}
L.~Wolz, S.G.~Murray, C.~Blake and J.S.~Wyithe, \emph{Intensity mapping
  cross-correlations ii: Hi halo models including shot noise},
  \href{https://doi.org/10.1093/mnras/sty3142}{\emph{MNRAS} {\bfseries 484}
  (2018) 1007–1020}.

\bibitem{2012bethermin}
M.~B\'ethermin, E.~Daddi, G.~Magdis, M.T.~Sargent, Y.~Hezaveh, D.~Elbaz et~al.,
  \emph{A unified empirical model for infrared galaxy counts based on the
  observed physical evolution of distant galaxies},
  \href{https://doi.org/10.1088/2041-8205/757/2/l23}{\emph{ApJ} {\bfseries 757}
  (2012) L23}.

\bibitem{Krolewski_2021}
A.~Krolewski, S.~Ferraro and M.~White, \emph{Cosmological constraints from
  {unWISE} and planck {CMB} lensing tomography},
  \href{https://doi.org/10.1088/1475-7516/2021/12/028}{\emph{Journal of
  Cosmology and Astroparticle Physics} {\bfseries 2021} (2021) 028}.

\bibitem{Mainzer_2011}
A.~Mainzer, J.~Bauer, T.~Grav, J.~Masiero, R.M.~Cutri, J.~Dailey et~al.,
  \emph{Preliminary results from neowise: An enhancement to thewide-field
  infrared survey explorerfor solar system science},
  \href{https://doi.org/10.1088/0004-637x/731/1/53}{\emph{The Astrophysical
  Journal} {\bfseries 731} (2011) 53}.

\bibitem{2013wise.rept....1C}
R.M.~{Cutri}, E.L.~{Wright}, T.~{Conrow}, J.W.~{Fowler}, P.R.M.~{Eisenhardt},
  C.~{Grillmair} et~al., ``{Explanatory Supplement to the AllWISE Data Release
  Products}.'' Explanatory Supplement to the AllWISE Data Release Products, by
  R. M. Cutri et al., Nov., 2013.

\bibitem{Mainzer_2014}
A.~Mainzer, J.~Bauer, R.M.~Cutri, T.~Grav, J.~Masiero, R.~Beck et~al.,
  \emph{Initial performance of theneowisereactivation mission},
  \href{https://doi.org/10.1088/0004-637x/792/1/30}{\emph{The Astrophysical
  Journal} {\bfseries 792} (2014) 30}.

\bibitem{Gorski_2005}
K.M.~Gorski, E.~Hivon, A.J.~Banday, B.D.~Wandelt, F.K.~Hansen, M.~Reinecke
  et~al., \emph{Healpix: A framework for high‐resolution discretization and
  fast analysis of data distributed on the sphere},
  \href{https://doi.org/10.1086/427976}{\emph{ApJ} {\bfseries 622} (2005) 759}.

\bibitem{2011planckcib}
{Planck Collaboration}, \emph{Planckearly results. xviii. the power spectrum of
  cosmic infrared background anisotropies},
  \href{https://doi.org/10.1051/0004-6361/201116461}{\emph{A\&A} {\bfseries
  536} (2011) A18}.

\bibitem{2015ApJ...809..153M}
P.G.~{Martin}, K.P.M.~{Blagrave}, F.J.~{Lockman}, D.~{Pinheiro
  Gon{\c{c}}alves}, A.I.~{Boothroyd}, G.~{Joncas} et~al., \emph{{GHIGLS: H I
  Mapping at Intermediate Galactic Latitude Using the Green Bank Telescope}},
  \href{https://doi.org/10.1088/0004-637X/809/2/153}{\emph{ApJ} {\bfseries 809}
  (2015) 153} [\href{https://arxiv.org/abs/1504.07723}{{\ttfamily
  1504.07723}}].

\bibitem{McClure_Griffiths_2009}
N.M.~McClure-Griffiths, D.J.~Pisano, M.R.~Calabretta, H.A.~Ford, F.J.~Lockman,
  L.~Staveley-Smith et~al., \emph{Gass: The parkes galactic all-sky survey. i.
  survey description, goals, and initial data release},
  \href{https://doi.org/10.1088/0067-0049/181/2/398}{\emph{The Astrophysical
  Journal Supplement Series} {\bfseries 181} (2009) 398–412}.

\bibitem{2010ApJS..188..488W}
B.~{Winkel}, P.M.W.~{Kalberla}, J.~{Kerp} and L.~{Fl{\"o}er}, \emph{{The
  Effelsberg-Bonn H I Survey: Data Reduction}},
  \href{https://doi.org/10.1088/0067-0049/188/2/488}{\emph{ApJS} {\bfseries
  188} (2010) 488} [\href{https://arxiv.org/abs/1005.4604}{{\ttfamily
  1005.4604}}].

\bibitem{mak2017measurement}
D.S.Y.~Mak, A.~Challinor, G.~Efstathiou and G.~Lagache, \emph{Measurement of
  cib power spectra over large sky areas from planck hfi maps}, {\emph{MNRAS}
  {\bfseries 466} (2017) 286}.

\bibitem{reischke2020information}
R.~Reischke, V.~Desjacques and S.~Zaroubi, \emph{The information content of
  cosmic infrared background anisotropies}, {\emph{MNRAS} {\bfseries 491}
  (2020) 1079}.

\bibitem{Yan_2019}
Z.~Yan, A.~Hojjati, T.~Tröster, G.~Hinshaw and L.v.~Waerbeke, \emph{An
  assessment of contamination in the thermal-sz map using cross-correlations},
  \href{https://doi.org/10.3847/1538-4357/ab40b2}{\emph{ApJ} {\bfseries 884}
  (2019) 139}.

\bibitem{chluba2017rethinking}
J.~Chluba, J.C.~Hill and M.H.~Abitbol, \emph{Rethinking cmb foregrounds:
  systematic extension of foreground parametrizations}, {\emph{MNRAS}
  {\bfseries 472} (2017) 1195}.

\bibitem{Maniyar_2019_isw}
A.~{Maniyar}, G.~{Lagache}, M.~{B{\'e}thermin} and S.~{Ili{\'c}},
  \emph{{Constraining cosmology with the cosmic microwave and infrared
  backgrounds correlation}},
  \href{https://doi.org/10.1051/0004-6361/201833765}{\emph{A \& A} {\bfseries
  621} (2019) A32} [\href{https://arxiv.org/abs/1809.04551}{{\ttfamily
  1809.04551}}].

\bibitem{Alonso_2018}
D.~Alonso, J.C.~Hill, R.~Hložek and D.N.~Spergel, \emph{Measurement of the
  thermal sunyaev-zel’dovich effect around cosmic voids},
  \href{https://doi.org/10.1103/physrevd.97.063514}{\emph{Phys. Rev. D}
  {\bfseries 97} (2018) }.

\bibitem{Alonso_2019}
D.~Alonso, J.~Sanchez and A.S.~and, \emph{A unified pseudo-$c_{\ell}$
  framework}, \href{https://doi.org/10.1093/mnras/stz093}{\emph{Monthly Notices
  of the Royal Astronomical Society} {\bfseries 484} (2019) 4127}.

\bibitem{Upham_2022}
R.E.~Upham, M.L.~Brown, L.~Whittaker, A.~Amara, N.~Auricchio, D.~Bonino et~al.,
  \emph{Euclid: Covariance of weak lensing $pseudo-c_{\ell}$ estimates:
  Calculation, comparison to simulations, and dependence on survey geometry},
  \href{https://doi.org/10.1051/0004-6361/202142908}{\emph{Astronomy \&
  Astrophysics} {\bfseries 660} (2022) A114}.

\bibitem{troster2021joint}
T.~Tröster, A.J.~Mead, C.~Heymans, Z.~Yan, D.~Alonso, M.~Asgari et~al.,
  \emph{Joint constraints on cosmology and the impact of baryon feedback:
  combining kids-1000 lensing with the thermal sunyaev-zeldovich effect from
  planck and act},  2021.

\bibitem{Efstathiou_2004}
G.~Efstathiou, \emph{Myths and truths concerning estimation of power spectra:
  the case for a hybrid estimator},
  \href{https://doi.org/10.1111/j.1365-2966.2004.07530.x}{\emph{MNRAS}
  {\bfseries 349} (2004) 603}.

\bibitem{Garc_a_Garc_a_2019}
C.~García-García, D.~Alonso and E.~Bellini, \emph{Disconnected
  pseudo-$c_{\ell}$ covariances for projected large-scale structure data},
  \href{https://doi.org/10.1088/1475-7516/2019/11/043}{\emph{JCAP} {\bfseries
  2019} (2019) 043}.

\bibitem{2019namaster}
D.~Alonso, J.~Sanchez and A.~Slosar, \emph{A unified pseudo-cl framework},
  \href{https://doi.org/10.1093/mnras/stz093}{\emph{MNRAS} {\bfseries 484}
  (2019) 4127–4151}.

\bibitem{2018cibng}
E.~Schaan, S.~Ferraro and D.N.~Spergel, \emph{Weak lensing of intensity
  mapping: The cosmic infrared background},
  \href{https://doi.org/10.1103/physrevd.97.123539}{\emph{Physical Review D}
  {\bfseries 97} (2018) }.

\bibitem{Takada_2013}
M.~Takada and W.~Hu, \emph{Power spectrum super-sample covariance},
  \href{https://doi.org/10.1103/physrevd.87.123504}{\emph{Phys. Rev. D}
  {\bfseries 87} (2013) }.

\bibitem{2018ssc}
F.~Lacasa, M.~Lima and M.~Aguena, \emph{Super-sample covariance approximations
  and partial sky coverage},
  \href{https://doi.org/10.1051/0004-6361/201630281}{\emph{A\&A} {\bfseries
  611} (2018) A83}.

\bibitem{2021ssc}
K.~Osato and M.~Takada, \emph{Super sample covariance of the thermal
  sunyaev-zel’dovich effect},
  \href{https://doi.org/10.1103/physrevd.103.063501}{\emph{Physical Review D}
  {\bfseries 103} (2021) }.

\bibitem{Chisari_2019}
N.E.~Chisari, D.~Alonso, E.~Krause, C.D.~Leonard, P.~Bull, J.~Neveu et~al.,
  \emph{Core cosmology library: Precision cosmological predictions for lsst},
  \href{https://doi.org/10.3847/1538-4365/ab1658}{\emph{ApJS} {\bfseries 242}
  (2019) 2}.

\bibitem{Ando_2017}
S.~Ando, A.~Benoit-L{\'{e}}vy and E.~Komatsu, \emph{Angular power spectrum of
  galaxies in the 2mass redshift survey},
  \href{https://doi.org/10.1093/mnras/stx2634}{\emph{MNRAS} {\bfseries 473}
  (2017) 4318}.

\bibitem{2014hfi}
P.A.R.~Ade, N.~Aghanim, C.~Armitage-Caplan, M.~Arnaud, M.~Ashdown,
  F.~Atrio-Barandela et~al., \emph{Planck2013 results. ix. hfi spectral
  response}, \href{https://doi.org/10.1051/0004-6361/201321531}{\emph{Astronomy
  \& Astrophysics} {\bfseries 571} (2014) A9}.

\bibitem{bethermin2012unified}
M.~B{\'e}thermin, E.~Daddi, G.~Magdis, M.T.~Sargent, Y.~Hezaveh, D.~Elbaz
  et~al., \emph{A unified empirical model for infrared galaxy counts based on
  the observed physical evolution of distant galaxies}, {\emph{ApJL} {\bfseries
  757} (2012) L23}.

\bibitem{1989A&A...221..221S}
P.~{Schneider}, \emph{{The number excess of galaxies around high redshift
  quasars}}, {\emph{A\&A} {\bfseries 221} (1989) 221}.

\bibitem{1989ApJ...339L..53N}
R.~{Narayan}, \emph{{Gravitational Lensing and Quasar-Galaxy Correlations}},
  \href{https://doi.org/10.1086/185418}{\emph{ApJl} {\bfseries 339} (1989)
  L53}.

\bibitem{hui2007anisotropic}
L.~Hui, E.~Gaztanaga and M.~LoVerde, \emph{Anisotropic magnification distortion
  of the 3d galaxy correlation. i. real space}, {\emph{Phys. Rev. D} {\bfseries
  76} (2007) 103502}.

\bibitem{ziour2008magnification}
R.~Ziour and L.~Hui, \emph{Magnification bias corrections to galaxy-lensing
  cross-correlations}, {\emph{Phys. Rev. D} {\bfseries 78} (2008) 123517}.

\bibitem{hilbert2009ray}
S.~Hilbert, J.~Hartlap, S.~White and P.~Schneider, \emph{Ray-tracing through
  the millennium simulation: Born corrections and lens-lens coupling in cosmic
  shear and galaxy-galaxy lensing}, {\emph{A\&A} {\bfseries 499} (2009) 31}.

\bibitem{Xu_2020}
J.~Xu, M.~Sun and Y.~Xue, \emph{Agns are not that cool: Revisiting the
  intrinsic agn far-infrared spectral energy distribution},
  \href{https://doi.org/10.3847/1538-4357/ab811a}{\emph{The Astrophysical
  Journal} {\bfseries 894} (2020) 21}.

\bibitem{1993ARA&A..31..473A}
R.~{Antonucci}, \emph{{Unified models for active galactic nuclei and
  quasars.}},
  \href{https://doi.org/10.1146/annurev.aa.31.090193.002353}{\emph{ARAA}
  {\bfseries 31} (1993) 473}.

\bibitem{Laigle_2016}
C.~Laigle, H.J.~McCracken, O.~Ilbert, B.C.~Hsieh, I.~Davidzon, P.~Capak et~al.,
  \emph{The cosmos2015 catalog: Exploring the 1<z<6 universe with half a
  million galaxies},
  \href{https://doi.org/10.3847/0067-0049/224/2/24}{\emph{The Astrophysical
  Journal Supplement Series} {\bfseries 224} (2016) 24}.

\bibitem{2018ApJS..234...23A}
R.J.~{Assef}, D.~{Stern}, G.~{Noirot}, H.D.~{Jun}, R.M.~{Cutri} and
  P.R.M.~{Eisenhardt}, \emph{{The WISE AGN Catalog}},
  \href{https://doi.org/10.3847/1538-4365/aaa00a}{\emph{APJS} {\bfseries 234}
  (2018) 23} [\href{https://arxiv.org/abs/1706.09901}{{\ttfamily 1706.09901}}].

\bibitem{jauzac_cosmic_2011}
M.~Jauzac, H.~Dole, E.~Le~Floc’h, H.~Aussel, K.~Caputi, O.~Ilbert et~al.,
  \emph{The cosmic far-infrared background buildup since redshift 2 at 70 and
  160 microns in the {COSMOS} and {GOODS} fields},
  \href{https://doi.org/10.1051/0004-6361/201015432}{\emph{A\&A} {\bfseries
  525} (2011) A52}.

\bibitem{cai_hybrid_2013}
Z.-Y.~Cai, A.~Lapi, J.-Q.~Xia, G.~De~Zotti, M.~Negrello, C.~Gruppioni et~al.,
  \emph{A hybrid model for the evolution of galaxies and {Active} {Galactic}
  {Nuclei} in the infrared},
  \href{https://doi.org/10.1088/0004-637X/768/1/21}{\emph{ApJ} {\bfseries 768}
  (2013) 21}.

\bibitem{roebuck_role_2016}
E.~Roebuck, A.~Sajina, C.C.~Hayward, A.~Pope, A.~Kirkpatrick, L.~Hernquist
  et~al., \emph{{THE} {ROLE} {OF} {STAR} {FORMATION} {AND} {AGN} {IN} {DUST}
  {HEATING} {OF} \textit{{Z}} = 0.3–2.8 {Galaxies}. {II}. {INFORMING} {IR}
  {AGN} {FRACTION} {ESTIMATES} {THROUGH} {SIMULATIONS}},
  \href{https://doi.org/10.3847/1538-4357/833/1/60}{\emph{ApJ} {\bfseries 833}
  (2016) 60}.

\bibitem{xu_agns_2020}
J.~Xu, M.~Sun and Y.~Xue, \emph{{AGNs} {Are} {Not} {That} {Cool}: {Revisiting}
  the {Intrinsic} {AGN} {Far}-infrared {Spectral} {Energy} {Distribution}},
  \href{https://doi.org/10.3847/1538-4357/ab811a}{\emph{ApJ} {\bfseries 894}
  (2020) 21}.

\bibitem{Foreman_Mackey_2013}
D.~Foreman-Mackey, D.W.~Hogg, D.~Lang and J.~Goodman, \emph{emcee: The mcmc
  hammer}, \href{https://doi.org/10.1086/670067}{\emph{Publications of the
  Astronomical Society of the Pacific} {\bfseries 125} (2013) 306}.

\bibitem{abbott2018dark}
T.~Abbott, F.B.~Abdalla, A.~Alarcon, J.~Aleksi{\'c}, S.~Allam, S.~Allen et~al.,
  \emph{Dark energy survey year 1 results: Cosmological constraints from galaxy
  clustering and weak lensing}, {\emph{Phys. Rev. D} {\bfseries 98} (2018)
  043526}.

\bibitem{heymans2020kids1000}
C.~Heymans, T.~Tr{\"o}ster, M.~Asgari, C.~Blake, H.~Hildebrandt, B.~Joachimi
  et~al., \emph{Kids-1000 cosmology: Multi-probe weak gravitational lensing and
  spectroscopic galaxy clustering constraints}, {\emph{A\&A} {\bfseries 646}
  (2021) A140}.

\bibitem{2020euclidprep}
A.~Blanchard, S.~Camera, C.~Carbone, V.F.~Cardone, S.~Casas, S.~Clesse et~al.,
  \emph{Euclid preparation},
  \href{https://doi.org/10.1051/0004-6361/202038071}{\emph{A\&A} {\bfseries
  642} (2020) A191}.

\bibitem{2019ApJ...883..203G}
Y.~{Gong}, X.~{Liu}, Y.~{Cao}, X.~{Chen}, Z.~{Fan}, R.~{Li} et~al.,
  \emph{{Cosmology from the Chinese Space Station Optical Survey (CSS-OS)}},
  \href{https://doi.org/10.3847/1538-4357/ab391e}{\emph{ApJ} {\bfseries 883}
  (2019) 203} [\href{https://arxiv.org/abs/1901.04634}{{\ttfamily
  1901.04634}}].

\bibitem{stacey2018ccatprime}
G.J.~Stacey, M.~Aravena, K.~Basu, N.~Battaglia, B.~Beringue, F.~Bertoldi
  et~al., \emph{Ccat-prime: Science with an ultra-widefield submillimeter
  observatory at cerro chajnantor},  2018.

\bibitem{Ade_2019}
P.~Ade, J.~Aguirre, Z.~Ahmed, S.~Aiola, A.~Ali, D.~Alonso et~al., \emph{The
  simons observatory: science goals and forecasts},
  \href{https://doi.org/10.1088/1475-7516/2019/02/056}{\emph{Journal of
  Cosmology and Astroparticle Physics} {\bfseries 2019} (2019) 056–056}.

\bibitem{harris2020array}
C.R.~Harris, K.J.~Millman, S.J.~van~der Walt, R.~Gommers, P.~Virtanen,
  D.~Cournapeau et~al., \emph{Array programming with {NumPy}},
  \href{https://doi.org/10.1038/s41586-020-2649-2}{\emph{Nature} {\bfseries
  585} (2020) 357}.

\bibitem{2020SciPy-NMeth}
P.~Virtanen, R.~Gommers, T.E.~Oliphant, M.~Haberland, T.~Reddy, D.~Cournapeau
  et~al., \emph{{{SciPy} 1.0: Fundamental Algorithms for Scientific Computing
  in Python}}, \href{https://doi.org/10.1038/s41592-019-0686-2}{\emph{Nature
  Methods} {\bfseries 17} (2020) 261}.

\bibitem{astropy:2018}
{Astropy Collaboration}, A.M.~{Price-Whelan}, B.M.~{Sip{\H{o}}cz},
  H.M.~{G{\"u}nther}, P.L.~{Lim}, S.M.~{Crawford} et~al., \emph{{The Astropy
  Project: Building an Open-science Project and Status of the v2.0 Core
  Package}}, \href{https://doi.org/10.3847/1538-3881/aabc4f}{\emph{The
  Astronomical Journal} {\bfseries 156} (2018) 123}
  [\href{https://arxiv.org/abs/1801.02634}{{\ttfamily 1801.02634}}].

\bibitem{Hunter:2007}
J.D.~Hunter, \emph{Matplotlib: A 2d graphics environment},
  \href{https://doi.org/10.1109/MCSE.2007.55}{\emph{Computing in Science \&
  Engineering} {\bfseries 9} (2007) 90}.

\bibitem{Zonca2019}
A.~Zonca, L.~Singer, D.~Lenz, M.~Reinecke, C.~Rosset, E.~Hivon et~al.,
  \emph{healpy: equal area pixelization and spherical harmonics transforms for
  data on the sphere in python},
  \href{https://doi.org/10.21105/joss.01298}{\emph{Journal of Open Source
  Software} {\bfseries 4} (2019) 1298}.

\bibitem{Lewis:2019xzd}
A.~Lewis, \emph{{GetDist: a Python package for analysing Monte Carlo samples}},
   \href{https://arxiv.org/abs/1910.13970}{{\ttfamily 1910.13970}}.

\bibitem{Rohatgi2020}
A.~Rohatgi, \emph{Webplotdigitizer: Version 4.5},  2021.

\bibitem{2008MNRAS.391..481S}
R.S.~{Somerville}, P.F.~{Hopkins}, T.J.~{Cox}, B.E.~{Robertson} and
  L.~{Hernquist}, \emph{{A semi-analytic model for the co-evolution of
  galaxies, black holes and active galactic nuclei}},
  \href{https://doi.org/10.1111/j.1365-2966.2008.13805.x}{\emph{MNRAS}
  {\bfseries 391} (2008) 481}
  [\href{https://arxiv.org/abs/0808.1227}{{\ttfamily 0808.1227}}].

\bibitem{2018MNRAS.477.1822M}
B.P.~{Moster}, T.~{Naab} and S.D.M.~{White}, \emph{{EMERGE - an empirical model
  for the formation of galaxies since z {\ensuremath{\sim}} 10}},
  \href{https://doi.org/10.1093/mnras/sty655}{\emph{MNRAS} {\bfseries 477}
  (2018) 1822} [\href{https://arxiv.org/abs/1705.05373}{{\ttfamily
  1705.05373}}].

\end{thebibliography}\endgroup
\bibliographystyle{JHEP}

% Please avoid comments such as "For a review'', "For some examples",
% "and references therein" or move them in the text. In general,
% please leave only references in the bibliography and move all
% accessory text in footnotes.

% Also, please have only one work for each \bibitem.

%\end{thebibliography}
\end{document}